\documentclass[prb,amsmath,amssymb,twocolumn,superscriptaddress,longbibliography]{revtex4-2}

\usepackage[pdftex]{hyperref}
\usepackage{dcolumn}
\usepackage{bm}
\usepackage{physics}
\usepackage{color}

\usepackage{amsmath}	
\usepackage{amsfonts}
\usepackage{braket}
\usepackage{amsthm}
\usepackage[legacycolonsymbols]{mathtools}

\usepackage{bbm}

\hypersetup{
    colorlinks=true,
    linkcolor=blue,
    citecolor=blue,
    urlcolor=blue,
}

\usepackage[export]{adjustbox}

\makeatletter
\let\Caption\@makecaption
\makeatother
\usepackage{subcaption}
\makeatletter
\let\@makecaption\Caption
\makeatother

\begin{document}


\title{Stability of quasiparticle creation and multiband quantum geometry in fractional Chern insulators under magnetic fields}

\author{Nozomi Higashino}
\affiliation{Quantum Matter Program, Graduate School of Advanced Science and Engineering, Hiroshima University,
Higashihiroshima, Hiroshima 739-8530, Japan}
 
\author{Yasuhiro Tada}%
\email[]{ytada@hiroshima-u.ac.jp}
\affiliation{Quantum Matter Program, Graduate School of Advanced Science and Engineering, Hiroshima University,
Higashihiroshima, Hiroshima 739-8530, Japan}
\affiliation{Institute for Solid State Physics, University of Tokyo, Kashiwa 277-8581, Japan}




\begin{abstract}
We study creation of quasiparticles in fractional Chern insulators (FCI) under magnetic fields. 
We consider two representative models, the Kapit-Mueller model and the checkerboard model, which have distinct band properties in terms of the quantum geometry.
The former satisfies the so-called ideal condition and well mimics the lowest Landau level, while the latter is not ideal for realization of FCI states.
It is found within exact diagonalization that both quasiholes and quasielectrons are stably created by the magnetic fields in the Kapit-Mueller model. 
On the other hand, stability of the quasiparticle creation depends on directions of the magnetic field in the checkerboard model. 
Although the quasielectron creation is stable under a magnetic field, the quasihole creation and the underlying FCI state are unstable for the opposite field direction, leading to a field-induced non-FCI state. 
We point out that this difference can be understood based on the multiband quantum geometry in the presence of the magnetic fields.

\end{abstract}

\maketitle


\section{\label{sec:level1}Introduction}
A fractional Chern insulator (FCI) is a strongly interacting variant of a Chern band insulator~\cite{PhysRevX.1.021014,PhysRevLett.106.236804,PhysRevB.85.075116,PhysRevB.91.035136,PhysRevLett.106.236802,PARAMESWARAN2013816,doi:10.1142/S021797921330017X,LIU2024515}.
It exhibits fractionally quantized Hall conductance and is regarded as a zero-field analogue of the fractional quantum Hall effect (FQHE) in strong magnetic fields.
Intuitively, one could expect an FCI state if the system has some similarity to the fractional quantum Hall system such as flat topological bands well separated from other bands and strong electron interactions.
More precisely, 
it was argued that an FCI can be realized on a band that mimics the lowest Landau level (LLL). 
Such a band is characterized by a flat energy dispersion, a non-zero Chern number, and well-defined quantum geometry which describes geometrical properties of the Bloch states. 
The quantum geometry provides the conditions of momentum space holomorphicity called the ideal condition and the emergent algebra of density operators. 
The relevance of the quantum geometry for stabilization of an FCI has been extensively discussed in various systems~\cite{PhysRevLett.127.246403, PhysRevResearch.5.023167,PhysRevB.90.165139,PhysRevB.104.045103,PhysRevB.96.165150}. 
The Bloch wavefunction can also be characterized by the vortexability which measures similarity of a Bloch wavefunction to the LLL wavefunction~\cite{PhysRevB.108.205144, PhysRevB.110.245112}.
These characterizations have been used also for FCIs in multiband systems~\cite{parker2021fieldtunedzerofieldfractionalchern,PhysRevLett.134.106502,liu2024theorygeneralizedlandaulevels,schoonderwoerd2022interactiondrivenplateautransitioninteger,PhysRevB.104.115126,PhysRevB.110.165142}.
Theoretically, there are various systems for FCIs, where some of them satisfy the ideal condition and vortexability, and some do not~\cite{PhysRevResearch.2.023237,PhysRevLett.105.215303,10.21468/SciPostPhys.12.4.118,PhysRevB.96.165150}.
Such quantitative difference among the models could be related to stability of the FCIs ~\cite{parker2021fieldtunedzerofieldfractionalchern,10.21468/SciPostPhys.12.4.118,PhysRevLett.128.176404,PhysRevB.109.245111,PhysRevB.110.165142,PhysRevLett.133.156504}.

The FCIs have been observed experimentally in moire materials~\cite{doi:10.1126/science.aan8458,Xie2021,Cai2023,Zeng2023,Park2023,PhysRevX.13.031037,Lu2024}.
Especially for MoTe$_2$ and rombohedral pentalayer graphene, the fractional anomalous quantum Hall effects at exactly zero fields and stable plateaus of the Hall resistance in magnetic fields have been reported.
Although the applied magnetic fields themselves are small, the corresponding fluxes per moire unit cell can be large because the moire periods are much longer than typical lattice constants~\cite{Moon2012,Dean2013,Ponomarenko2013,Hunt2013}.
Therefore, even small magnetic fields may have non-trivial effects on FCIs, especially on the quantum Hall plateaus.
The existence of a plateau of the quantum Hall conductance is one of the most characteristic and fundamental properties in FQHE.
The standard understanding of this phenomenon is that
quasiparticles with fractional charges are created when the magnetic field is varied in an FQH state, but they are spatially localized by disorder and consequently the quantized Hall effect robustly remains~\cite{yoshioka2002quantum,RevModPhys.80.1083}.
Similar mechanisms would work also in FCIs in magnetic fields.
However, there are various systems for FCIs which are distinct in terms of the quantum geometry and vortexability.
It is not trivial how such differences could influence stability of the quantized Hall effect in FCIs under magnetic fields.
Especially, the magnetic field can affect the underlying band structure and, consequently the quantum geometry could change possibly leading to instability of FCIs in some systems.
This is in sharp contrast to the creation of quasiparticles by changing the electron filling of FCIs in the absence of magnetic fields, where the underlying band structures are kept intact~\cite{PhysRevX.1.021014,PhysRevB.91.045126,PhysRevB.99.045136,PhysRevB.87.205136}.

In this study, we investigate creation of quasiparticles and stability of FCI states under magnetic fields in two representative models which have distinct band properties, namely the Kapit-Mueller model and the checkerboard model~\cite{PhysRevLett.105.215303,PhysRevLett.106.236803}. 
The former model satisfies the ideal condition, while the latter does not.
We demonstrate by the exact diagonalization that the two models show different behaviors under magnetic fields. 
Both of the quasiholes and quasielectrons are stable in the Kapit-Mueller model, while the quasiholes are unstable in the checkerboard model.
The difference in the stability can be attributed to the quantum geometries of multiband structures under magnetic fields.

The remainder of this paper is organized as follows. In Sec.~\ref{sec:level2}, we introduce the quantum geometry and the two models considered in this study. 
In Sec.~\ref{sec:level3}, a summary of the key results is given. 
In Sec.~\ref{sec:level4}, we show the exact diagonalization results for both models. 
In Sec.~\ref{sec:level5}, we discuss  relation between the FCI stability and the multiband structure. 
Finally, a conclusion is given in Sec.~\ref{sec:level6}.

\section{\label{sec:level2}Models}
\subsection{Brief overview of quantum geometry}
Similarity of a Bloch band with the LLL is a key to realize an FCI, and there are various models with different degrees of the similarity~\cite{PhysRevLett.127.246403, PhysRevResearch.5.023167,PhysRevB.90.165139}.
In this study, we consider two representative models which have distinct band properties in terms of the quantum geometry.
Here, we provide a brief overview of the quantum geometry.
The quantum geometric tensor for a single band is given by
\begin{equation}
    \begin{split}
        \eta_{\alpha \beta}(\bm{k}) &= A\mel{\partial_{k_{\alpha}}u_{\bm{k}}}{1-{\mathcal P}(\bm{k})}{\partial_{k_{\beta}}u_{\bm{k}}}\\
        &= g_{\alpha \beta}(\bm{k}) - \frac{i}{2}\mathcal{F}_{\alpha \beta}(\bm{k}),
    \end{split}
\end{equation}
where $\alpha,\beta\in\{x,y\}$ and $\bm{k}$ is a wave vector in the Brillouin zone. $A$ is the area of the Brillouin zone {which has been introduced for later convenience.} 
$\ket{u_{\bm{k}}}$ is the cell-periodic function for the band which we focus on, and  ${\mathcal P}(\bm{k})=\ket{u_{\bm{k}}}\bra{u_{\bm{k}}}$ is the corresponding projection operator. 
$g(\bm{k})$ and $\mathcal{F}(\bm{k})$ are Fubini-Study metric and Berry curvature, respectively. 
They are given as real and imaginary parts of the quantum geometric tensor, and known to satisfy the following inequality,
\begin{equation}
    \tr[g(\bm{k})]\geq |\mathcal{F}_{xy}(\bm{k})|,
\end{equation}
for arbitrary $\bm{k}$ in the Brillouin zone. 
It is known that a holomorphic wavefunction in the momentum space is given by a single-particle eigenfunction which saturates the inequality and has a positive-definite Berry curvature~\cite{PhysRevLett.127.246403, PhysRevResearch.5.023167}. 
Such a holomorphic single-particle wavefunction can lead to the FCI in the presence of short-range interactions. Hence, this geometric condition is called the ideal condition and the corresponding wavefunctions are known as generalization of the LLL. 
Moreover, a single-particle band satisfying the ideal condition is classified as a vortexable band from a perspective of the vortexability which extends the ideal condition to a larger family \cite{PhysRevB.108.205144, PhysRevB.110.245112}, and it has been suggested that these frameworks can also apply to multiband systems \cite{parker2021fieldtunedzerofieldfractionalchern,PhysRevLett.134.106502,liu2024theorygeneralizedlandaulevels}. 
The mutliband quantum geometry will be discussed in Sec.~\ref{sec:level5}.

\subsection{Two lattice models}
\begin{figure}[ht]
    \centering
    \begin{subfigure}{0.47\linewidth}
        \centering
        \textbf{(a) Kapit-Mueller model}
        \includegraphics[width=\linewidth]{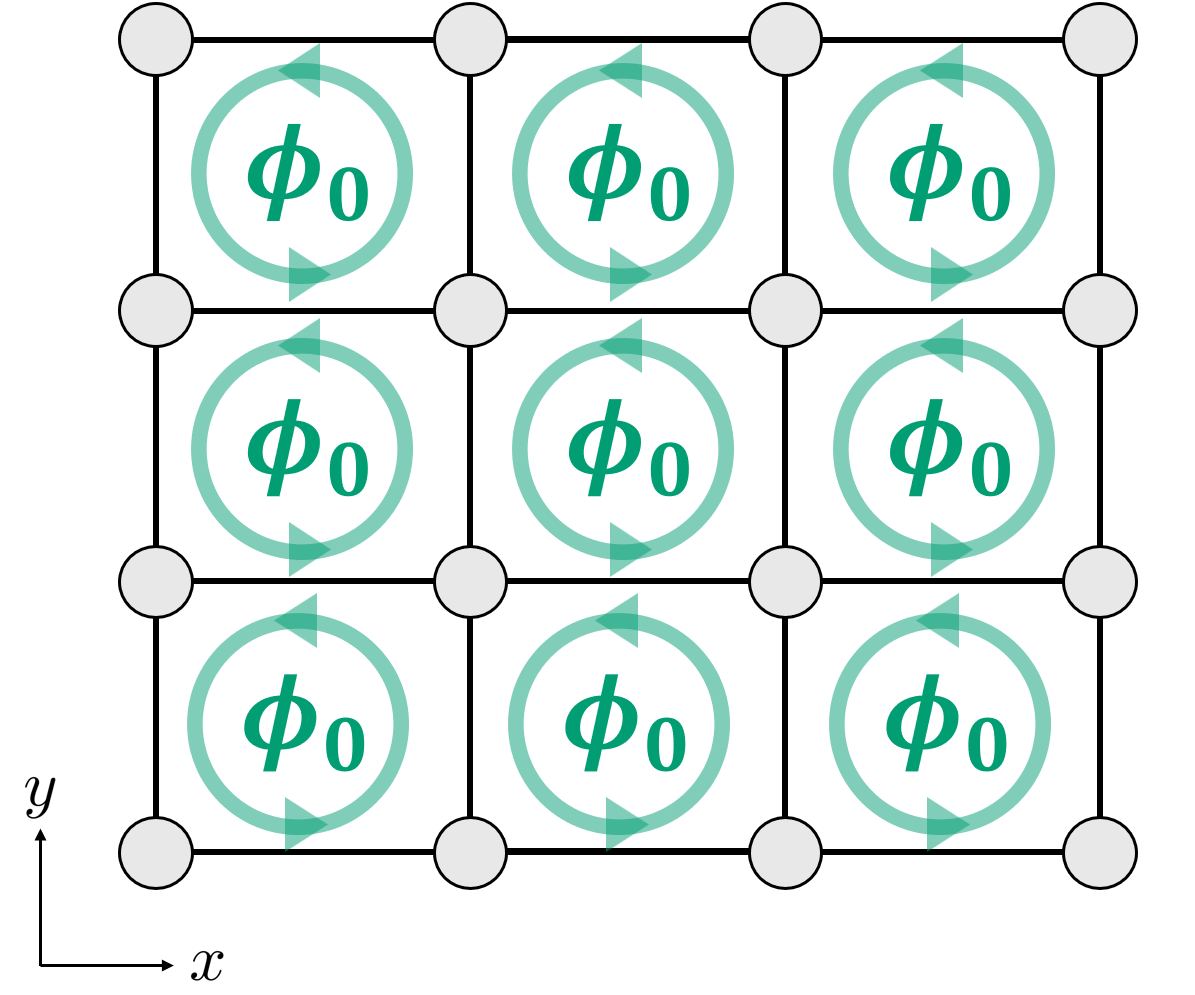}
        \label{latKM}
    \end{subfigure}
    \begin{subfigure}{0.44\linewidth}
        \centering
        \textbf{(b) Checkerboard model}
        \includegraphics[width=\linewidth]{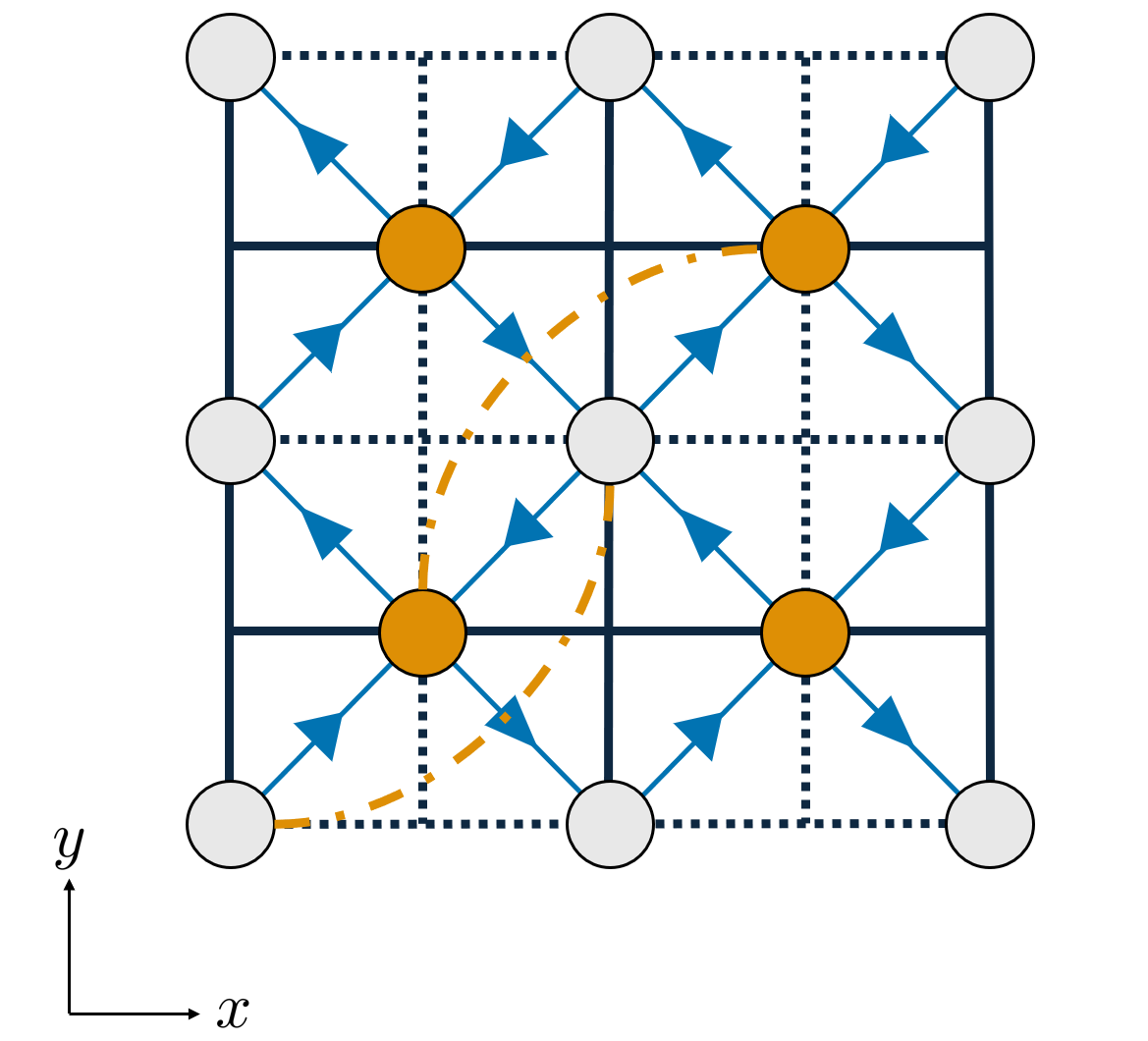}
        \label{latCB}
    \end{subfigure}
    \caption{Schematic pictures of (a) the Kapit-Muller model and (b) the checkerboard model under zero magnetic fields. 
    For the Kapit-Mueller model, $\phi_0$ is a model parameter which is distinguished from an applied magnetic field. For the checkerboard model, the direction of the arrows on the NN hopping corresponds to the sign of the phase factor $\psi_{jk}=\pm 1$. The NNN hopping strength $t_1(t_2)$ is specified by solid(dashed) lines. The NNNN hopping strength is $t''$ represented by the dashed curves.}
    \label{fig:lat}
\end{figure}

We introduce two models to understand the quasiparticle creation and the stability of the FCIs against magnetic fields: the Kapit-Mueller model~\cite{PhysRevLett.105.215303,10.21468/SciPostPhys.12.4.118} and the checkerboard model~\cite{PhysRevLett.106.236803}. 
Both models have flat bands with the Chern number $|C_\text{s}|=|(1/2\pi A)\int d^2k\mathcal{F}_{xy}|=1$ and realize the FCIs in the absence of magnetic fields, but they have different quantum geometries. 
In presence of interactions, FCI states are driven in these models. 
In this study, we consider spinless fermions with the nearest neighbor interaction
\begin{equation}
    H_{\text{int}} = \sum_{\langle j,k \rangle} U n_jn_k,
\end{equation}
where $U$ is the strength of the interaction and $\langle \dots\rangle$ means the nearest-neighbor sites. 
The operator $n_j$ is the number operator $n_j=c_j^{\dagger}c_j$ on site $j$,
where $c_j$ is the fermion annihilation operator.
We use the periodic boundary condition throughout this study.
{It is noted that the FCI states are stable for interaction strength larger than the single-particle band gap, which is commonly seen in various models~\cite{Kourtis2014}.}

The Hamiltonian of the Kapit-Mueller model under a magnetic field is,
\begin{align}
    &H_{\text{KM}} = \sum_{j,k}J(z_j,z_{k})c_{j}^{\dag}c_{k}+\text{h.c.} + H_{\text{int}},\\
    J(z_{j},z_{k}) &= (-1)^{x+y+xy}e^{-\frac{\pi}{2}(1-\phi_0)|z|^2}G_{jk}(\phi_0+\phi),
\end{align}
where $z_{k}=x_{k}+iy_{k}$ is the lattice complex coordinate for the site $k$ and $z=z_j-z_k=x+iy$.
$\phi_0$ is a model parameter which is distinguished from the applied flux and is chosen to be $\phi_0 = 1/3$ {so that the model can efficiently describe the $\nu=1/3$-Laughlin state}. 
For this parameter, there are three energy bands in the absence of the magnetic field, and the lowest energy band is exactly flat and has a Chern number $C_s=1$. 
In addition, it satisfies the ideal condition and thus the single-particle Bloch function of this band is a holomorphic function.
$\phi$ is the applied flux per plaquette, and there is no distinction between the flux and the magnetic field in this study since the lattice constant is the length unit. 
Note that, since $\phi$ is not included in the Gaussian factor in the hopping integral $J(z_j,z_k)$, the Poisson summation rule~\cite{PhysRevLett.105.215303,shen2025exactparenthamiltonianslandau} for hopping integrals does not hold at $\phi\neq0$ and the ideal condition is not necessarily satisfied under the magnetic field.
Nevertheless, deviation from the ideal condition is small as will be discussed later in Sec.~\ref{sec:level5}.
$G_{jk}$ is the phase factor corresponding to the magnetic field $\phi$ in addition to the model parameter $\phi_0=1/3$. 
We use the string gauge~\cite{PhysRevLett.83.2246,doi:10.7566/JPSJ.86.103701,PhysRevLett.127.237204,pgLSM},
\begin{align}
    &G_{jk}(\phi) = e^{iA'_{jk}(\phi)+iA''_{jk}(\phi)},\label{eq:Gjk}\\
    &A'_{jk}(\phi) = 2\pi\phi\frac{(x_j+x_k)(y_j-y_k)}{2},\\
    &A''_{jk}(\phi) = -2\pi\phi \frac{y_j+y_k}{2}L_x\delta_x,
\end{align}
where $L_x,L_y$ are the linear system sizes in each direction. $\delta_x=\pm 1$ for the hopping between $x=0$ and $x=L_x-1$ and $\delta_x=0$ otherwise. 
In this gauge, one can describe the flux per square plaquette $\phi=N_{\phi}/N_s$ with the number of the sites $N_s$, where $N_{\phi}$ is the number of the flux quanta applied to the entire system.
The fermion filling for the lowest energy band at $\phi=0$ is $\nu=N/N_{\text{orb}}=1/3$, where $N$ is the number of the fermions and $N_{\text{orb}}$ is the number of the single-particle states in the lowest energy band.
$N_{\text{orb}}=N_s/3$ at $\phi=0$ and it changes as $\phi$ is introduced.
The ground state at $\phi=0$ is the $\nu=1/3$ Laughlin state on the lowest energy band.

The Hamiltonian of the checkerboard model~\cite{PhysRevLett.106.236803} is
\begin{align}
    &H_{\text{CB}} = \sum_{j,k}t_{jk}G_{jk}(\phi)c_{j}^{\dag}c_{k} + H_{\text{int}}\\
    &t_{jk} = \left\{
    \begin{aligned}
        &te^{i\psi_{jk}}, \quad &\text{:NN hopping}\\
        &t'_{jk}, \quad &\text{:NNN hopping}\\
        &t'', \quad &\text{:NNNN hopping}\\
        &0, \quad &\text{:otherwise}
    \end{aligned}
    \right.
\end{align}
The NN, NNN and NNNN hopping mean the nearest-neighbor, the second  nearest-neighbor and the third nearest-neighbor hopping respectively.
In this study, we set $t=1,\psi_{jk} = \pm \pi/4$, whose sign is determined by the direction of the arrow in Fig.\ref{fig:lat}(b), and $t'_{jk}=t_1, t_2$ which is identified by choosing a bipartite site and $t_1=-t_2=1/(2+\sqrt{2})$. In addition, we set $t''=1/(2+2\sqrt{2})$ to realize a nearly flat Chern band with $C_\text{s}=-1$ at zero magnetic field, $G_{jk}(\phi=0)=1$. 
This system exhibits the $\nu=1/3$ Laughlin state on the lowest energy band ~\cite{Sheng2011}. 
However, because this is a two-band model, its Berry curvature must have at least one zero point in the Brillouin zone, and the trace condition does not hold exactly ~\cite{PhysRevB.104.045104,PhysRevB.96.165150}.  
Therefore, the checkerboard model does not satisfy the ideal condition in contrast to the Kapit-Mueller model.
The phase factor $G_{jk}(\phi)$ describes the Peierls phase. 
The concrete expression of $G_{jk}$ for the checkerboard model is essentially the same as that in the Kapit-Mueller model (Eq.~\eqref{eq:Gjk})~\cite{pgLSM}. 
The fermion filling in the absence of the magnetic field is $\nu=N/N_{\text{orb}}=1/3$, where $N_{\text{orb}}=N_s/2$ at $\phi=0$.

\section{\label{sec:level3}Summary of results}
\begin{figure}[ht]
    \centering
    \begin{subfigure}{1.0\linewidth}
        \centering
        \textbf{(a) Kapit-Mueller model}
        \includegraphics[width=\linewidth]{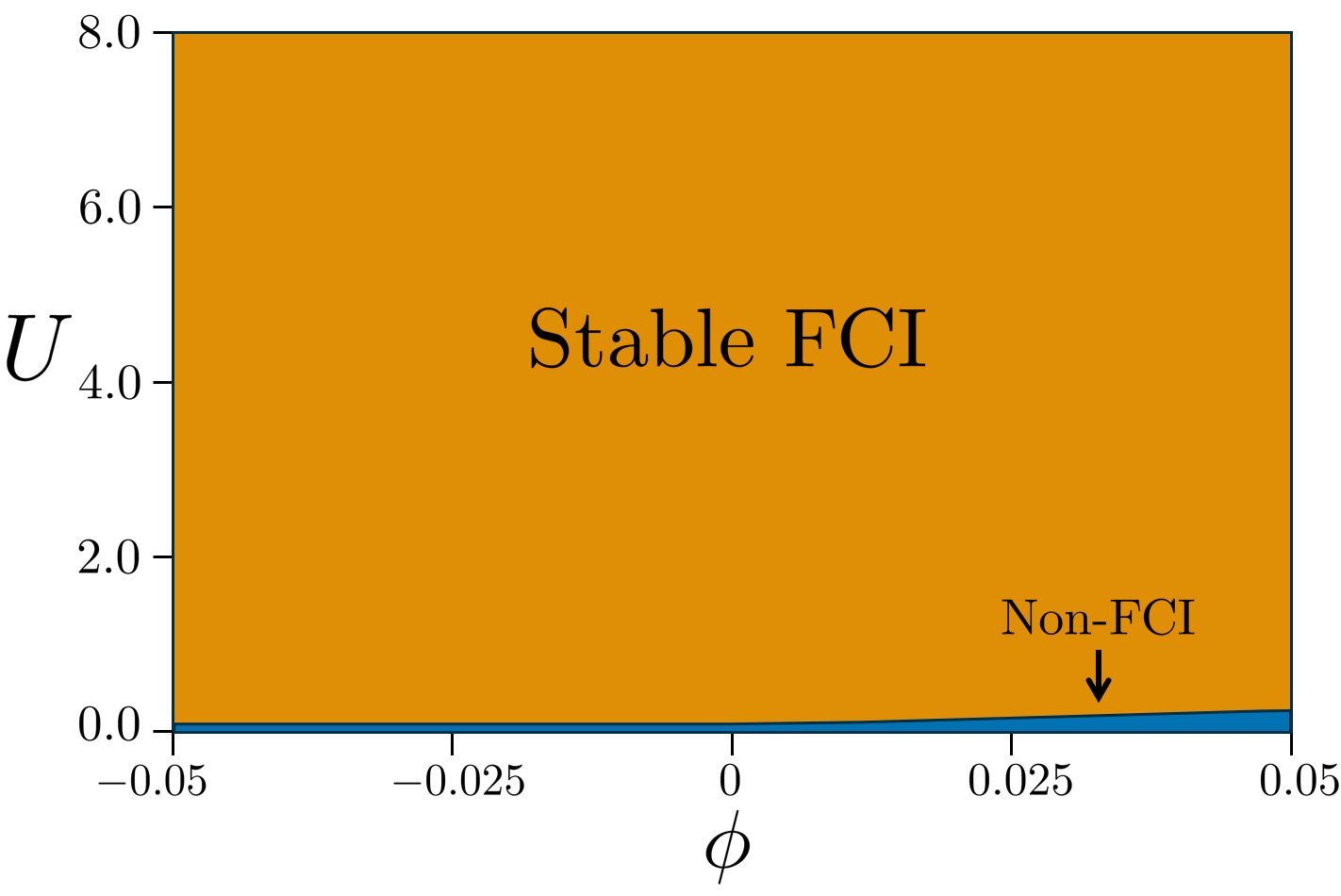}
    \end{subfigure}\\
    \begin{subfigure}{1.0\linewidth}
        \centering
        \textbf{(b) Checkerboard model}
        \includegraphics[width=\linewidth]{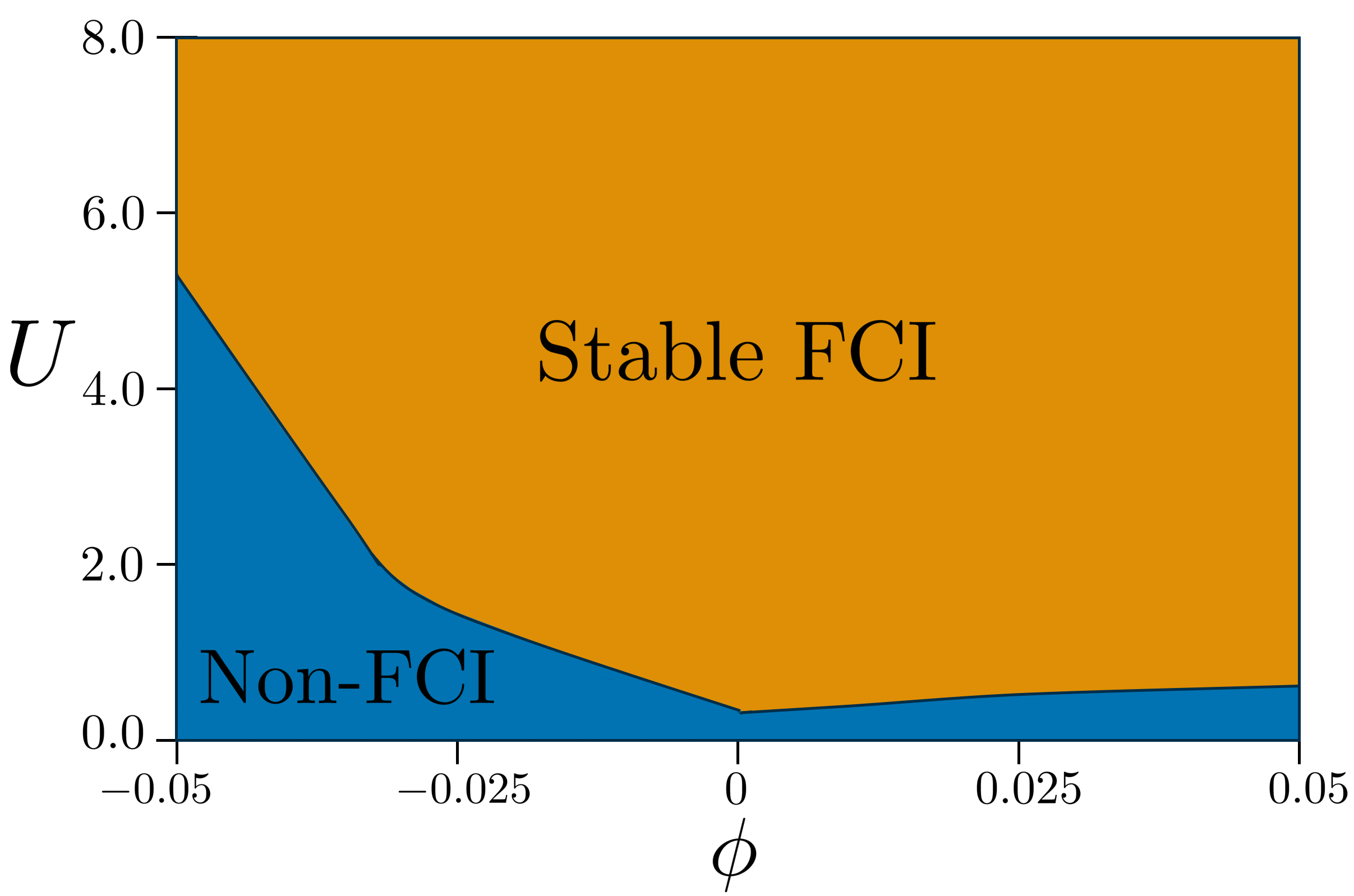}
    \end{subfigure}
    \caption{Expected phase diagrams for (a) the Kapit-Mueller model and (b) the checkerboard model with the pinning potential $V_p\neq0$. $\phi$ is the applied magnetic field and $U$ is the strength of the nearest-neighbor interaction. The strength of the pinning potential is chosen so that the created quasiparticles are well localized. It is assumed that effects of interactions between the localized quasiparticles are negligible in a thermodynamically large system.}
    \label{fig:Schematic}
\end{figure}
 Before presenting detailed calculation results, we first summarize the key results here.
 quasiparticles can be created by the magnetic field $\phi$, and they can be localized by an additional pinning potential $V_p$ corresponding to disorder.
 We show expected phase diagrams of the two models with the pinning potential $V_p\neq0$ in Fig.~\ref{fig:Schematic} based on the exact diagonalization and the quantum geometry under magnetic fields. {We have used different values of $V_p$ in the exact diagonalization and details of the phase diagram could depend on the strength of $V_p$, although it is qualitatively unchanged.}
 Generally, a magnetic field increases or decreases the number of (nearly degenerate) single-particle states $N_{\text{orb}}$ according to the Streda's formula~\cite{Streda1982,Streda1983,Widom1982,MacDonald1983}, $N_{\text{orb}}\to N_{\text{orb}}+C_sN_{\phi}$, where $C_\text{s}$ is the Chern number of the band considered and $N_{\phi}$ is the number of the added flux for the entire system. 
 If $N_{\text{orb}}$ increases (decreases) under a magnetic field, the filling $\nu=N/N_{\text{orb}}$ decreases (increases), which can lead to quasihole (quasielectron) creation.
 In the Kapit-Mueller model, since the Chern number of the lowest energy band is $+1$, adding a positive magnetic field  induces quasihole creation, while a negative magnetic field leads to quasielectron creation. 
 In the checkerboard model, the Chern number of the lowest energy band is $-1$, and a positive (negative) magnetic field corresponds to quasielectron (quasihole) creation. 
 The quasiholes can be localized by a pinning potential $V_p>0$ while the quasielectrons are by $V_p<0$ in our notation.
Within the exact diagonalization for limited system sizes in the present study, the number of the created quasiparticles is $N_{\text{qp}}=1$ in each model.
In a thermodynamically large system, the number of the flux quanta is $|N_{\phi}|\gg 1$ for a fixed magnetic field $\phi$ and there will be $|N_{\phi}|$ quasiparticles which are localized by a pinning potential.
In Fig.~\ref{fig:Schematic}, we have assumed that the quasiparticles are well localized and effects of interactions between the pinned quasiparticles are negligible.

 As seen in Fig.~\ref{fig:Schematic}(a) for the Kapit-Mueller model, the FCI state is stable under the magnetic field irrespective of its direction.
 More precisely, except for very small interactions, quasiparticles are stably created by the magnetic field and then localized by the pinning potential. Consequently, the fractionally quantized Hall conductance is maintained over a wide range of parameters, which means that the FCI phase is stable. 
 The phase diagram is almost symmetric for $+\phi$ and $-\phi$, and thus the creations of quasiholes ($\phi>0$) and quasielectrons ($\phi<0$) are both stable.
 This is a reasonable result, since the Kapit-Mueller model is a lattice model which well mimics the lowest Landau level in the continuum system and basic properties are similar to the corresponding FQHE.
 
 On the other hand, in the checkerboard model, the quasielectron creation under $\phi>0$ is stable except for small interactions, while the quasihole creation for $\phi<0$ is unstable even for moderately strong interactions. 
 In the latter case, the FCI state is destabilized by the magnetic field and a non-FCI state is realized. 
 The non-FCI state at very small interactions is smoothly connected to the non-interacting Fermi liquid at $U=0$.
 However, it is difficult to fully clarify nature of the non-FCI state within the present calculations, and exact identification of the non-FCI state is left for a future study.
We stress that the asymmetry with respect to $\pm \phi$ is a characteristic feature of a lattice system which is absent in the standard FQHE in the continuum model.

 In the following sections, we will show the exact diagonalization results and argue that the qualitative features of the phase diagrams can be understood based on the quantum geometry in the presence of the magnetic field.

\section{\label{sec:level4}Exact diagonalization}
\subsection{\label{sec:level4-1}Characterization of FCI}
 In this section, we study the FCI with use of the exact diagonalization.
 First, we briefly explain how to characterize an FCI state within the exact diagonalization.
 To examine stability of the FCI states against magnetic fields, we apply one flux quantum $N_{\phi}=\pm 1$ uniformly over the entire system and discuss quasiparticles induced by the magnetic field based on the ``counting-rule" on the energy spectra \cite{PhysRevX.1.021014,PhysRevLett.100.246802,PhysRevB.85.075128}. 
 Previous works examined quasihole and quasielectron creations in FCIs at $\phi=0$ based on the counting-rule, when the number of unit cells is increased or decreased~\cite{PhysRevB.91.045126,PhysRevB.99.045136,PhysRevB.87.205136}. 
 When the system size is changed with a fixed number $N$ of the fermions, the number of the nearly degenerate single-particle states $N_{\text{orb}}$ and the filling $\nu=N/N_{\text{orb}}$ varies, while the single-particle band structure itself is unchanged.
 Instead of varying the system size, a change in the number of the single-particle orbits can be induced by the magnetic field according to the Streda's formula, $N_{\text{orb}}\to N_{\text{orb}}+C_sN_{\phi}$. 
 In contrast to the system-size induced change of $N_{\text{orb}}$, the system does not keep the (magnetic) translation symmetry under the additional one flux quantum $N_{\phi}=\pm1$~\cite{PhysRevLett.127.237204}. 
 However, one can still regard the nearly degenerate single-particle orbits as a nearly flat band and discuss the stability of the FCI in a manner similar to those in the previous studies at $\phi=0$. 
 
  Since the single-particle Chern number is $|C_s|=1$ in the present models, a single flux quantum leads to the change $N_{\text{orb}}\to N_{\text{orb}}\pm1$. This gives rise to creation of one quasihole or quasielectron with the fractional charge $|e^*|=1/3$. 
  In case of the quasihole creation, the ground state degeneracy $\#_{\text{qh}}$ under the magnetic field obeys the counting rule given by ~\cite{PhysRevX.1.021014}
\begin{equation}
    \#_{\text{qh}} = N_{\text{orb}}\frac{(N_{\text{orb}}-2N-1)!}{N!(N_{\text{orb}}-3N)!},
    \label{counting-hole}
\end{equation}
where $N_{\text{orb}}$ is the number of the low energy single-particle states in the presence of the magnetic field.
For quasielectron creation, no simple formula for the ground state degeneracy $\#_{\text{qe}}$ is known, 
but we can verify the counting rule of the quasielectrons by referring to energy spectra of the corresponding FQHE states.
More precisely, we calculate $\#_{\text{qe}}$ for the LLL in the electron gas model and also for fermions in the Kapit-Mueller model with the Poisson summation rule which is a lattice realization of the LLL~\cite{PhysRevLett.105.215303,shen2025exactparenthamiltonianslandau}.
We numerically find that $\#_{\text{qe}}=11$ for $(N=4,N_{\text{orb}}=11)$, $\#_{\text{qe}}=14$ for $(N=5,N_{\text{orb}}=14)$,
and $\#_{\text{qe}}=17$ for $(N=6,N_{\text{orb}}=17)$.

The quasiparticles created by the uniform magnetic field are extended over the entire system.
They can be spatially localized in the presence of an additional pinning potential $V_p$ which couples with the quasiparticle charge~\cite{yoshioka2002quantum,RevModPhys.80.1083}.
To this end, we first perform the exact diagonalization to obtain low energy states $\ket{\Psi_1},\ket{\Psi_2},\cdots,\ket{\Psi_{N_\text{cut}}}$ without the pinning potential~\cite{PhysRevB.91.045126,PhysRevB.99.045136}, where
$N_\text{cut}=\#_{\text{qh}}$ or $\#_{\text{qe}}$ is a cut-off for the low energy states.
Then, we take $V_p$ into account within this low energy sector by diagonalizing the projected Hamiltonian with the pinning potential, 
\begin{align}
\tilde{H} = PHP + PV_pn_{j_0}P,
\end{align}
where $H$ is either $H_\text{KM}$ or $H_\text{CB}$.
$P$ is the projection onto the low energy states $\{\ket{\Psi_n}\}_{n=1}^{N_{\text{cut}}}$ and $j_0$ is the position of the pinning potential.
The projection enables us to efficiently focus on the low energy sector and also greatly reduces computational costs for the analysis of the pinning potential~\cite{PhysRevB.91.045126,PhysRevB.99.045136}.
When the quasiparticle created by $\phi\neq0$ is localized by the pinning potential, it does not contribute to the Hall conductance.
Correspondingly, the ground state degeneracy of $\tilde{H}$ will be three fold for the present $\nu=1/3$ system.
In this sense, the ground states of $\tilde{H}$ belong to the same FCI phase of $H$ at $\phi=0$.
However, the quasiparticle creation can be unstable and a magnetic field may destabilize the underlying FCI state.
In this case, the counting rule does not hold and the ground states of $\tilde{H}$ are no longer FCIs.

We can further characterize the ground states of $\tilde{H}$ as FCIs by calculating the many-body Chern number $C$,
with use of the Fukui-Hatsugai-Suzuki's method \cite{PhysRevB.31.3372,doi:10.1143/JPSJ.74.1674,PhysRevLett.122.146601}. 
The many-body Chern number $C$ is defined as
\begin{equation}
    C = \frac{1}{2\pi}\int_{0}^{2\pi}\int_{0}^{2\pi}F(\theta_{x},\theta_{y})d\theta_{x}d\theta_{y},
    \label{eq:mbC}
\end{equation}
where $F(\vec{\theta})=\pdv{A_{y}}{\theta_{x}}-\pdv{A_{x}}{\theta_{y}}$, $A_{\mu}(\vec{\theta})=i\Phi^{\dag}\pdv{\Phi}{\theta_{\mu}}$, and $\Phi(\vec{\theta})=(\ket{\Psi_{1}(\vec{\theta})},\dots,\ket{\Psi_{q}(\vec{\theta})})^T$ is the ground state multiplet with $q=3$. 
$\vec{\theta}=(\theta_{x},\theta_{y})$ represents a twisting phase for which $c_{n_{x}+L_{x},n_{y}}=e^{i\theta_{x}}c_{n_{x},n_{y}}$ and $c_{n_{x},n_{y}+L_{y}}=e^{i\theta_{y}}c_{n_{x},n_{y}}$. 
We divide the parameter space $\{(\theta_x,\theta_y)\}=[0,2\pi]\times[0,2\pi]$ into $L_{\theta}\times L_{\theta}$ discrete points with $L_{\theta}=20$, which is large enough to accurately compute $C$. 
Note that the Fukui-Hatsugai-Suzuki's method gives a quantized value of $C$ by construction even for a gapless state in a finite size system, and it does not give any physical implications in such a case.

To supplement our argument, we also examine the one-plaquette Chern number defined as
\begin{align}
\mathcal{C}(\vec{\theta})=\frac{1}{2\pi}L_{\theta}^{2}F(\vec{\theta}) 
\end{align}
in the discretized parameter space. 
It is known that, in a gapped system, $\mathcal{C}(\vec{\theta})$ is approximately uniform in the parameter space and it is close to $C$ given by Eq.~\eqref{eq:mbC}  \cite{PhysRevLett.122.146601,koma2015topologicalcurrentfractionalchern}.
To verify the uniformity, we examine the variance of $\mathcal{C}(\theta_{x},\theta_{y})$,
\begin{equation}
    \mathcal{V}[\mathcal{C}] = \sqrt{\frac{1}{L_{\theta}^2}\sum_{ \theta_{x},\theta_{y}}\qty[\mathcal{C}(\theta_{x},\theta_{y})-C]^2}.
\end{equation}
The variance is exponentially small if the ground states are FCIs, while it can be large for gapless states.
In this way, we can identify FCI states within the exact diagonalization for small system sizes.
It turns out that, even if the ground state of $\tilde{H}$ is an FCI, $\mathcal{V}[\mathcal{C}]$ diverges when we use $N_{\text{cut}}< \#_{\text{qh}},\#_{\text{qe}}$.
When $N_{\text{cut}}= \#_{\text{qh}},\#_{\text{qe}}$,  the variance $\mathcal{V}[\mathcal{C}]$ is well suppressed in the FCI phase. {This is because it is necessary to include all the states $\ket{\Psi_1},\cdots,\ket{\Psi_{\#{\rm qh(qe)}}}$ with the spatially extended quasiparticle at $V_p=0$ for its localization under $V_p\neq0$.}

In this study, we consider the systems with the number of the fermions $N=4\sim6$.
This is relatively small compared to those in the previous studies where quasiparticles are created by the increase of the system sizes~\cite{PhysRevX.1.021014,PhysRevB.91.045126,PhysRevB.99.045136,PhysRevB.87.205136}.
Technically, one flux quantum breaks the (magnetic) translation symmetry and the size of the Hilbert space cannot be reduced by the translation symmetry~\cite{PhysRevLett.127.237204}.
However, it turns that we can discuss physical properties of the systems with $N=4\sim6$ by carefully paying attention to finite size effects as will be discussed in the following sections.
Furthermore, we have confirmed that quasiparticles are stably created by the increase of the system sizes for $N=4,5$ in consistent with the previous studies~\cite{PhysRevB.91.045126,PhysRevB.99.045136,PhysRevB.87.205136}.
{See also Appendix~\ref{app:unitcell}.}
This implies that finite size effects are small for these system sizes.

\subsection{\label{sec:level4-2}Kapit-Mueller model}
\begin{figure}[h]
    \centering
        \begin{subfigure}{0.49\linewidth}
        \centering
        \textbf{(a) $N_{\phi}=1,V_{p}=0.0$} \\
        \includegraphics[width=\linewidth]{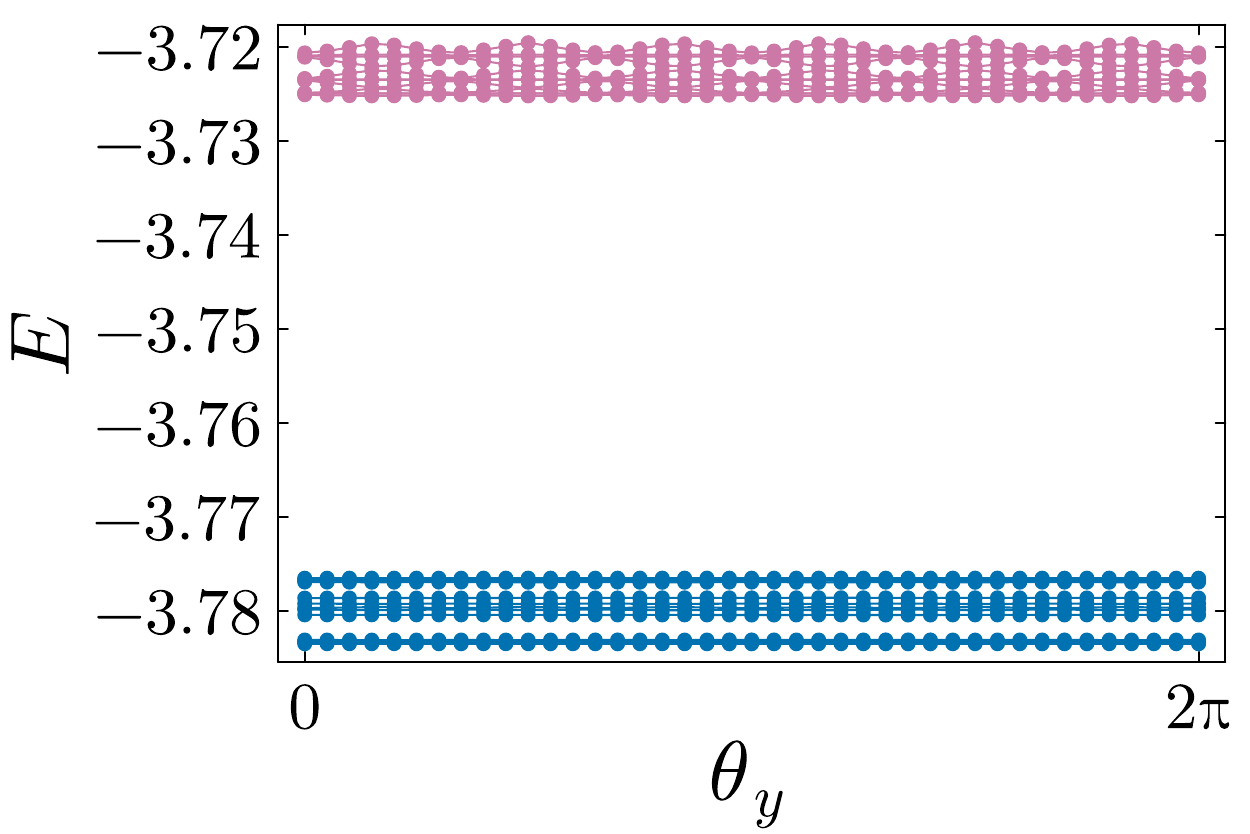}
    \end{subfigure}
    \begin{subfigure}{0.49\linewidth}
        \centering
        \textbf{(b) $N_{\phi}=1,V_{p}=1.0$} \\
        \includegraphics[width=\linewidth]{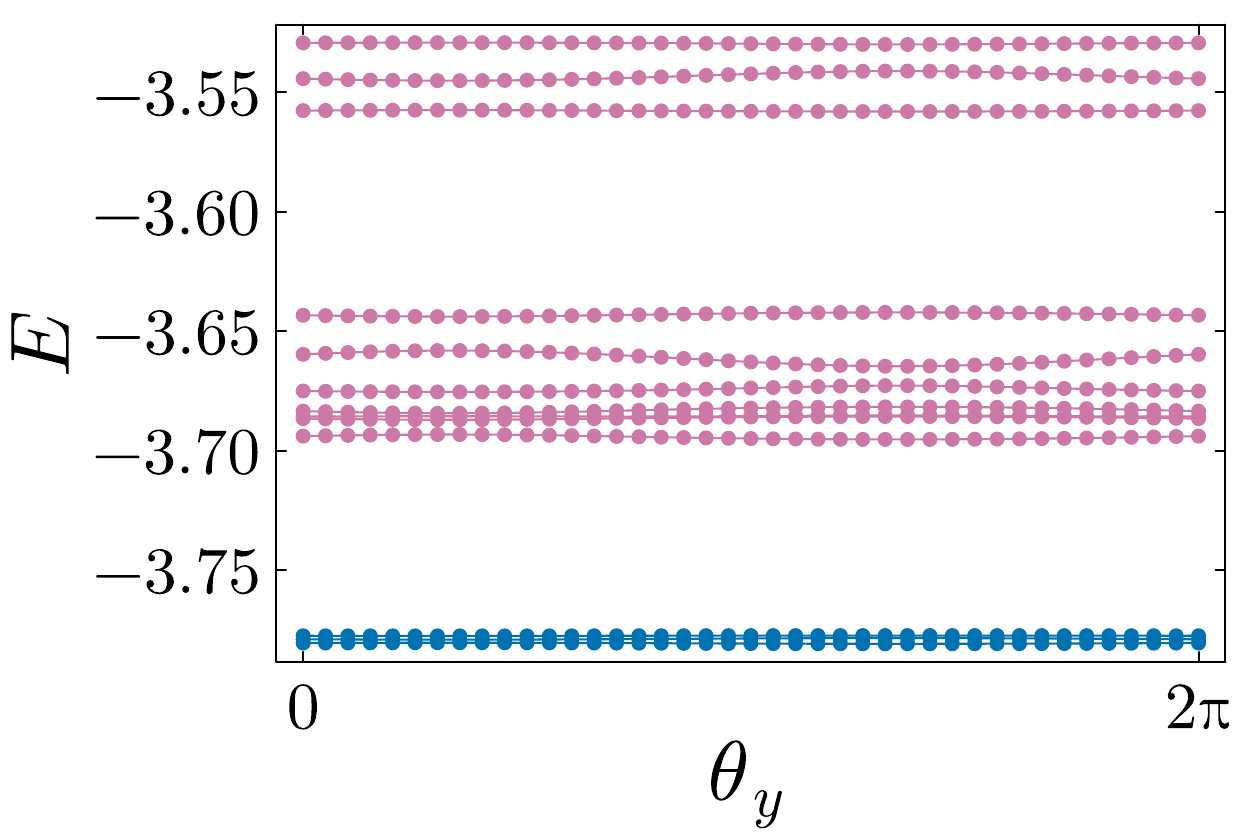}
    \end{subfigure}\\
    \begin{subfigure}{0.49\linewidth}
        \centering
        \textbf{(c) $N_{\phi}=-1,V_{p}=0.0$} \\
        \includegraphics[width=\linewidth]{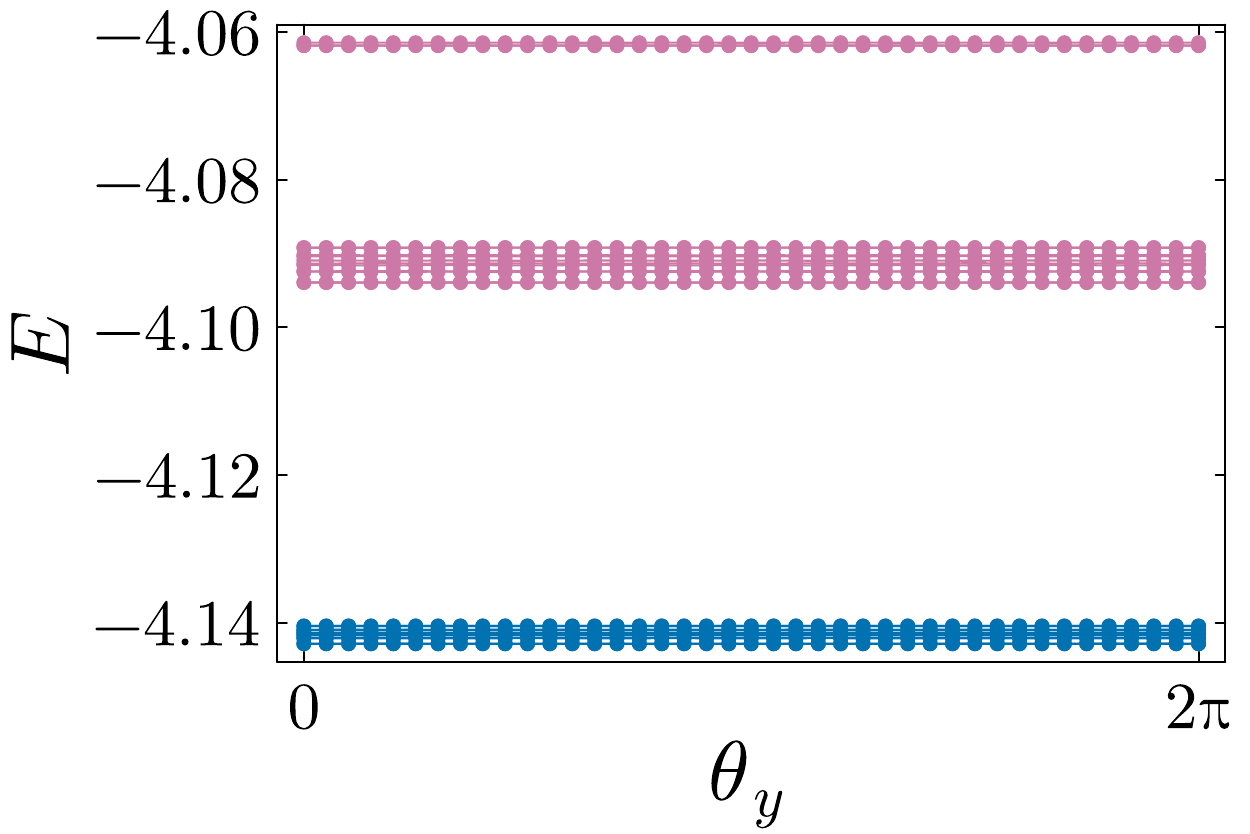}
    \end{subfigure}
    \begin{subfigure}{0.49\linewidth}
        \centering
        \textbf{(d) $N_{\phi}=-1,V_{p}=-1.0$} \\
        \includegraphics[width=\linewidth]{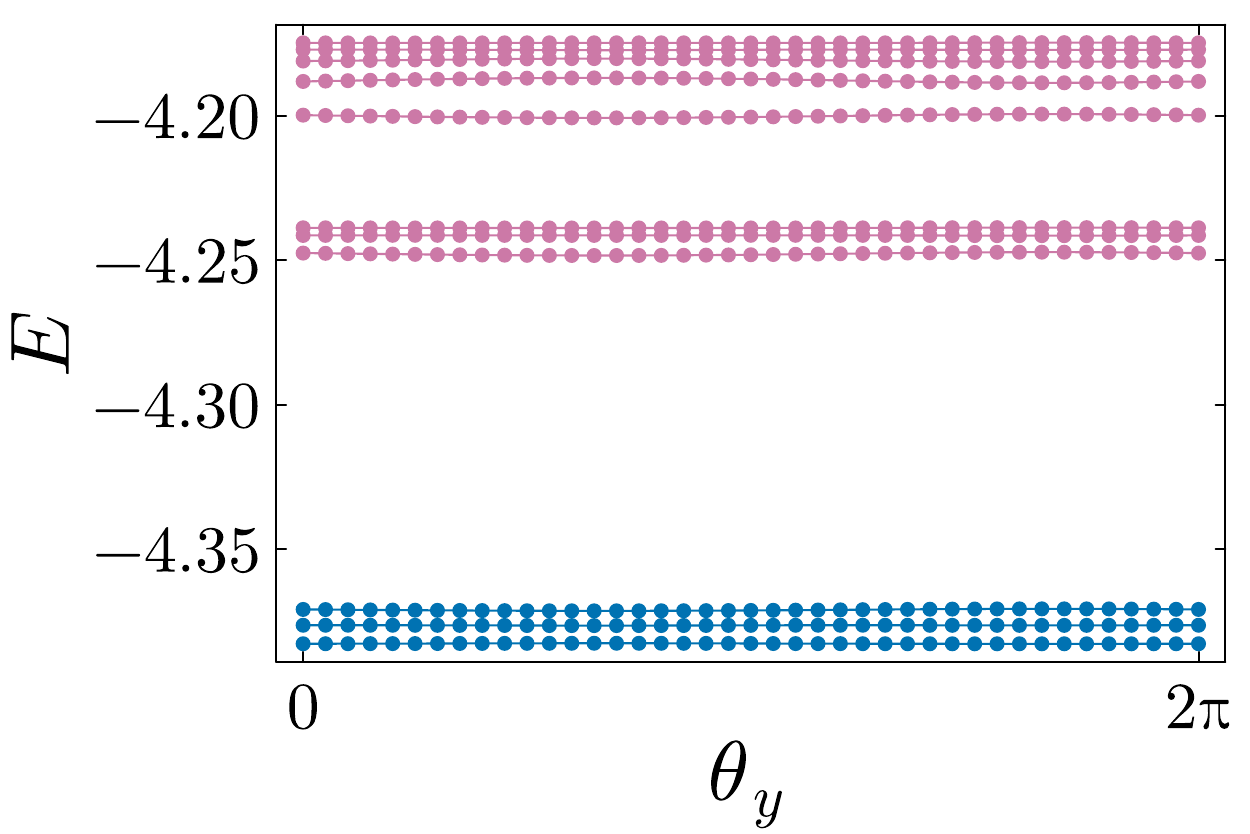}
    \end{subfigure}
        \caption{Spectral flow in the Kapit-Mueller model for $N_s=36, N=4, U=1.0$ under the magnetic fields $N_{\phi}=\pm 1$. 
        The flux is $N_{\phi}=+1$ in (a) and (b). The pinning potential is (a) $V_{p}=0.0$ and (b) $V_{p}=1.0$.  
        Similarly, (c) $N_{\phi}=-1, V_p=0$ and (d) $N_{\phi}=-1, V_p=-1.0$.  
        Blue curves in (a) and (c) represent energy levels predicted by the counting-rule, $\#_{\text{qh}}=13$ and $\#_{\text{qe}}=11$. 
        Blue curves in (b) and (d) represent the lowest three energy states corresponding to the topological degeneracy.}
        \label{fig:KMSF}
\end{figure}
\begin{figure}[t]
    \centering
    \begin{subfigure}{0.98\linewidth}
        \centering
        \includegraphics[width=\linewidth]{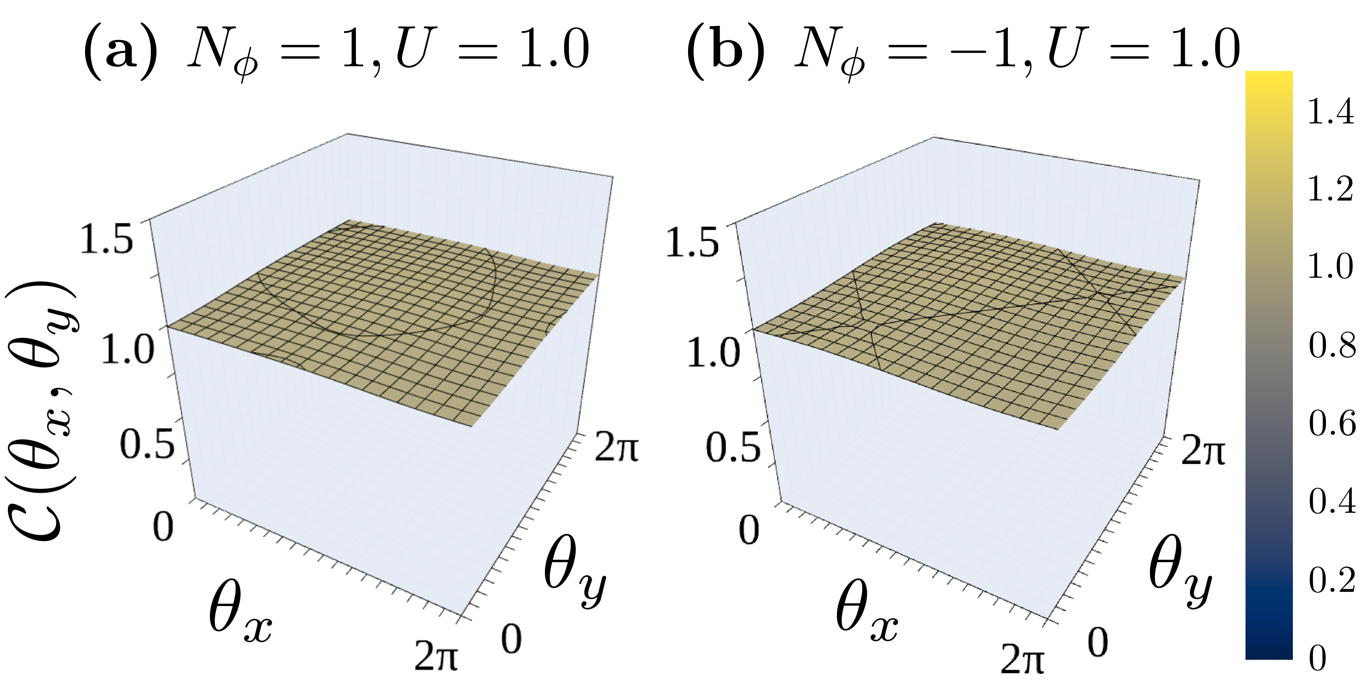}
    \end{subfigure}\\
    \begin{subfigure}{0.90\linewidth}
    \textbf{(c)}
        \includegraphics[width=\linewidth]{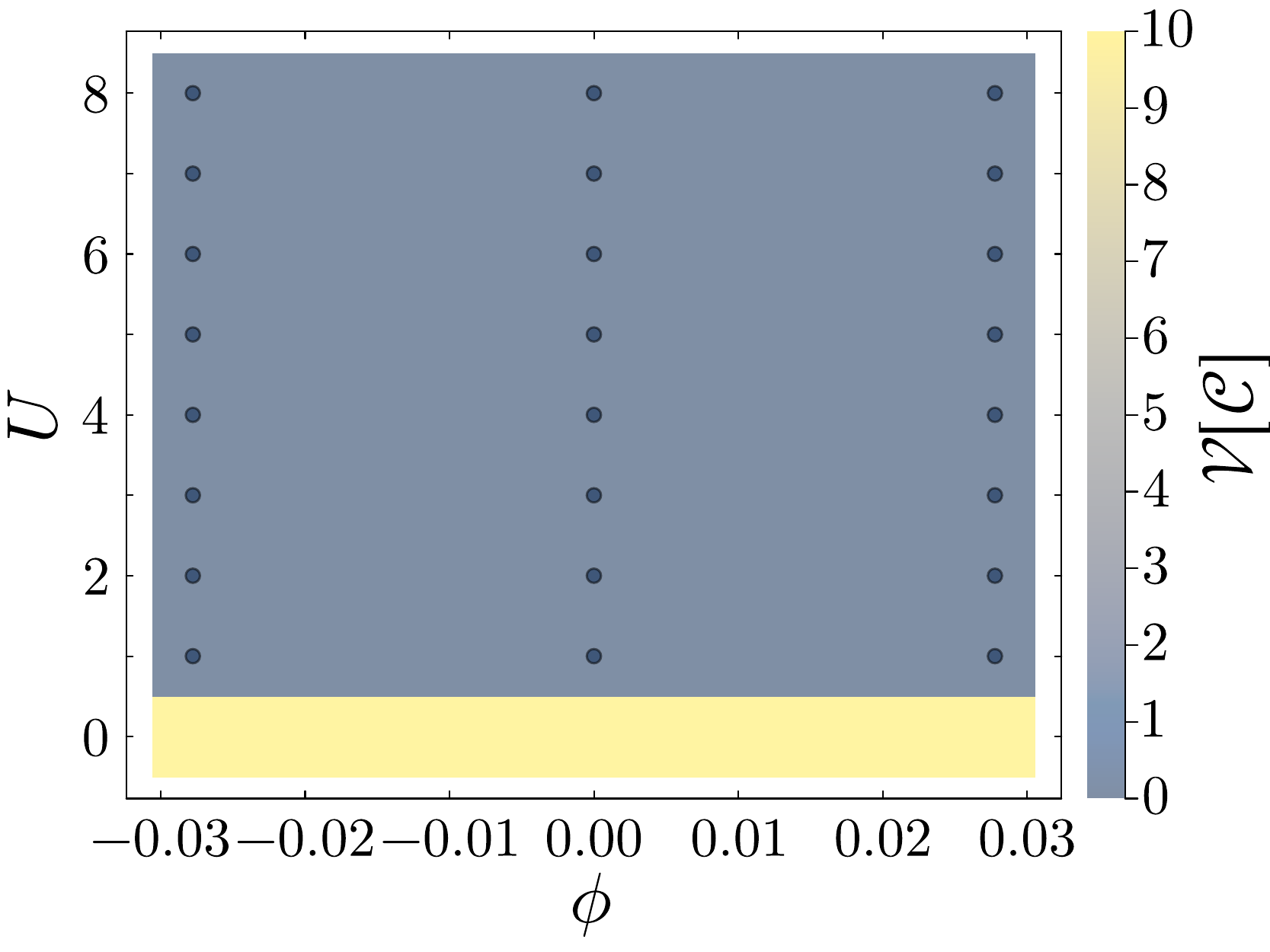}
    \end{subfigure}
    \caption{One-plaquette Chern number $\mathcal{C}$ in the Kapit-Mueller model with $U=1.0, N=4$ at (a) $N_{\phi}=+1, V_p=-1.0$ and (b) $N_{\phi}=-1, V_p=+1.0$. 
    (c) Variance $\mathcal{V}[\mathcal{C}]$ for ${|V_p|=1,} U=1\sim8$ and the flux density $N_{\phi}=0,\pm 1$. The yellow region with large $\mathcal{V}[\mathcal{C}]$ corresponds to the non-FCI state around the non-interacting limit $U=0$.
    }
     \label{fig:KMOPCN}
\end{figure}

 We first discuss the quasihole creation in the Kapit-Mueller model.
 In Fig.~\ref{fig:KMSF}, we show spectral-flow {as a function of $\theta_y$} for the Kapit-Mueller model with one additional positive flux quantum $N_{\phi}=1$. {Essentially the same results can be obtained also as a function of $\theta_x$, since the boundary twisting has only negligibly small effects in a gapped system in the thermodynamic limit~\cite{koma2015topologicalcurrentfractionalchern}.}
 The counting-rule for the quasiholes is $\#_{\text{qh}}=13$ for the present system with $N=4, N_{s}/3=12, C_{\text{s}}N_{\phi}=1$, and
 correspondingly the lowest 13 energy levels are shown in blue in Fig.~\ref{fig:KMSF}(a). 
 There exists a clear energy gap between the set of the blue curves and others, which means that the counting-rule is satisfied. 
 Figure \ref{fig:KMSF}(b) represents spectral flow for the system with the pinning potential $V_{p}=1.0$, where 
 the blue curves represent the three lowest energy states.
 Now these three fold quasi-degenerate states are well separated from other states, and they are expected to exhibit the fractionally quantized Hall conductance as mentioned in the previous section~\cite{yoshioka2002quantum,RevModPhys.80.1083}.
 Indeed, we find that the many-body Chern number for the three fold quasi-degenerate ground states for $V_p=1.0$ is $C=1$ corresponding to the $\nu=1/3$ FQHE. 
 In addition, the one-plaquette Chern number $\mathcal{C}(\vec{\theta})$ is nearly flat in the parameter space as shown in Fig.~\ref{fig:KMOPCN}(a). 
 These results are fully consistent with the qualitative discussions in the previous section. 
 Namely, the quasiholes obeying the counting rule are created by the magnetic field and then get localized by the pinning potential, which consequently leads to the FCI ground state for $\phi\neq0$ and $V_p\neq0$. 
 Qualitatively same results are obtained for other values of the interaction strength, $U=1.0\sim8.0$. 
 The FCI state is unstable and a non-FCI state is realized only around the non-interacting limit $U=0$.
 {We have also performed the same calculations for different values of the pinning potential, $V_p=0.1$ and $V_p=10$. For the small pinning potential, the energy scale of the quasihole states is smaller than the energy gap between the ground states and the excited states at $V_p=0$, for which our projection scheme is well justified. We find that the qualitative behaviors do not change for both $V_p=0.1,10$ (not shown). Furthermore, we have checked that essentially the same results can be obtained even when we take the full Hilbert space into account without the projection onto the low energy sector.}

 Next we consider the quasielectron creation in the presence of one additional negative flux quantum $N_{\phi}=-1$.
 Spectral flow with no pinning potential at $N=4,U=1.0$ is shown in Fig.~\ref{fig:KMSF}(c). 
 As mentioned in the previous section, the counting rule for the quasielectrons is $\#_{\text{qe}}=11$.
 It is found that there are 11 fold quasi-degeneracy for the present Kapit-Mueller model, and thus the quasielectrons are stably created by the mangnetic field. 
 Similarly to the quasihole case, spectral flow with the pinning potential (Fig.~\ref{fig:KMSF}(d)) also shows the three fold quasi-degenerate ground states which are separated from other excited states with a gap. 
 Note that, in the quasielectron case, a negative pinning potential $V_{p}=-1.0$ is necessary to localize the quasielectron, since it has a positive fractional charge as opposed to the quasiholes. 
 The many-body Chern number of the three fold quasi-degenerate ground states is found to be $C=1$, and the one-plaquette Chern number is uniform as shown in Fig.~\ref{fig:KMOPCN} (b). 
 Similar results are obtained for $U=1.0\sim8.0$ {and $V_p=-0.1,-10$. Calculations without the projection also give essentially same results.}
 
Finally, we show the variance $\mathcal{V}[\mathcal{C}]$ for various interaction strength and the flux $N_{\phi}=0,\pm1 (\phi=0,\pm0.028)$ in Fig.~\ref{fig:KMOPCN} (c). 
The variance is very small and therefore the FCI ground states of $\tilde{H}$ are stable even for the small system with $N=4$. 
For larger systems, finite size effects will be further suppressed and one flux quantum corresponds to smaller magnetic fields.
Therefore, the ground states of $\tilde{H}$ for such systems will also be stable FCIs.
We have numerically confirmed that this is indeed the case for $N=5$.
Based on these observations, we expect the quantum phase diagram of the Kapit-Mueller model shown in Fig.~\ref{fig:Schematic}(a) in the previous section.
The phase diagram has been extrapolated to the region $|\phi|>0.03$ based on the analyses of the quantum geometry under magnetic fields in Sec.~\ref{sec:level5}.

\subsection{\label{sec:level4-3}Checkerboard model}
\begin{figure}[tb]
    \centering
    \begin{subfigure}{0.49\linewidth}
        \centering
        \textbf{(a) $N=6,U=1.0$} \\
        \includegraphics[width=\linewidth]{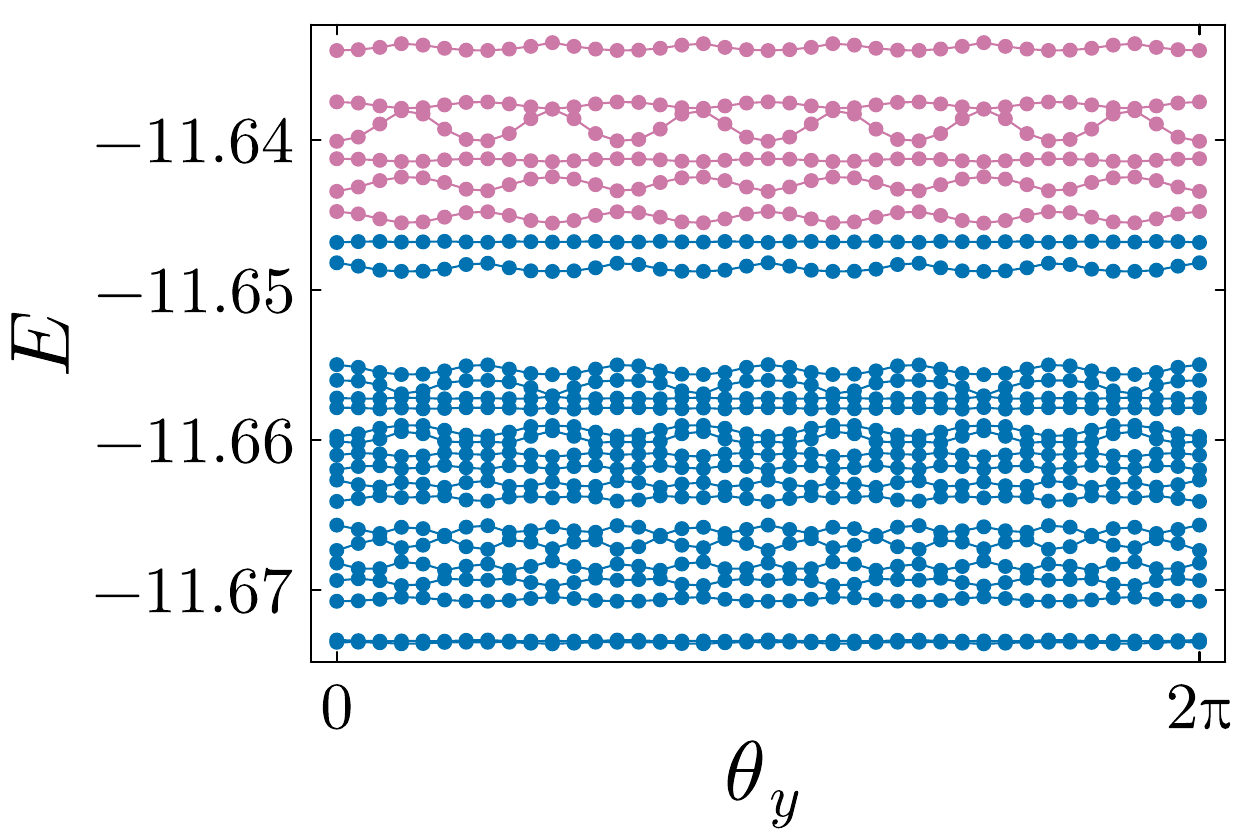}
    \end{subfigure}
    \begin{subfigure}{0.49\linewidth}
        \centering
        \textbf{(b) $V_{p}=10.0$}
        \includegraphics[width=\linewidth]{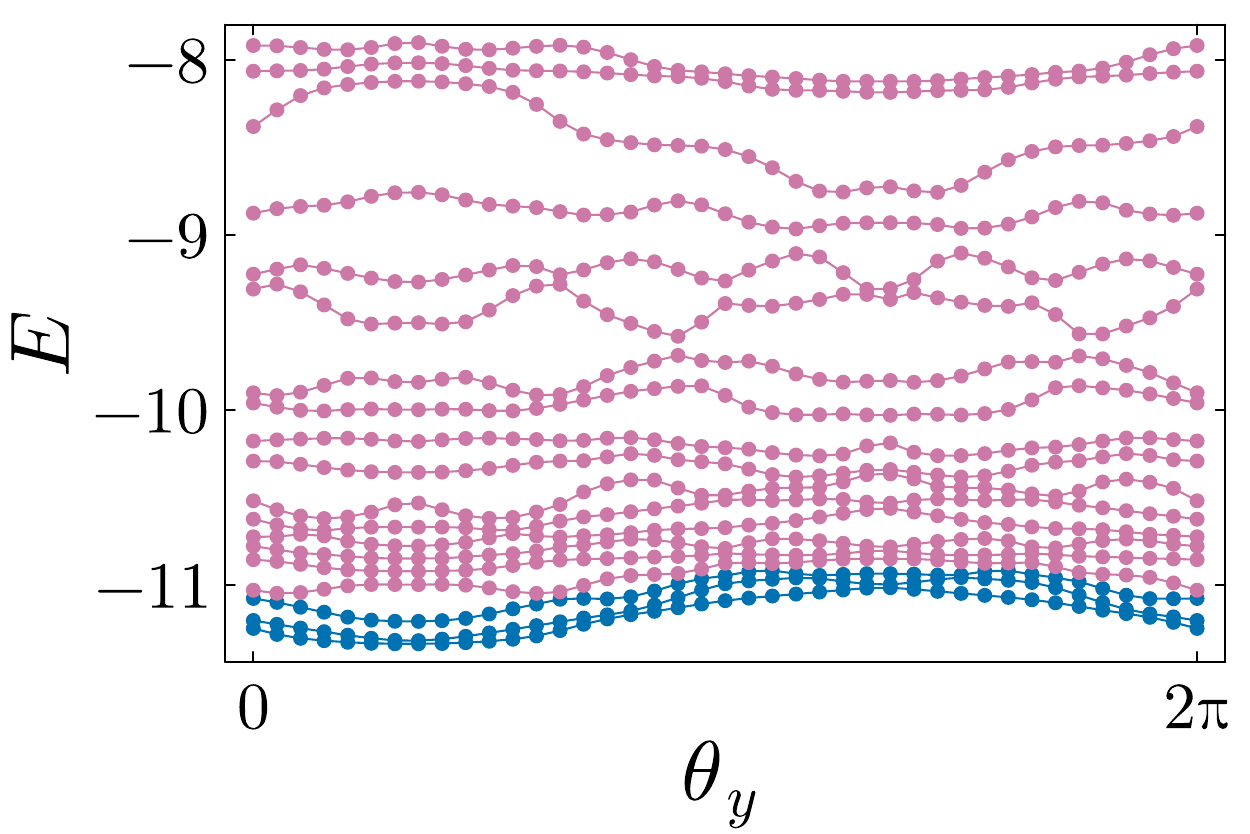}
    \end{subfigure}\\
    \begin{subfigure}{0.49\linewidth}
        \centering
        \textbf{(c) $N=6,U=2.0$}
        \includegraphics[width=\linewidth]{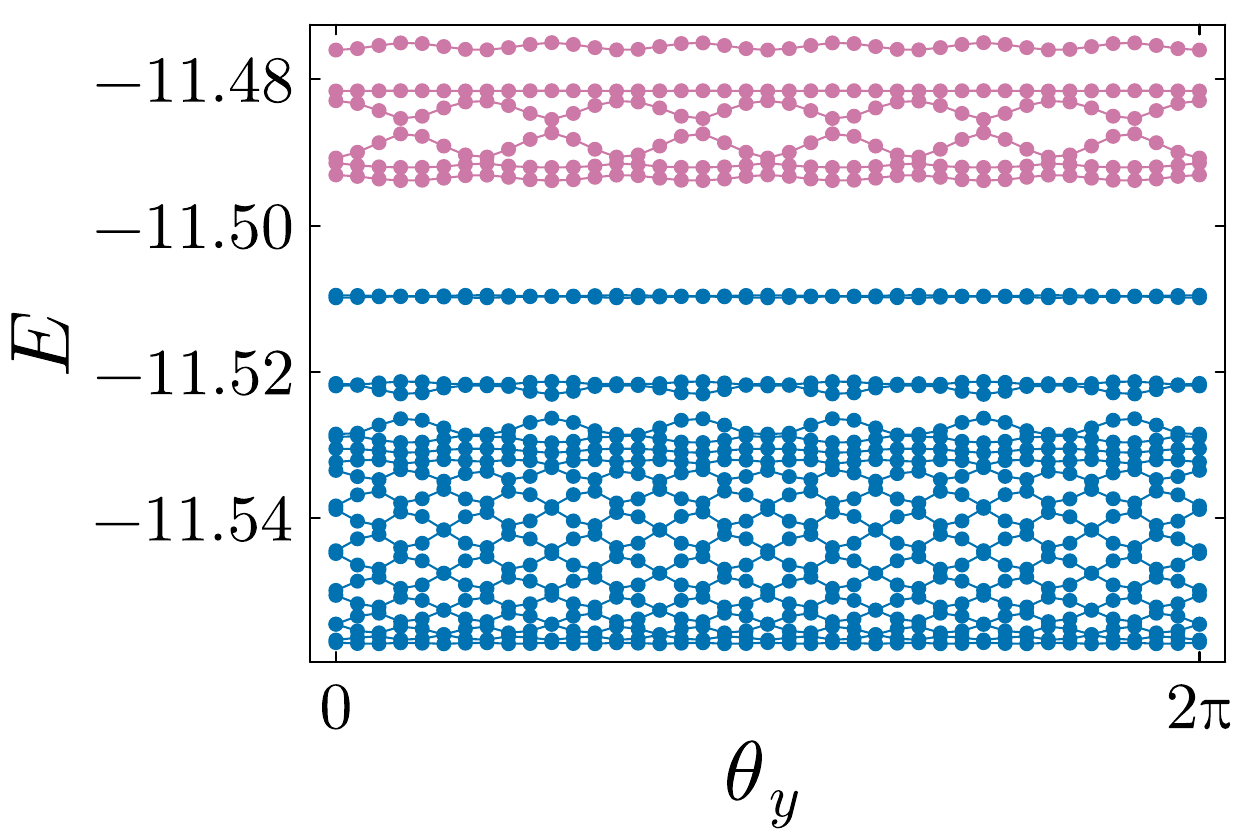}
    \end{subfigure}
    \begin{subfigure}{0.49\linewidth}
        \centering
        \textbf{(d) $V_{p}=10.0$}
        \includegraphics[width=\linewidth]{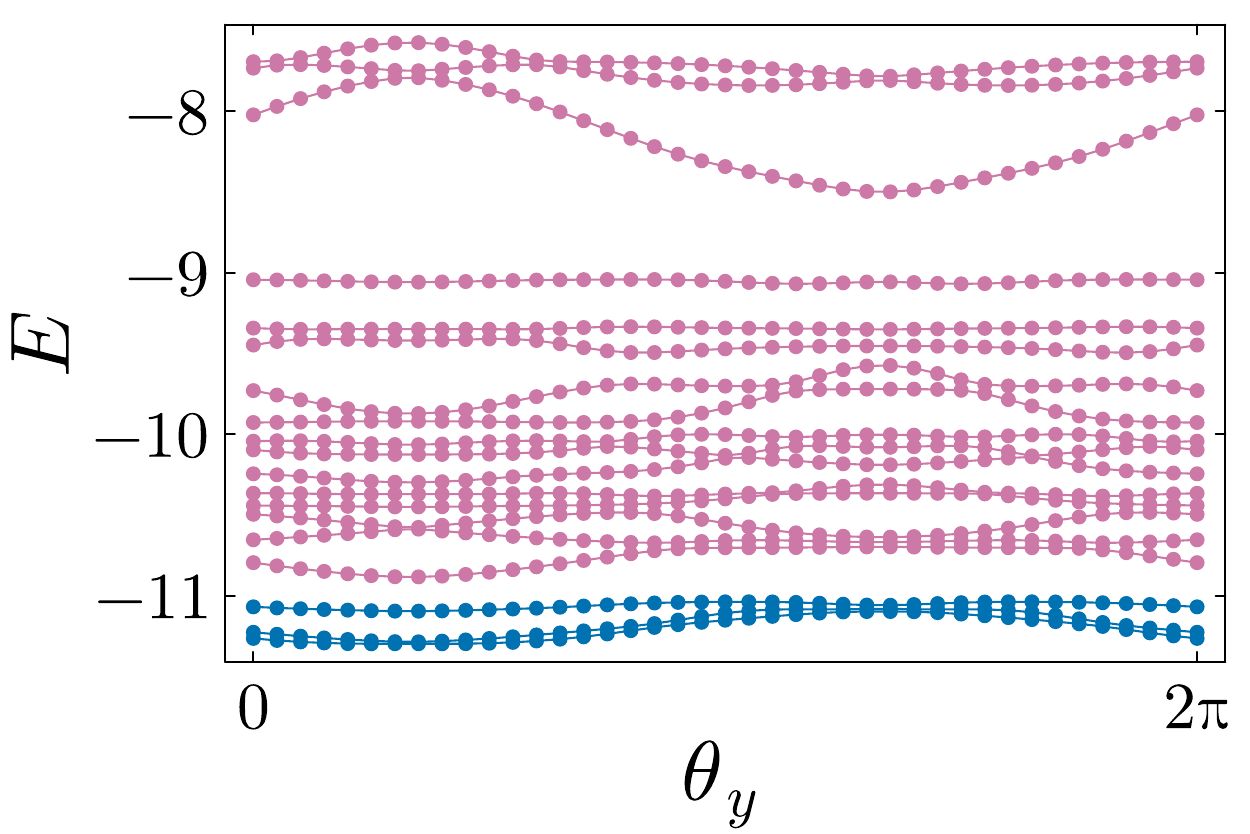}
    \end{subfigure}\\
    \begin{subfigure}{0.49\linewidth}
        \centering
        \textbf{(e) $N=6,U=3.0$}
        \includegraphics[width=\linewidth]{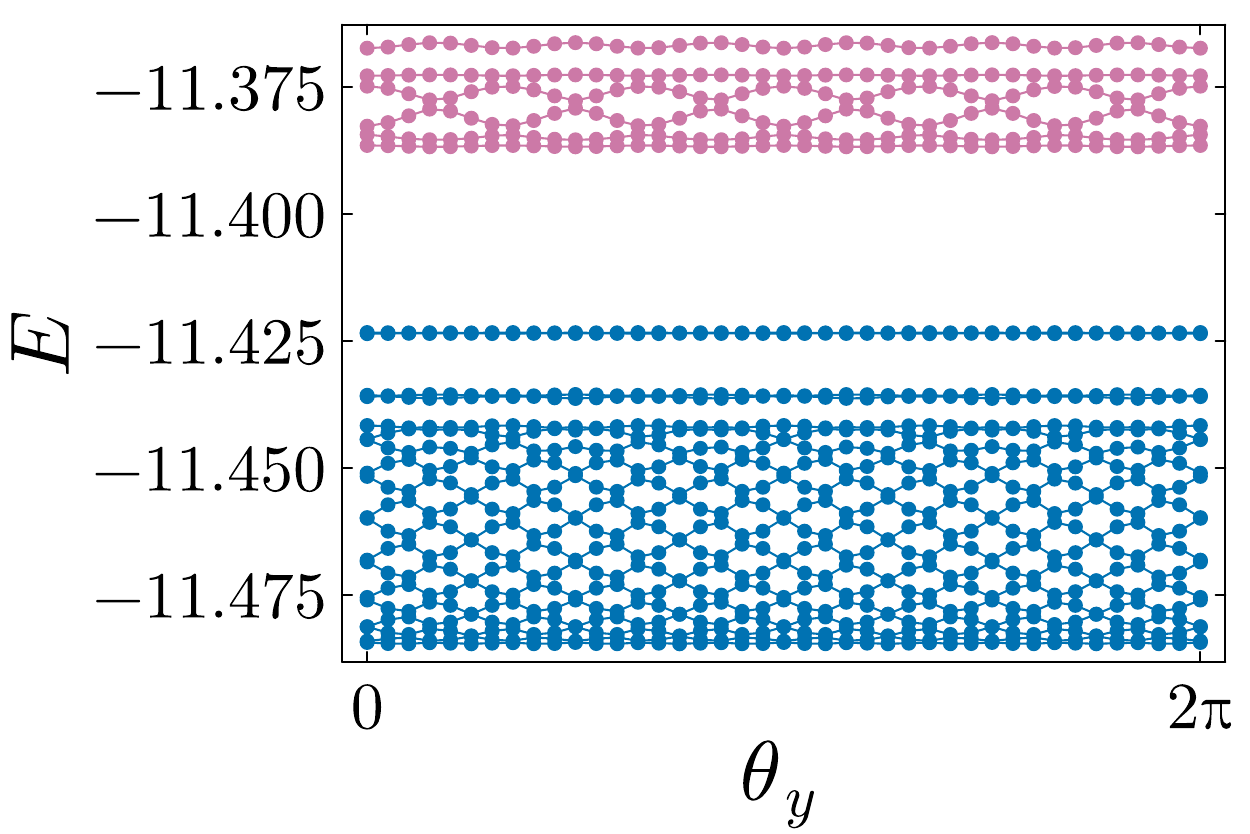}
    \end{subfigure}
    \begin{subfigure}{0.49\linewidth}
        \centering
        \textbf{(f) $V_{p}=10.0$}
        \includegraphics[width=\linewidth]{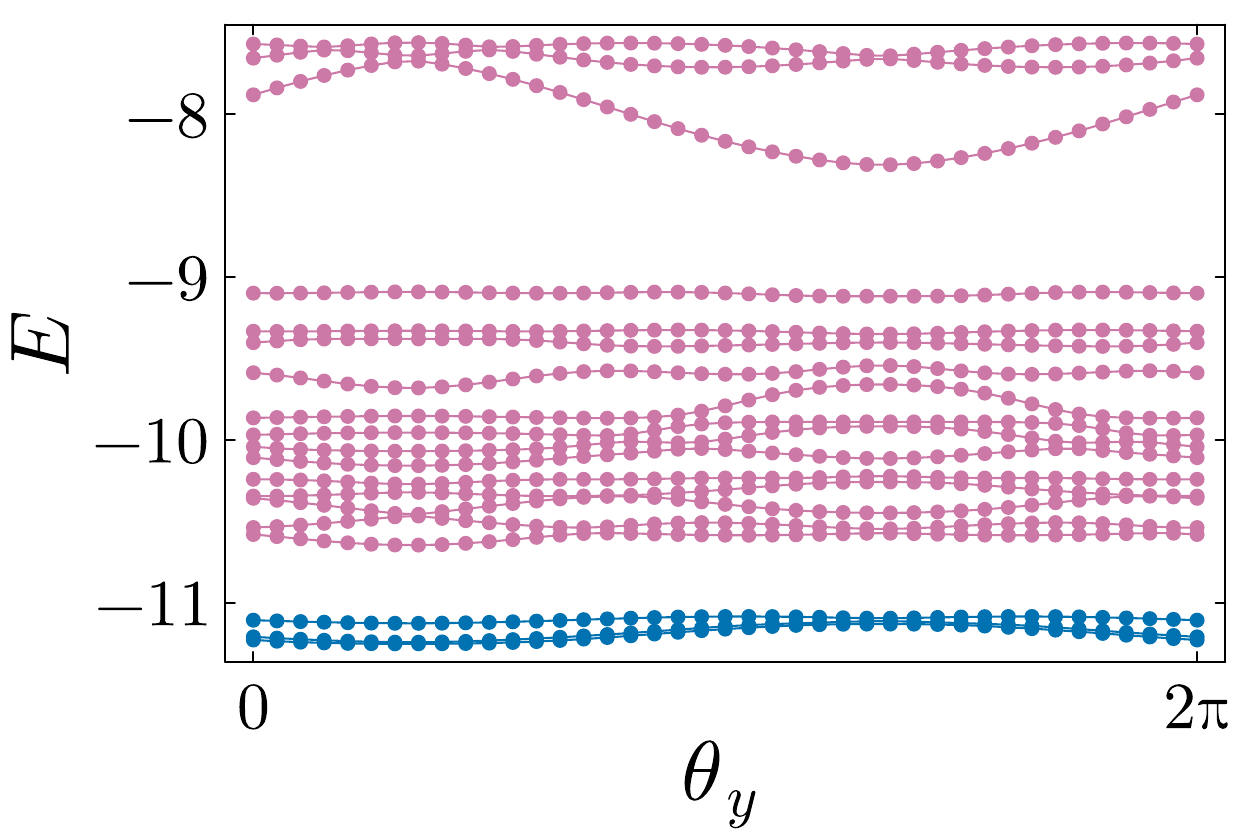}
    \end{subfigure}\\
    \caption{Spectral flow in the checkerboard model for the system $N_s=36$, $N=6$, and $N_{\phi}=-1$.
    The parameters are (a) $U=1, V_p=0$, (b) $U=1, V_p=10$, (c) $U=2, V_p=0$, (d) $U=2, V_p=10$, (e) $U=3, V_p=0$, and (f) $U=3, V_p=10$.
    Blue curves represent in (a), (c), (d) are states expected from the counting rule $\#_{\text{qh}}=19$. 
    Blue curves in (b), (d), (f) are the lowest three energy levels.}
    \label{fig:KSSF1}
\end{figure}

\begin{figure}
    \centering
    \begin{subfigure}{0.98\linewidth}
        \centering
        \includegraphics[width=\linewidth]{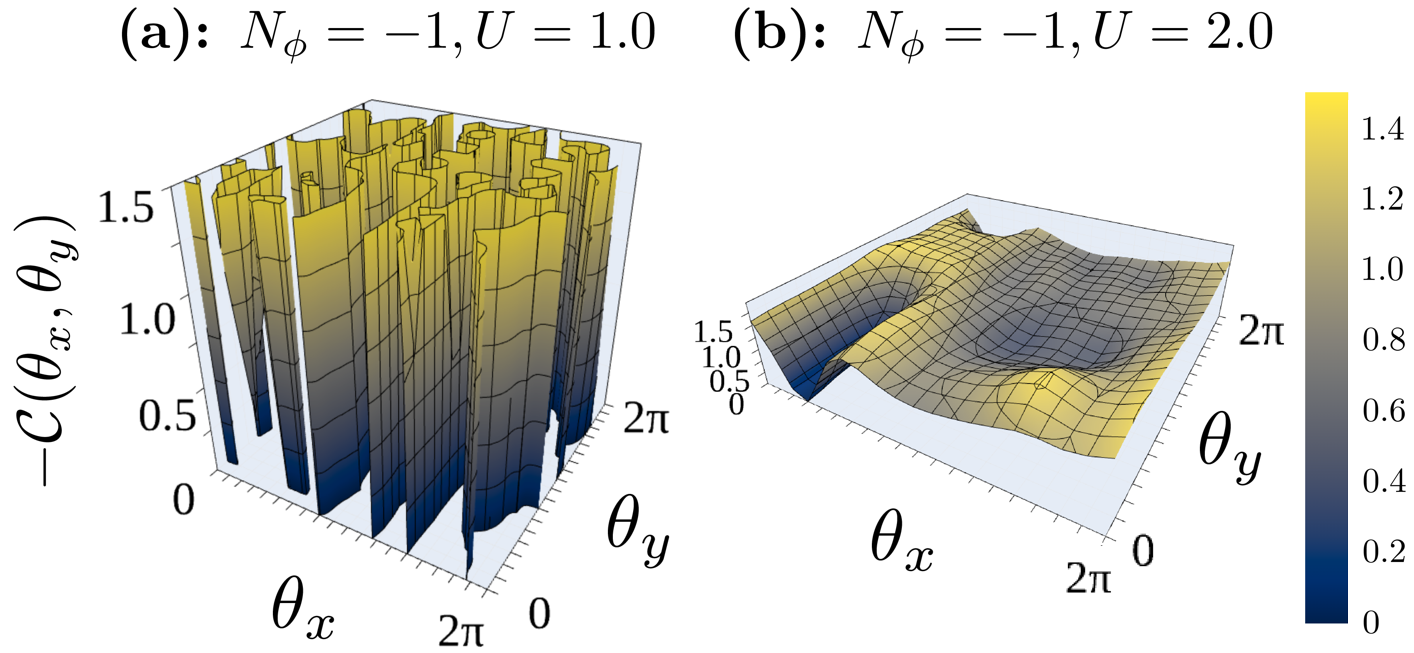}
    \end{subfigure}\\
    \begin{subfigure}{0.98\linewidth}
        \centering
        \includegraphics[width=\linewidth]{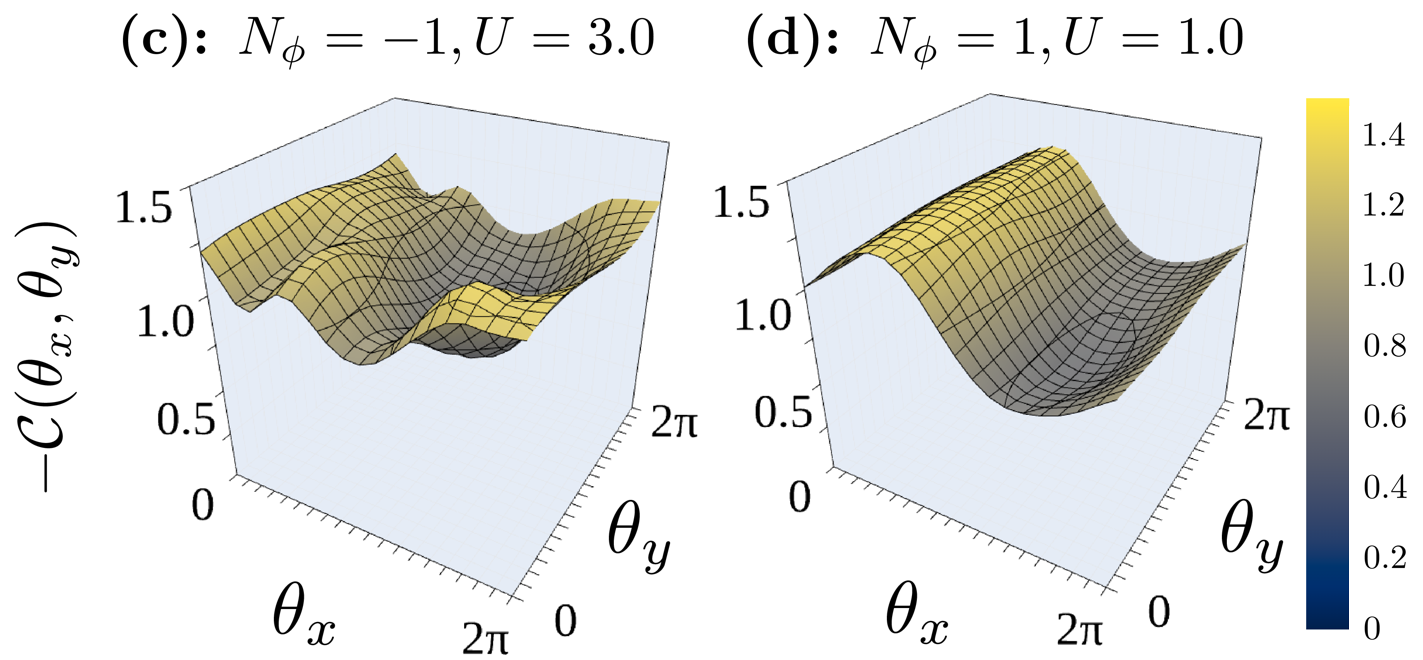}
    \end{subfigure}
    \caption{One-plaquette Chern number $\mathcal{C}$ for the three lowest energy states of $\tilde{H}$ in the checkerboard model at (a) $N=6, N_{\phi}=-1, U=1$, (b) $N=6, N_{\phi}=-1, U=2$, (c) $N=6, N_{\phi}=-1, U=3$, (d) $N=6, N_{\phi}=+1, U=1$. {The pinning potential is $V_p=10$ for the quasiholes and $V_p=-1$ for the quasielectrons.}}
    \label{fig:KSOPCN1}
\end{figure}

     We present numerical results of the spectral flow for the quasihole creation in the checkerboard model at $N=6, U=1.0$ in Fig.~\ref{fig:KSSF1}(a). 
     The blue curves in Fig.\ref{fig:KSSF1}(a) represent the energy levels expected from the counting rule, $\#_{\text{qh}}=19$ for $N=6, N_s/2=18, C_sN_{\phi}=1$. 
     However, at $U=1.0$, there is no clear energy gap between the blue curves and others, and it is difficult to identify the ground state sector in contrast to the Kapit-Mueller model in the previous section.
Consequently, there are no clear three fold quasi-degenerate ground states with the pinning potential as shown in Fig.~\ref{fig:KSSF1}(b).
Note that the pinning potential is sufficiently large $V_p=10$ so that density distribution is well localized around the site $j_0$, and we have confirmed the absence of the quasi-degeneracy for other values of the pinning potential, {$V_p=0.1,1$}.
This implies that quasiholes are not created and the underlying FCI state is no longer stable in the presence of the magnetic field.
Indeed, the one-plaquette Chern number for the three lowest energy states in Fig.~\ref{fig:KSSF1}(b) diverges and has large fluctuations (Fig.~\ref{fig:KSOPCN1}(a)). 
Since the one-plaquette Chern number is nearly flat in the $\theta$-space in a general gapped system in the thermodynamic limit~\cite{PhysRevLett.122.146601,koma2015topologicalcurrentfractionalchern}, the strong divergence implies that the present system with $U=1$ does not have the three fold topological quasi-degeneracy {well separated from excited states}.
Therefore, the ground state is a non-FCI state.

\begin{figure*}
    \centering
    \begin{subfigure}{0.49\linewidth}
        \centering
        \textbf{(a): Many-Body Chern Number}
        \includegraphics[width=\linewidth]{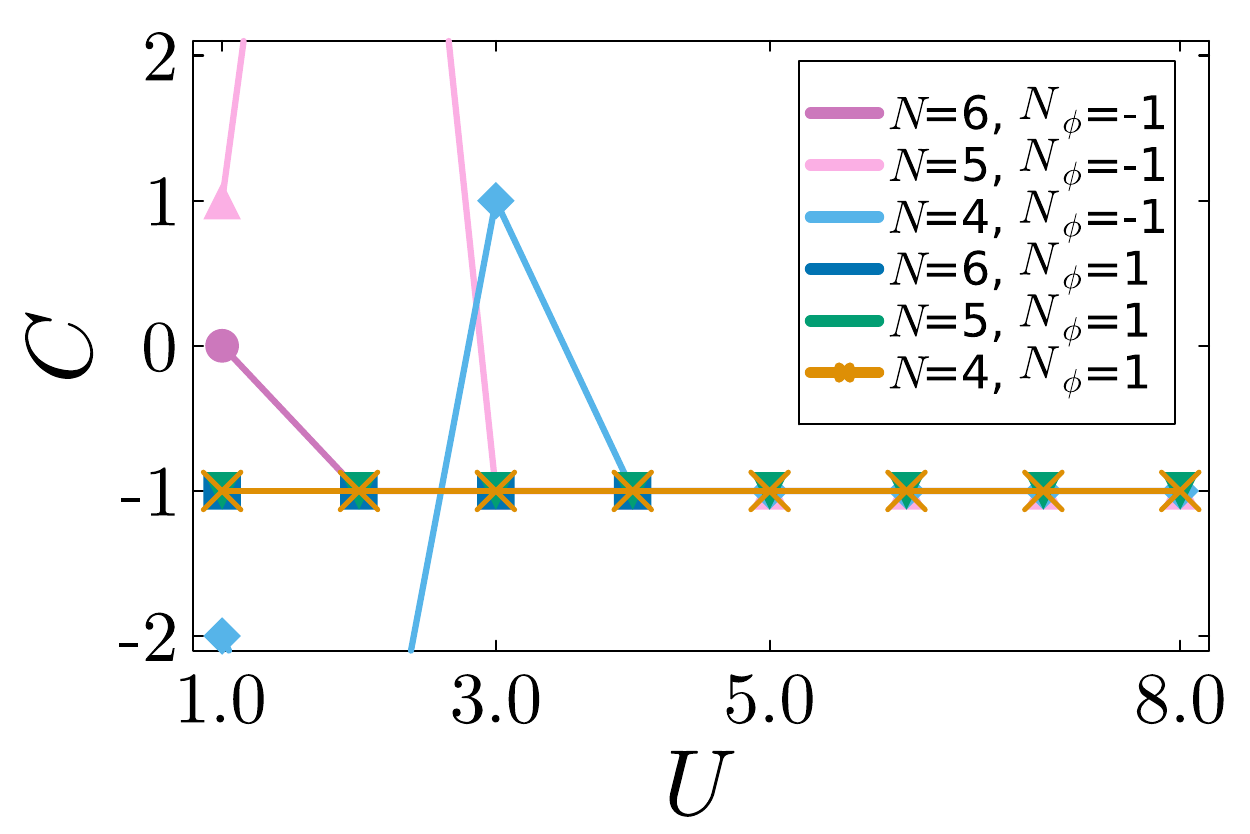}
    \end{subfigure}
    \begin{subfigure}{0.49\linewidth}
        \centering
        \textbf{(b): Variance of one-plaquette Chern number}
        \includegraphics[width=\linewidth]{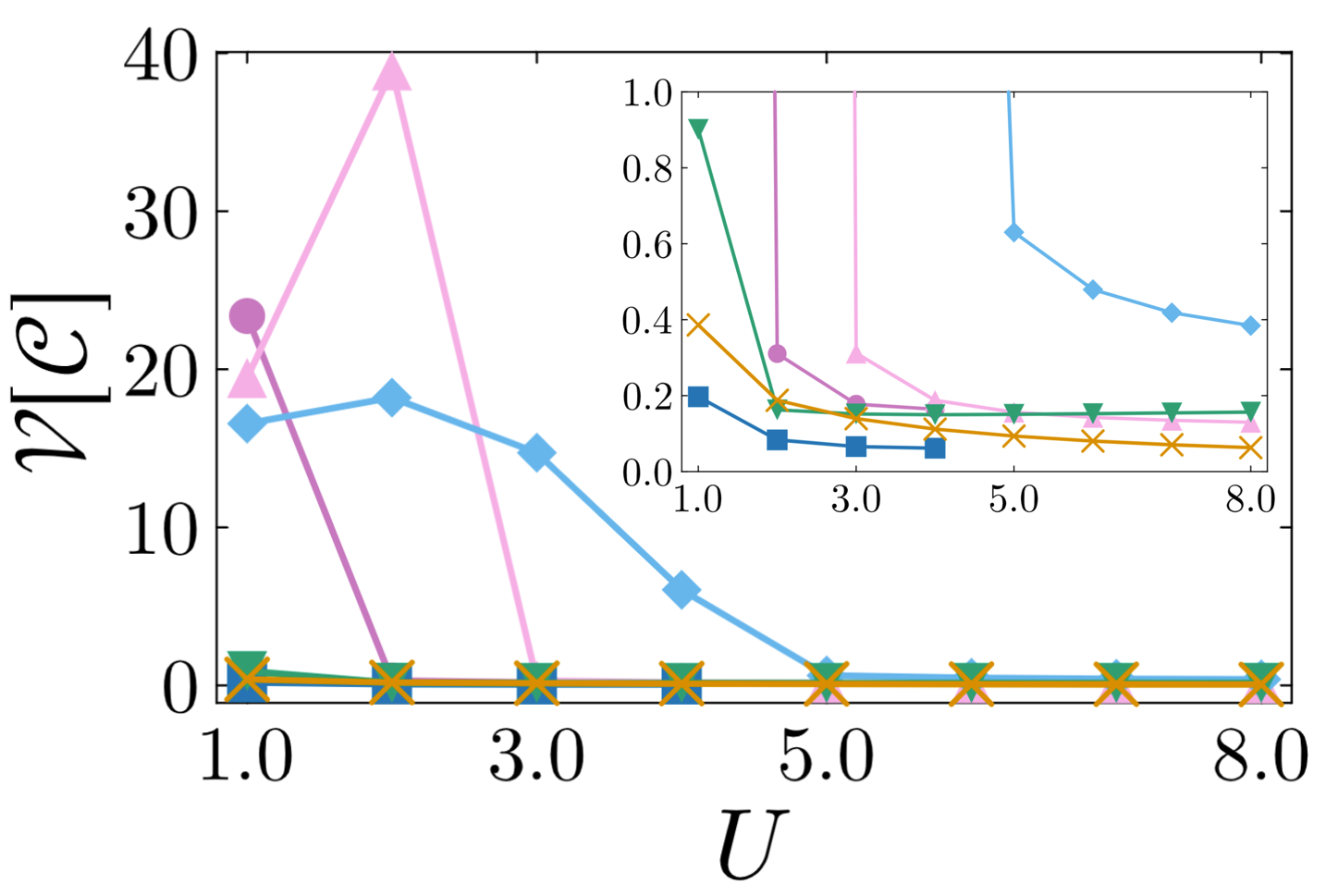}
    \end{subfigure}
    \caption{(a) Many-body Chern number $C$ for various system sizes and fluxes.
    (b) Variance $\mathcal{V}[\mathcal{C}]$. The inset is an enlarged view so that the diverging behaviors are clearly seen. {The pinning potential is $V_p=10$ for the quasiholes and $V_p=-1$ for the quasielectrons.}}
    \label{fig:KSOPCN2}
\end{figure*}

    However, as we increase the interaction strength from $U=1.0$ to $U=2.0,3.0$, an energy gap develops in the spectral flow at $V_p=0$ and the low energy 19 states become separated from others as seen in Fig.~\ref{fig:KSSF1}(c),(e). {This implies that the spatially extended quasihole states may be created by the magnetic field when the interaction is strong.}
    Furthermore, in the presence of the pinning potential, we find that there exists a clear gap between the three lowest energy levels and other excitations. 
    The many-body Chern number for these three fold quasi-degenerate states is $C=-1$ and the variance $\mathcal{V}[\mathcal{C}]$ for $U=2.0\sim3.0$ is suppressed as shown in Figs.~\ref{fig:KSOPCN1} and ~\ref{fig:KSOPCN2}.
These results suggest that there is a quantum phase transition from the non-FCI state to the stable FCI state as $U$ increases. 
The phase transition point is located around $U=2.0$, where the diverging behaviors of the variance $\mathcal{V}[\mathcal{C}]$ get suppressed and it becomes rather flat as seen in Fig.~\ref{fig:KSOPCN2}(b).
To be more precise, a quantum phase transition is well defined only in the thermodynamic limit.
In a thermodynamically large system, the number of the fluxes is $|N_{\phi}|\gg1$ for a fixed magnetic field $\phi$ and there will be $|N_{\phi}|$ quasiparticles localized by a pinning potential corresponding to disorder.
We have simply assumed that effects of interactions between the pinned quasiparticles are negligible.
This enables us to discuss the phase transition based on the exact diagonalization for the small system with $|N_{\phi}|=1$.
{The above discussions are based on the results with the strong pinning potential $V_p=10$, but we have confirmed that the qualitative results that the FCI state is unstable at weak interactions is unchanged for other values of the pinning potential, $V_p=0.1,1$. We have also checked that qualitatively similar results can be obtained without the projection onto the low energy sector. However, it is difficult to quantitatively evaluate the critical interaction strength, and it will be an interesting future problem.}

Similar phase transitions are found for the systems with different numbers of the fermions, $N=4,5$. 
The applied flux per plaquette is $\phi=-1/N_s\simeq -0.042$ for $N=4$ and $\phi\simeq -0.033$ for $N=5$, which enables us to discuss magnetic field dependence of the quantum phase transition. 
Similarly to the $N=6$ case, as $U$ increases,
energy gaps develop between the $\#_{\text{qh}}$ lowest energy states and other states at $V_p=0$, and also between the three lowest energies and excited states for $V_p\neq0$ (see Appendix~\ref{app:cb45} for details).
We find that the strength of the interactions $U_c$ required to stabilize the FCI increases as $N$ decreases. 
Besides, the variance of the Chern number is large for small interactions as seen in Fig.~\ref{fig:KSOPCN2}(b).
Based on these numerical results,
we suppose that the phase transition occurs at $U_c=2.0\sim3.0$ for $N=5$ and $U_c=4.0\sim5.0$ for $N=4$. 
This in turn implies that the critical interaction $U_c$ increases as the negative magnetic field increases.
There would be some finite size effects for smaller system sizes, but we can confirm that it is not relevant in the present system.
First, when the system size is changed, quasiholes are stably created for $N=5, U=1.0$ at $\phi=0$ (Appendix~\ref{app:unitcell}), which means that the absence of the FCI state for $N=5, U=1.0$ at $\phi\simeq -0.033$ is due to the magnetic field. 
Second, as will be discussed in the following, the FCI state becomes stable when the direction of the magnetic field is changed, $\phi\to -\phi$.
Therefore, the stability of the FCI state is dominated by the magnetic field.

\begin{figure}[t]
    \centering
    \begin{subfigure}{0.49\linewidth}
        \centering
          \textbf{(a) $N=6,U=1.0$}
        \includegraphics[width=\linewidth]{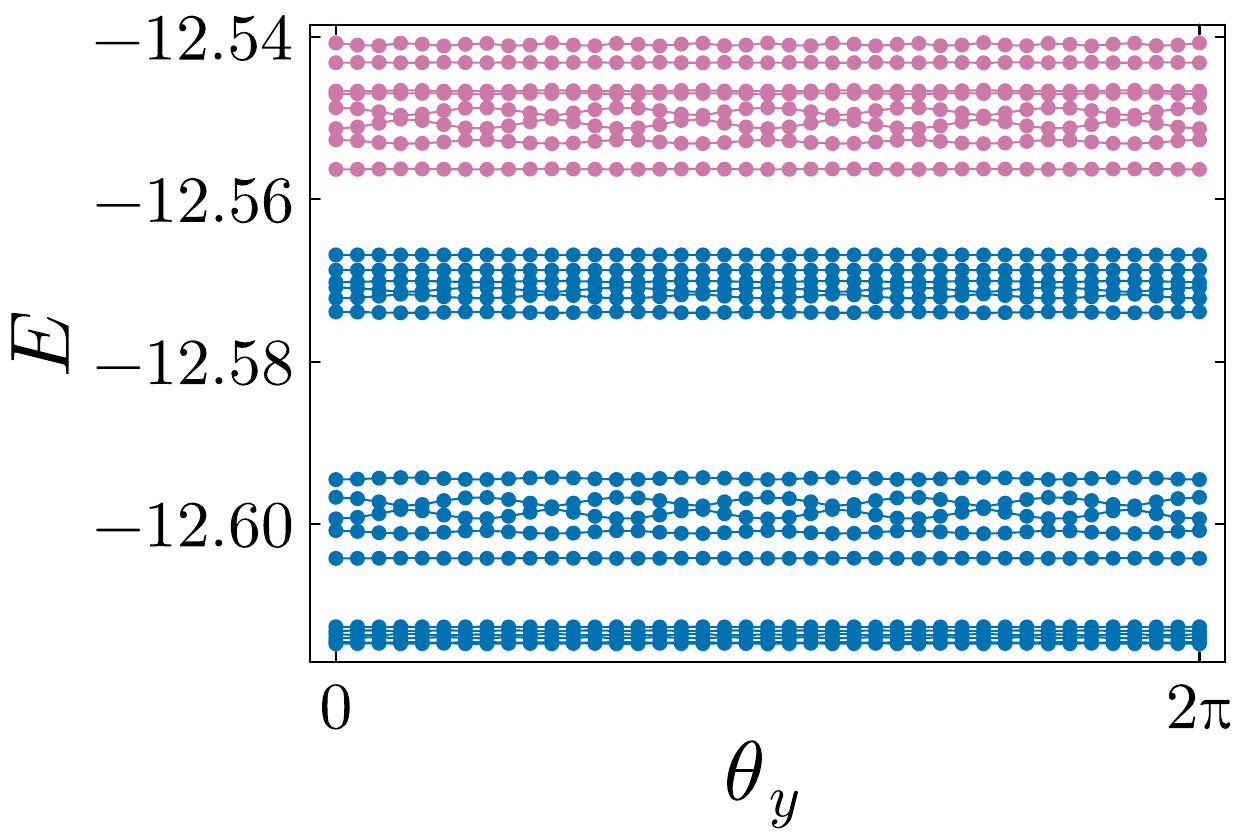}
    \end{subfigure}
    \begin{subfigure}{0.49\linewidth}
        \centering
        \textbf{(b) $V_{p}=-1.0$}
        \includegraphics[width=\linewidth]{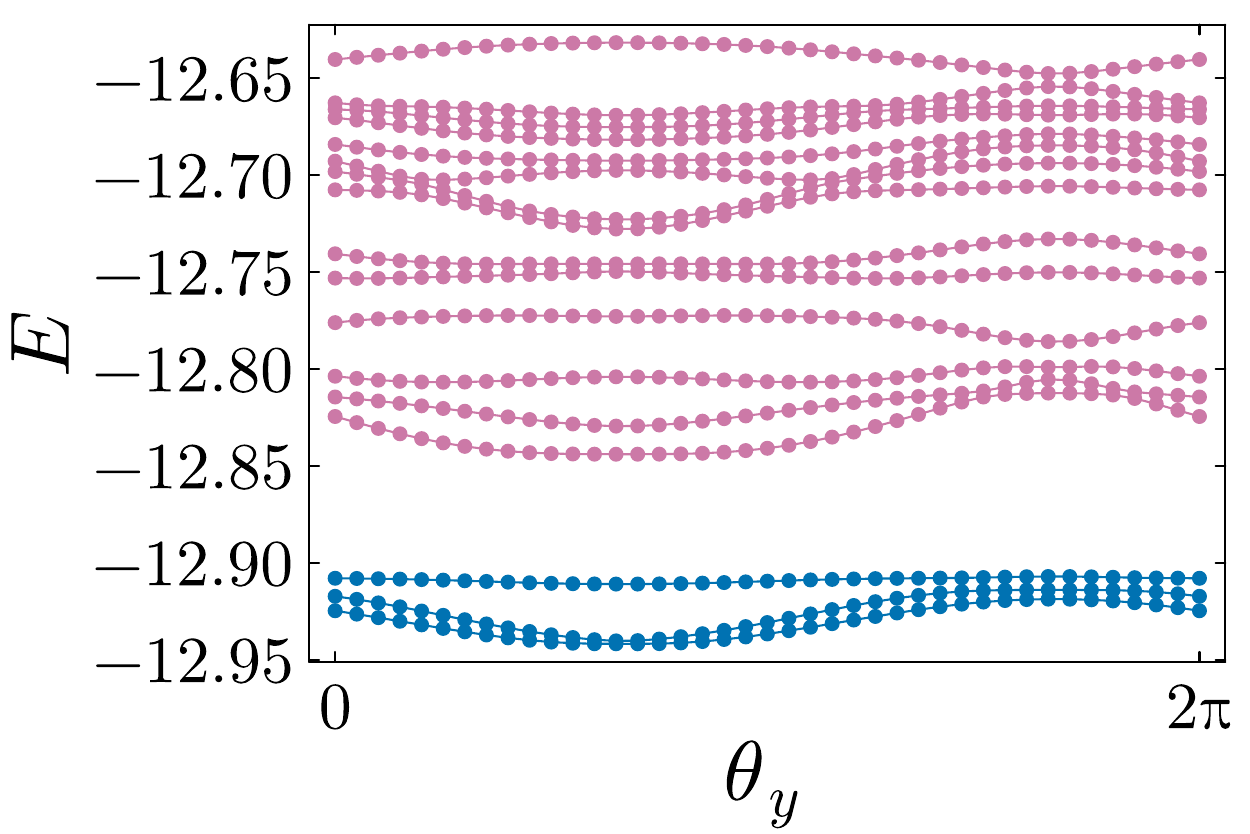}
    \end{subfigure}\\
    \caption{Spectral flow in the checkerboard model for $N_{\phi}=1,N_s=36,N=6$ at (a) $V_p=0$ and (b) $V_p=-1$.
    The blue curves in (a) are the $\#_{\text{qe}}=17$ states. 
    The blue curves in (b) are the lowest three energy states. }
    \label{fig:KSSF3}
\end{figure}

    In contrast to the quasihole case, creation of the quasielectrons is stable even for small interactions. 
    As shown in Fig.~\ref{fig:KSSF3} and Fig.~\ref{fig:KSSF4} in Appendix~\ref{app:cb45}, the counting rule derived from the numerical results for the FQHE and the three fold quasi-degeneracy are realized for $N=4,5,6$ at $U\approx1.0$.
    Moreover, the many-body Chern number $C=-1$ is robust even at the small interaction $U=1.0$ for all $N=4,5,6$. 
    Note that the variance ${\mathcal V}[C]$ is slightly large at $U=1.0$ for $N=4$ and $5$, but this may be due to finite size effects and ${\mathcal V}[C]$ is quickly suppressed for larger $U$. 
    Thus, the stable FCI region is extended over a wide range of interaction strength and magnetic fields for the quasielectron creation.
\begin{figure}[htbp]
    \centering
     \includegraphics[width=1.0\linewidth]{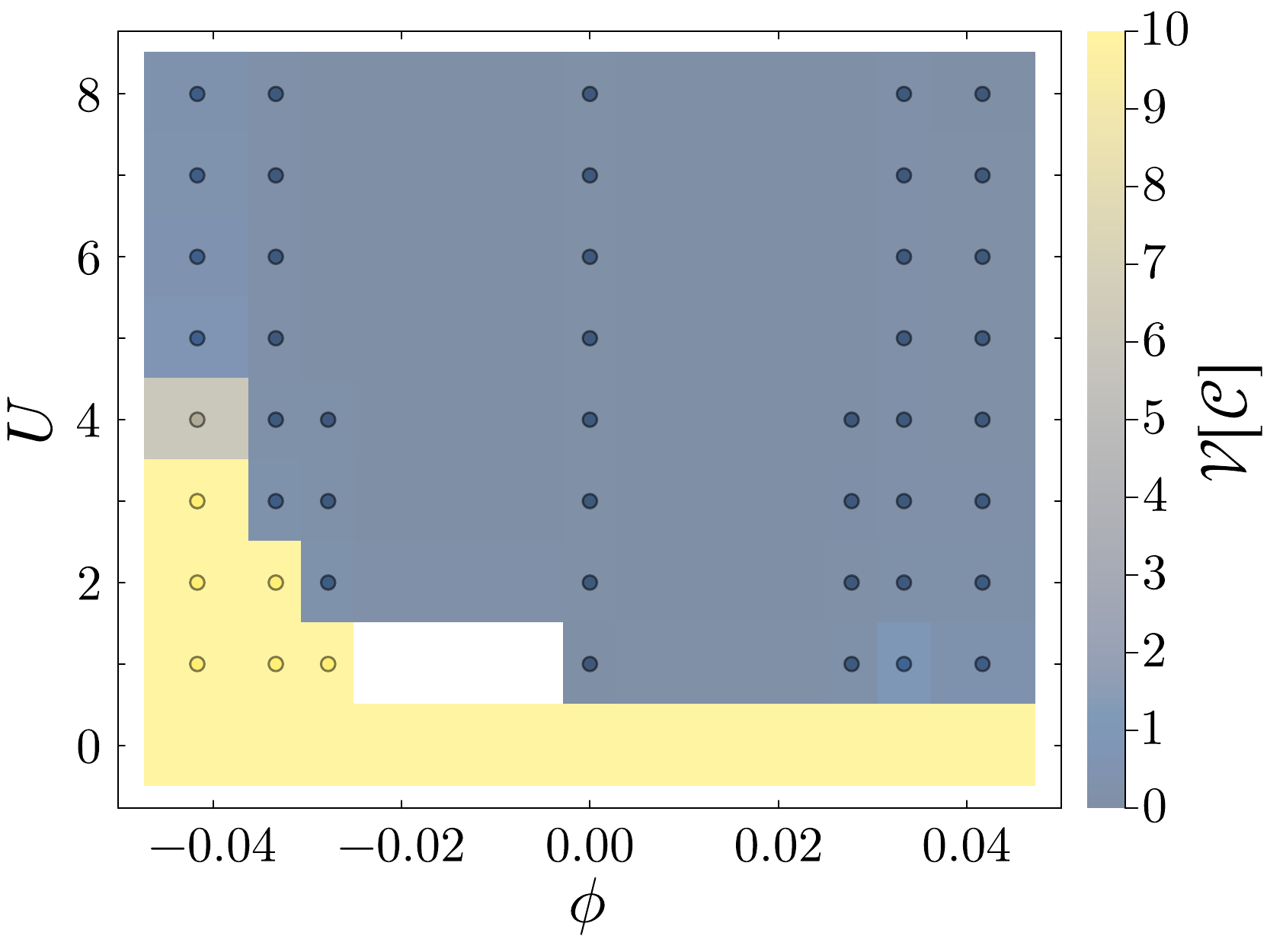}
     \caption{Variance $\mathcal{V}[\mathcal{C}]$ in the checkerboard model as a function of the flux $\phi$ and the interaction $U$. 
     The many-body Chern nubmer is $C=-1$ in the region with $\mathcal{V}[\mathcal{C}]\simeq0$, while it is not well-defined when $\mathcal{V}[\mathcal{C}]$ is not negligibly small. 
     The variance in the white-colored region cannot be interpolated from that in neighboring regions. {The pinning potential is $V_p=10$ for the quasiholes and $V_p=-1$ for the quasielectrons.}}
     \label{fig:colormapKS}
\end{figure}

In Fig.~\ref{fig:colormapKS}, we summarize the numerical results of the variance ${\mathcal V}[C]$ as an indicator for the FCI state.
The many-body Chern number for the three fold quasi-degenerate ground states with $V_p\neq0$ is $C=-1$ and it has small fluctuation ${\mathcal V}[C]$ in the FCI state, while the Chern number is not well defined and the variance diverges in the non-FCI state.
The flux $|\phi|=1/N_s\simeq 0.042,0.033, 0.028$ corresponds to the different numbers of the fermions, $N=4,5,6$, respectively. 
We see that the creation of the quasiparticles and the underlying FCI state are stable for a wide range of the parameters, while the non-FCI state is realized around the non-interaction region $U=0$ and under negative fluxes even with moderate interactions.
Based on these results, we obtain the expected phase diagram Fig.~\ref{fig:Schematic}(b) in Sec.~\ref{sec:level3}.
It is difficult to clarify details of the non-FCI state within the present calculations, and further discussions are left for a future study.

\section{\label{sec:level5}MultiBand Quantum Geometry}
\begin{figure*}[t]
    \centering
    \begin{subfigure}{0.49\linewidth}
        \centering
        \textbf{(a) Kapit-Mueller model}
        \includegraphics[width=\linewidth]{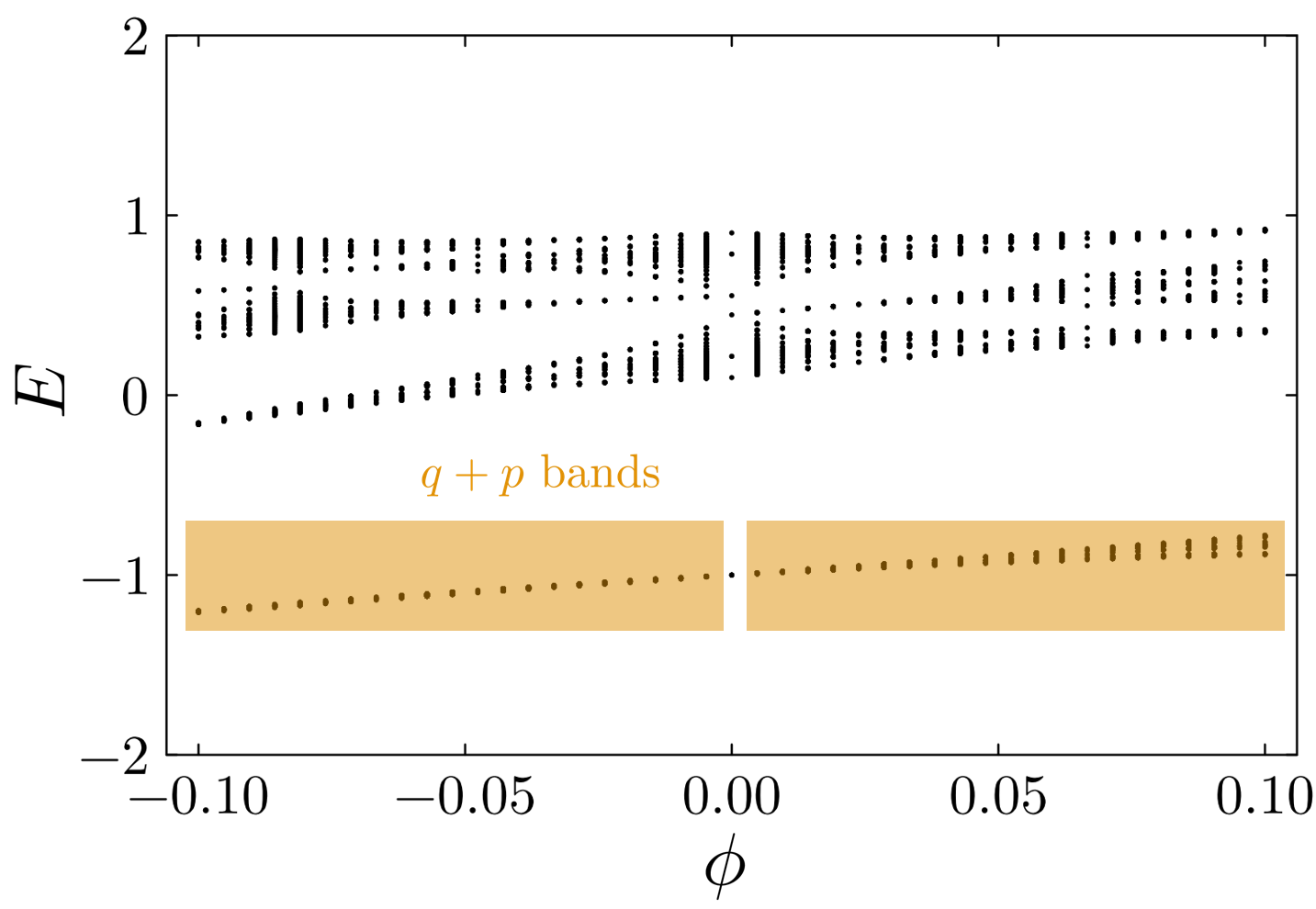}
    \end{subfigure}
    \begin{subfigure}{0.49\linewidth}
        \centering
        \textbf{(b) Checkerboard model}
        
        \includegraphics[width=\linewidth]{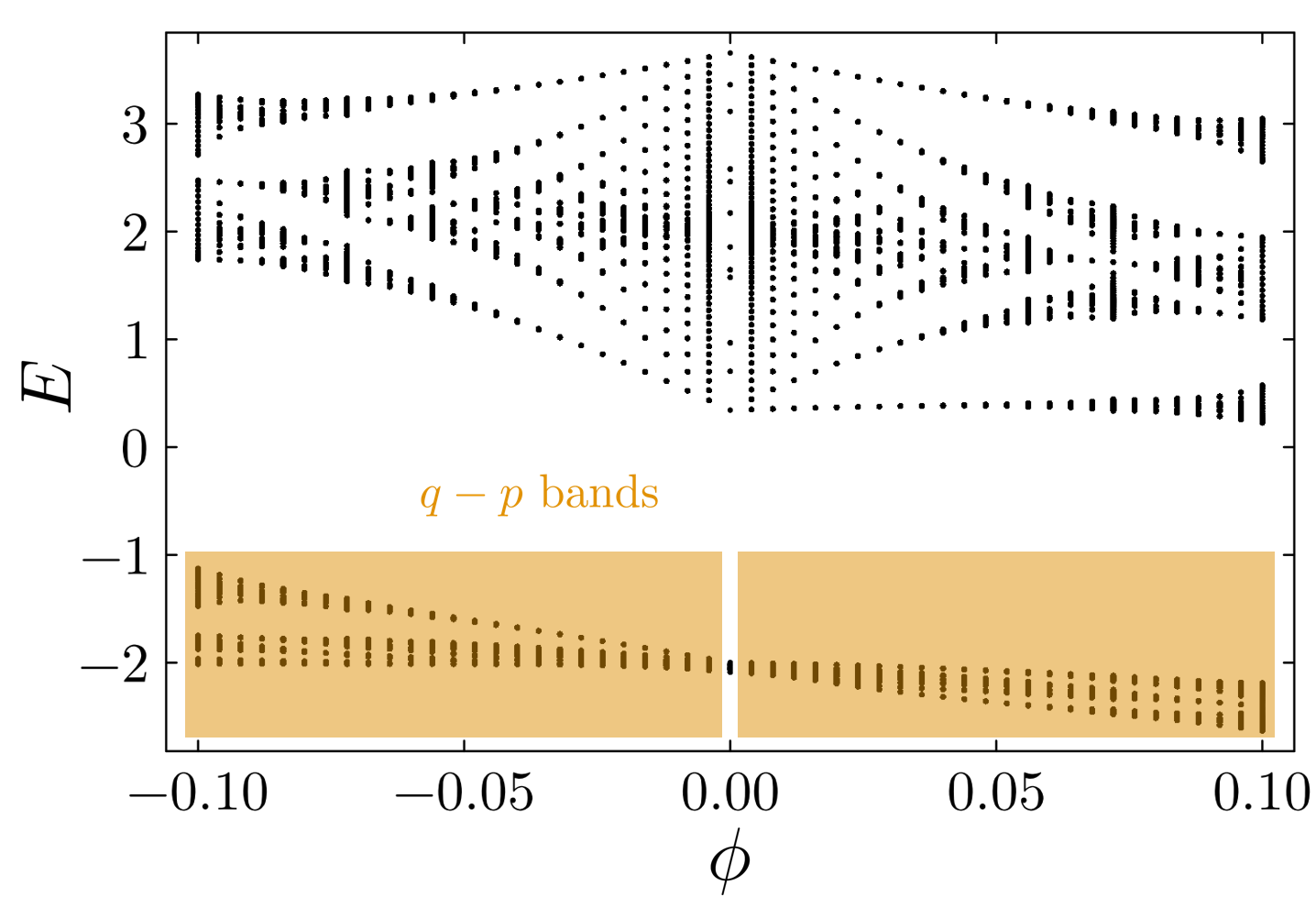}
        \end{subfigure}
    \caption{Hofstadter butterflies of the single-particle states for (a) the Kapit-Mueller model and (b) the ckeckerboard model. 
    The colored regions in correspond to the multibands connected to the single flatbands in the absence of magnetic fields. In this region, the number of the bands is derived by the Streda's formula as $q+p$ in the Kapit-Mueller model and $q-p$ in the checkerboard model.}
    \label{fig:MB}
\end{figure*}
The exact diagonalization in the previous section suggests that stability of the FCI states to the magnetic fields varies depending on models.
In terms of the quantum geometry, the Kapit-Mueller model satisfies the ideal condition at $\phi=0$ while the checkerboard model does not, which is related to the stability of the FCIs in the absence of magnetic fields.
Motivated by this result, 
here we discuss relation between band structures and the stability of the FCI states under magnetic fields.

In the original framework of the FQHE, a magnetic field merely changes the number of the single-particle orbits as $N_{\text{orb}}\to N_{\text{orb}}+C_sN_{\phi}$ and does not affect single-particle properties, which leads to the plateaus in the Hall conductance. 
For systems with their own band structures, however, the magnetic field changes $N_{\text{orb}}$ and simultaneously can affect single-particle states.
This is because the single-particle states are generally not given by Landau level wavefunctions and they depend on applied magnetic fields in a non-trivial way.
Note that modification of single-particle properties is absent when the system size is changed instead of applying magnetic fields~\cite{PhysRevX.1.021014,PhysRevB.91.045126,PhysRevB.99.045136,PhysRevB.87.205136}.

A single-particle energy band generally splits into several bands under a magnetic field, and correspondingly FCI states will be constructed on the multiband. 
If the energy scale of the multiband is smaller than the interaction energy scale and thus they can be effectively regarded as a single band, the Laughlin state can be realized for multiband systems~\cite{PhysRevB.104.115126,schoonderwoerd2022interactiondrivenplateautransitioninteger}. 
The condition for the realization of FCIs was proposed in the previous studies~\cite{parker2021fieldtunedzerofieldfractionalchern,PhysRevLett.134.106502,liu2024theorygeneralizedlandaulevels}, which is described by the multiband quantum geometric tensor, 
\begin{equation}
        \eta_{\alpha \beta}^{a b}(\bm{k}) = N_{\text{b}}A\mel{\partial_{k_\alpha}u^{b}_{\bm{k}}}{1-\mathcal{P}(\bm{k})}{\partial_{k_\beta}u^{a}_{\bm{k}}},
\end{equation}
where $\mathcal{P}(\bm{k})=\sum_{a=1}^{N_{\text{b}}}\ket{u^{a}_{\bm{k}}}\bra{u^{a}_{\bm{k}}}$ is the projection operator onto the target $N_{\text{b}}$ bands which we focus on. 
{The factor $N_{\rm b}A$ is redundant, but it has been introduced for later convenience in Appendix~\ref{app:flatness}.}
The multiband quantum metric and Berry curvature are given as the symmetric and anti-symmetric parts of $\eta$,
\begin{equation}
g_{\alpha \beta}^{a b}(\bm{k}) = \frac{\eta_{\alpha \beta}^{a b}+\eta_{\beta \alpha}^{a b}}{2}, \quad\mathcal{F}^{a b}(\bm{k}) = i\sum_{\alpha \beta}\epsilon^{\alpha \beta}\eta_{\alpha \beta}^{a b},
\end{equation}
where $\epsilon^{\alpha \beta}$ is the Levi-Civita symbol.
We extend $C_s$ to the single-particle Chern number of the multiband given by
\begin{equation}
    C_s = \frac{1}{2\pi}\text{Tr}\qty[\mathcal{F}],
\end{equation}
where Tr is the trace over the momentum and the band index,
\begin{equation}
\text{Tr}\qty[\mathcal{O}] = (N_{\text{b}}A)^{-1}\sum_{a=1}^{N_{\text{b}}}\int d^2k \mathcal{O}^{aa}(\bm{k}).
\end{equation}
The multiband condition to realize Laughlin states is the ideal condition extended to a multiband system, which also ensures that single-particle wavefunctions are holomorphic in the momentum space. The ideal condition is described by the following equations,
\begin{equation}
        \text{tr} g^{a a} = \mathcal{F}^{a a}\quad\text{for  }1\le a \le N_{\text{b}},\bm{k} \in \text{BZ},
\end{equation}
where $\text{tr}$ is a trace over the indices $\alpha, \beta$~\cite{parker2021fieldtunedzerofieldfractionalchern,PhysRevLett.134.106502,liu2024theorygeneralizedlandaulevels}. 
{This is a set of ideal conditions for each constituent single-band.}
Since the multiband quantum geometry plays fundamental roles for realization of the FCI state, 
we naively expect that the multiband quantum geometry under magnetic fields is relevant also for stability of the quasiparticle creation. 
More precisely, our expectation is that if a clean system with a ``commensurate" filling (i.e. $\nu=1/3$) has good multiband quantum geometry under a magnetic field $\phi\neq0$ and thus exhibits an FCI ground state, creation of spatially extended quasiparticles induced by the ``incommensurate" filling (i.e. $\nu=1/3+\delta \nu$ according to the Streda's formula) is stable at the same magnetic field $\phi$. 
The itinerant quasiparticles will be localized if there is a pinning potential.
It turns out that this conjecture is indeed helpful for understanding the exact diagonalization results from the perspective of band properties.

Firstly, we present Hofstadter butterflies for the two models in Fig.~\ref{fig:MB}(a), (b). 
In the Kapit-Mueler model,
the lowest energy multibands clearly remain for $|\phi|\leq 0.1$. 
On the other hand, in the checkerboard model, it seems that the lowest energy levels split into distinct bands especially for $\phi<0$.
However, these bands should be regarded as a multiband based on the Streda's formula, $N_{\text{orb}}\to N_{\text{orb}}+C_sN_{\phi}$.
When there is a flux $\phi_{uc}=p/q$ ($p,q$ are coprime) per unit cell of the checkerboard lattice containing two sites, the total number of the bands in the magnetic Brillouin zone is $2q$.
The number of low energy bands which are connected from the original lowest energy band at $\phi_{uc}=0$ is $(N_{\text{orb}}+C_sN_{\phi})/(N_s/2q)=q-p$, and this numerically matches the number of the single-particle levels in the energy range $E<0$ at $\phi\neq0$.
Therefore, the $q-p$ split bands form the multiband on which the FCI state is realized.
Similarly, in the Kapit-Mueller model, the number of the low energy bands is $q+p$ because of $C_s=1$. 
Note that the unit cell of Kapit-Muller (checkerboard) model contains three (two) sites, and thus $\phi_{uc}=3\phi (\phi_{uc}=2\phi)$ in the Kapit-Muller (checkerboard) model, where $\phi$ is the flux per square plaquette.

Bandwidth of single-particle states is generally important for a multiband FCI driven by interactions. 

\begin{figure}[t]
    \centering
        \begin{subfigure}{0.99\linewidth}
        \centering
        \includegraphics[width=\linewidth]{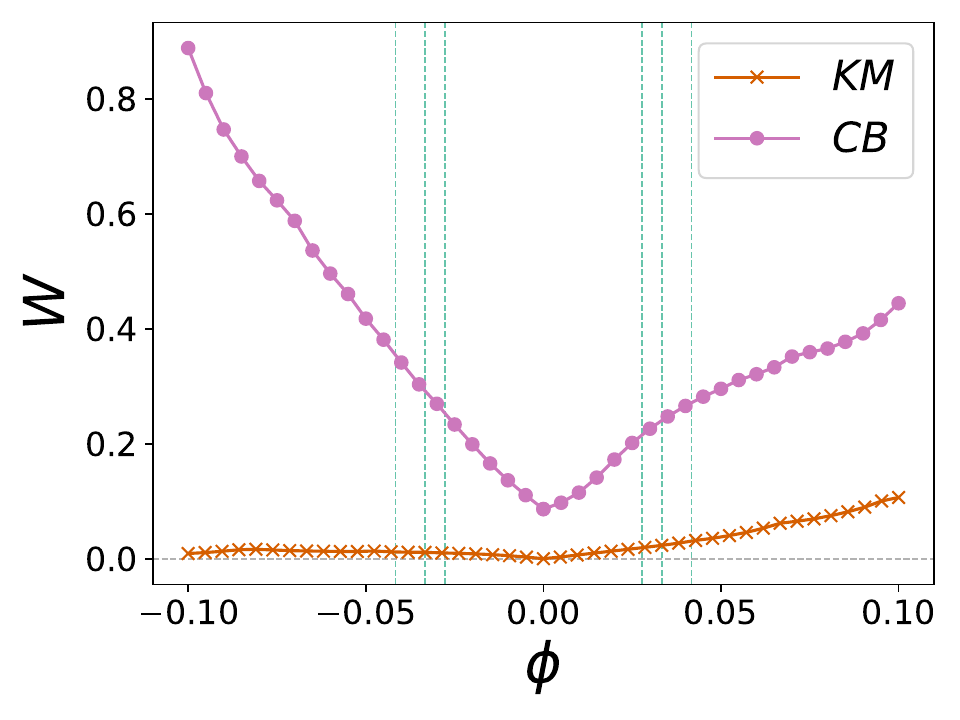}
        \end{subfigure}
    \caption{Width of the colored multiband in Fig.~\ref{fig:MB} for the Kapit-Mueller model (KM) and the checkerboard model (CB). The dashed vertical lines correspond to the values of $\phi$ used in the exact diagonalization.}
    \label{fig:MB2}
\end{figure}

We show the width $W$ of the multiband in Fig.~\ref{fig:MB2}, which is defined as the difference between the maximum and minimum energy values within the multiband. 
In the Kapit-Mueller model, the bandwidth $W$ is nearly unchanged for $\phi<0$ and its change is non-zero but small for $\phi>0$.
In the checkerboard model, the bandwidth varies largely at $\phi\neq0$. 
The negative magnetic field which induces quasihole creation enhances the multiband width more strongly than the positive magnetic field corresponding to the quasielectron creation. 
This indicates that a larger interaction is required to realize the multiband FCI for the negative strong magnetic field. However, within the range of $\phi$ considered in the exact diagonalization (the dashed vertical lines in Fig.~\ref{fig:MB2}), the difference of the bandwidth for $\pm\phi$ is relatively small. 
It would not be relevant for the asymmetric stability of the FCI under $\phi>0$ and $\phi<0$ found in the exact diagonalization.

\begin{figure}[t]
    \centering
        \includegraphics[width=0.99\linewidth]{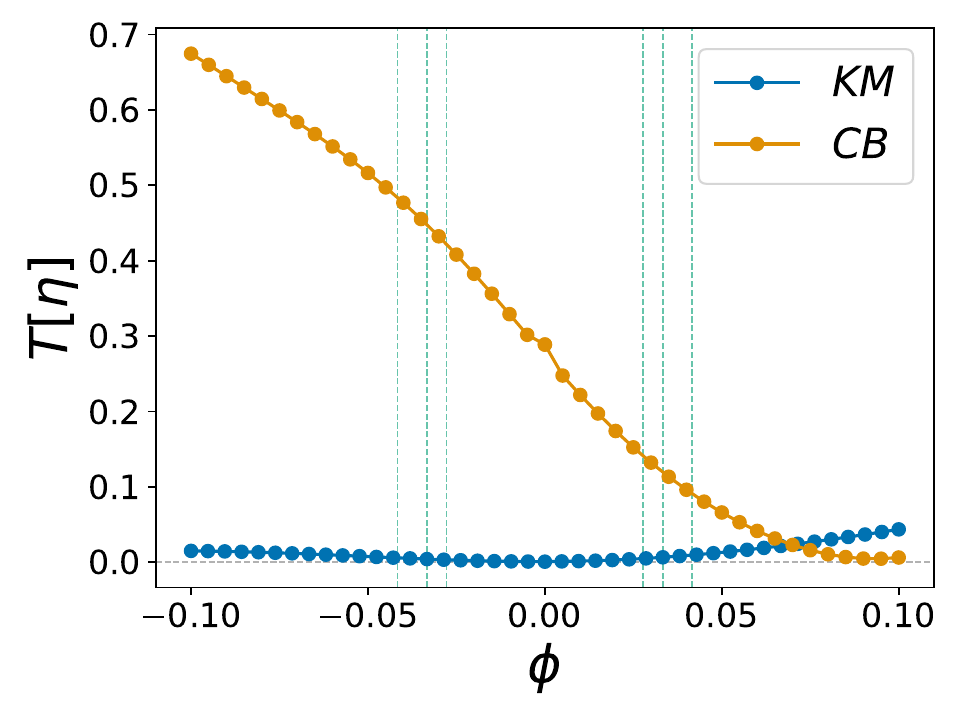}
    \caption{$T[\eta]$ for the Kapit-Mueller model (KM) and the checkerboard model (CB). The dashed vertical lines correspond to the values of $\phi$ used in the exact diagonalization.}
    \label{fig:MB3}
\end{figure}

On the other hand, the quantum geometry is shows characteristic behaviors depending on the field direction in the checkerboard model and hence is useful for understanding the stability of the quasiparticle creation, as discussed below.
The breakdown of the trace condition can be characterized by the following quantity introduced in the previous study~\cite{parker2021fieldtunedzerofieldfractionalchern}, 
\begin{equation}
    T[\eta] \coloneqq \text{Tr}[\text{tr}g]-\text{Tr}[\mathcal{F}].
    \label{trcond}
\end{equation}
We show $T[\eta]$ for the two models in Fig.~\ref{fig:MB3} which are calculated for the magnetic Brillouin zones of sufficiently large systems. 
In the Kapit-Mueller model, 
the trace condition is almost satisfied for all $|\phi|\leq 0.1$.
The band properties favor the multiband FCI under magnetic fields.
This is consistent with the stable quasiparticle creation by magnetic fields in the Kapit-Mueller model as seen in the exact diagonalization in the previous section.
Besides, the small $T[\eta]$ implies that the FCI phase is extended to a moderately large $|\phi|$.

In the checkerboard model, $T[\eta]$ decreases with the addition of positive magnetic fields $\phi>0$, whereas they increase for $\phi<0$. 
The multiband trace condition $T[\eta]=0$ is almost fully achieved around $\phi = 0.08\sim0.10$. 
This implies enhanced stability of the multiband FCI state under $\phi>0$, although the bandwidth $W$ increases to some extent.
On the other hand, $T[\eta]$ significantly increases for $\phi<0$.
These band properties suggest that the FCI at $\phi<0$ is less stable than that at $\phi>0$, which is consistent with the behaviors of the quasiparticle creation discussed in the previous section.
In this sense, the stability of the quasiparticle creation by magnetic fields can be understood based on the multiband quantum geometry.
Note that the relation between the quasiparticle creation and the multiband quantum geometry may not be valid for a clean system under strong magnetic fields, where there is a finite density of quasiparticles and interactions between them.
Stability of quasiparticles with a finite density would be an interesting future problem.

\section{\label{sec:level6}Summary}
In this work, we have discussed the stability of the quasiparticle creation in fractional Chern insulators under magnetic fields $\phi$. 
We considered two models with distinct quantum geometric properties, namely the Kapit -Mueller model and the checkerboard model. 
We demonstrated by the exact diagonalization that the Kapit-Mueller model with $\phi\neq0$ exhibits the stable quasiparticle creation. 
On the other hand, the FCI state in the checkerboard model is unstable under $\phi<0$ and the non-FCI state is realized even for moderate interactions, although the quasielectron creation is stable. 
The model dependence in the stability of the quasiparticles can be attributed to the difference in the multiband quantum geometry under the magnetic field. 
Therefore, multiband geometric structures together with the multiband bandwidth are essential ingredients to host stable quasiparticle creation.

\begin{acknowledgments}
We are grateful to Emil J. Bergholtz, Nobuyuki Okuma, and Manato Fujimoto for fruitful discussions. 
This work is supported by JSPS KAKENHI Grant No. 22K03513.
\end{acknowledgments}

\appendix

\section{quasiparticle creation by increasing unit cells}
\label{app:unitcell}
\begin{figure}[htbp]
    \centering
    \begin{subfigure}{0.49\linewidth}
        \centering
        \textbf{(a) $N_s=32,N=5,U=1.0$}
        \includegraphics[width=\linewidth]{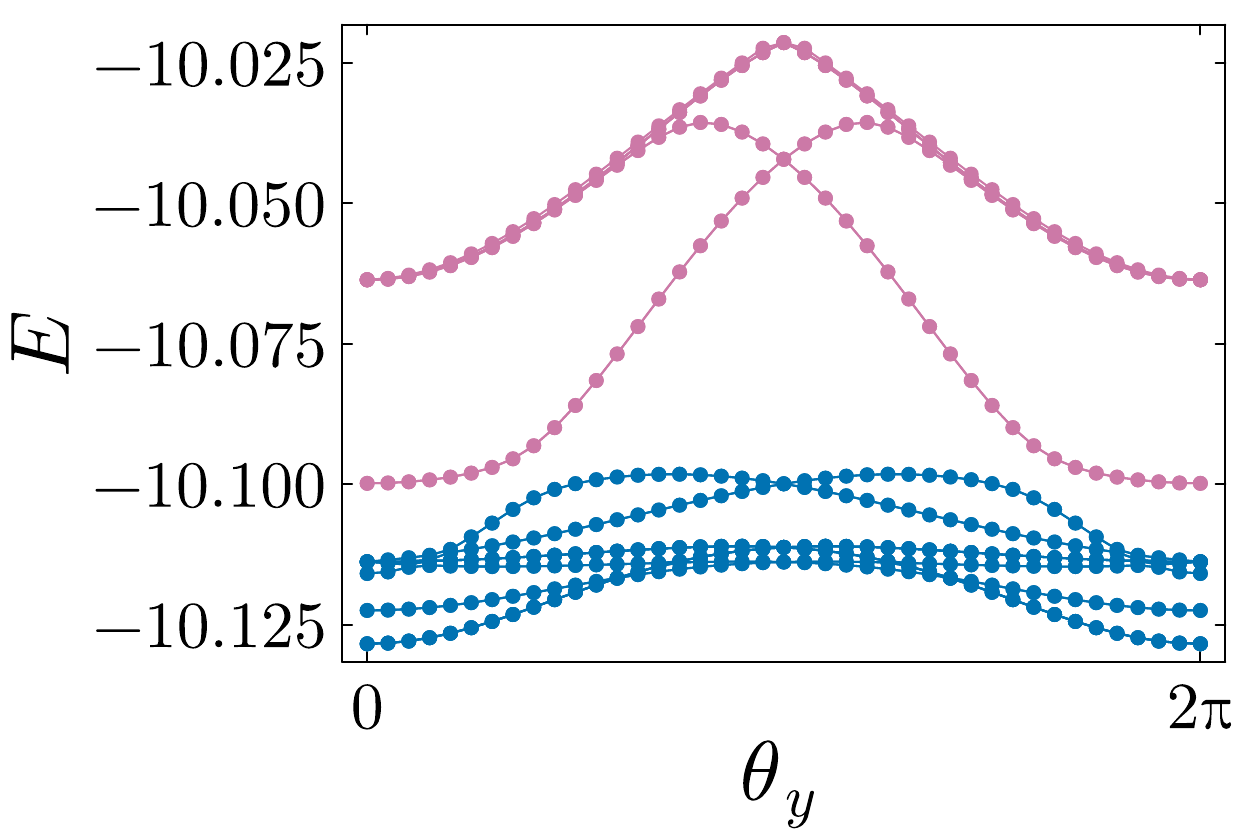}
    \end{subfigure}
    \begin{subfigure}{0.49\linewidth}
        \centering
        \textbf{(b) $V_{p}=1.0$}
        \includegraphics[width=\linewidth]{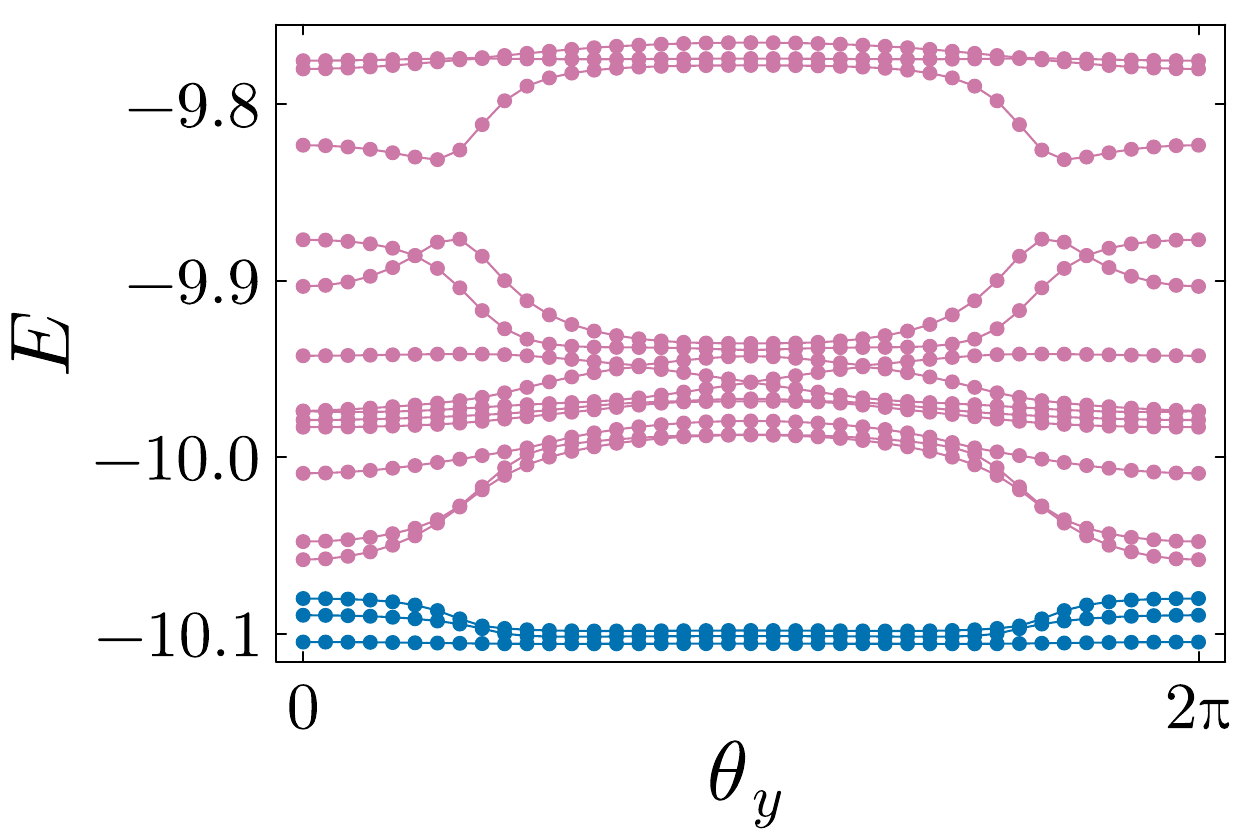}
    \end{subfigure}\\
    \begin{subfigure}{0.60\linewidth}
        \centering
        \textbf{(c)}
        \includegraphics[width=\linewidth]{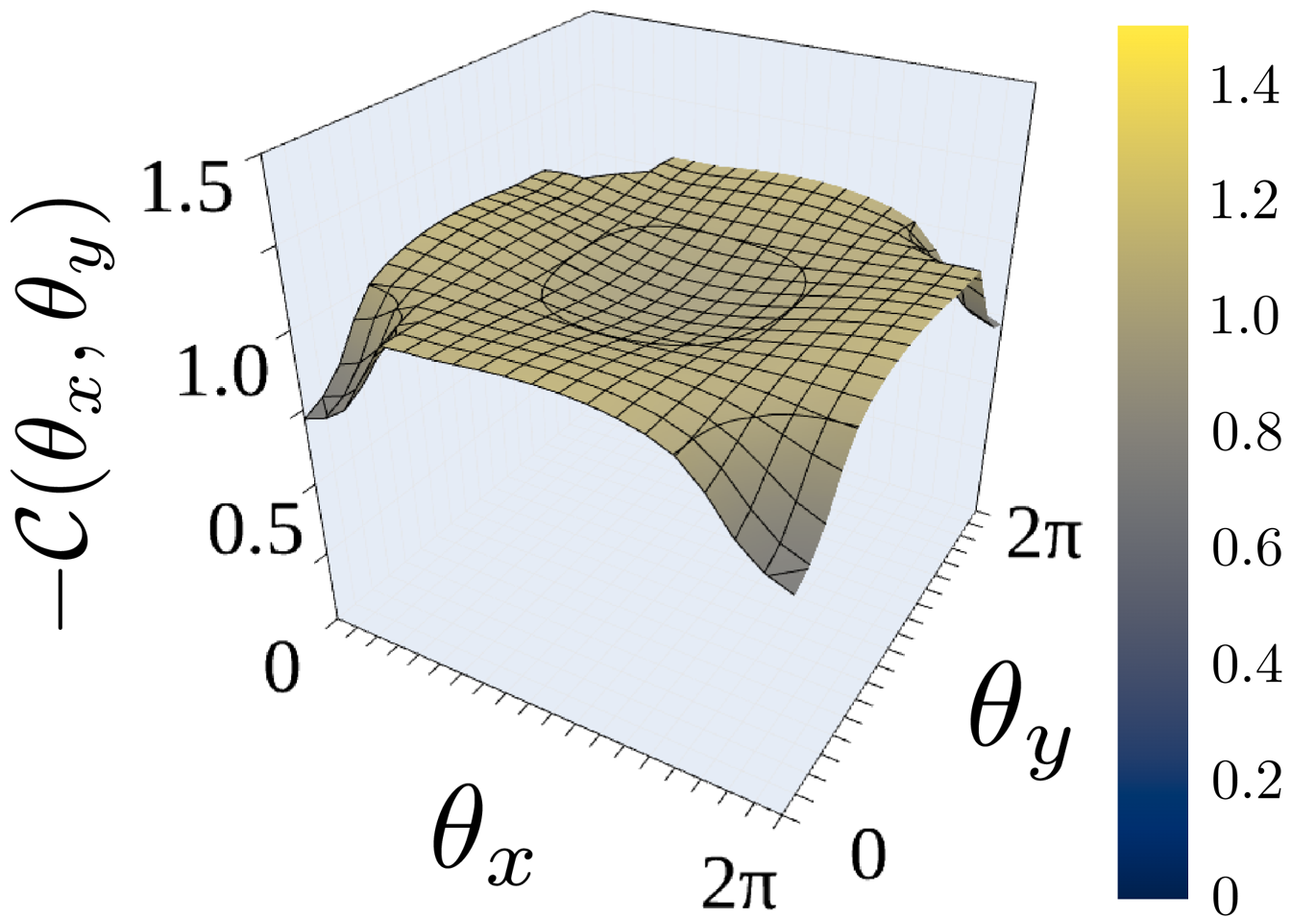}
    \end{subfigure}
    \caption{Spectral flow in the checkerboard model with $N=5,N_s=32,U=1.0$ at (a) $V_p=0$ and (b) $V_p=1$. Blue curves represent (a) the counting-rule $\#_{\text{qh}}=16$, and (b) the lowest three energy levels. (c) The one-plaquette Chern number for the three lowest energy states in (b). }
    \label{fig:AppendixA}
\end{figure}
Here, we quickly discuss the quasiparticle creation in the checkerboard model at $\phi=0$ caused by a change of the system size, where the band properties are kept intact. 
We consider the system with $N=5,U=1.0$ and increase one unit cell as $N_s=30\to 32$.
The spectral flow is shown in Fig.\ref{fig:AppendixA}.
The counting rule at $V_p=0$ is $\#_{\text{qh}}=16$ and they lead to the three fold quasi-degenerate ground state at $V_p=1$.
The one-plaquette Chern number in the parameter space is smooth and flat as shown in Fig.\ref{fig:AppendixA} (c), along with the many-body Chern number $C=-1$. 
These results suggest that finite size effects for the quasiparticle creation are negligible even for the small system size $N=5$.

In addition, we have also confirmed creation of the quasiparticles in the Kapit-Muller model with $N=4$ by increasing and decreasing one unit cell. 
It is found that the creation of the quasiparticles against magnetic fields is stable and the finite size effects are safely negligible.

\section{Spectral flow in the checkerboard model with $N=4,5$}
\label{app:cb45}
\begin{figure}[h]
    \centering
    \begin{subfigure}{0.49\linewidth}
        \centering
         \textbf{(a) $N=5,U=2.0$}
        \includegraphics[width=\linewidth]{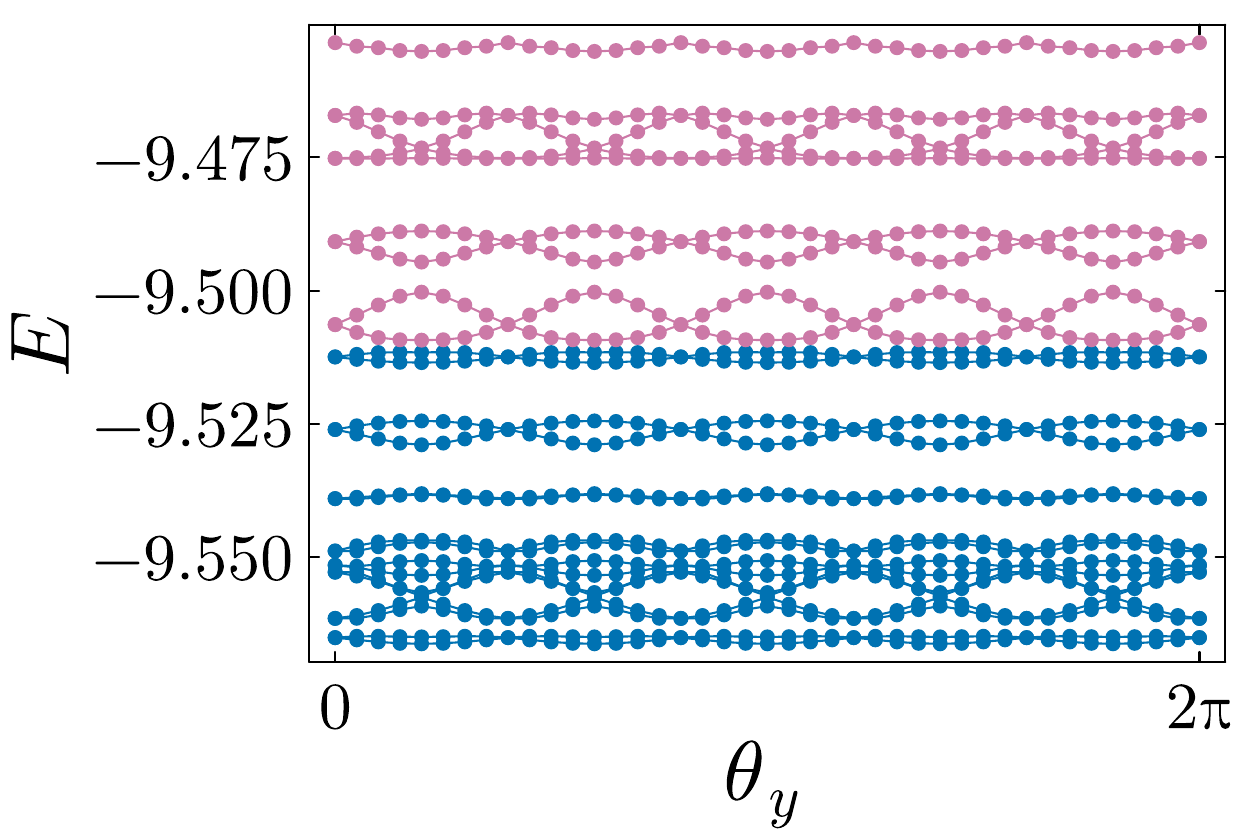}
    \end{subfigure}
    \begin{subfigure}{0.49\linewidth}
        \centering
         \textbf{(b) $V_{p}=10.0$}
        \includegraphics[width=\linewidth]{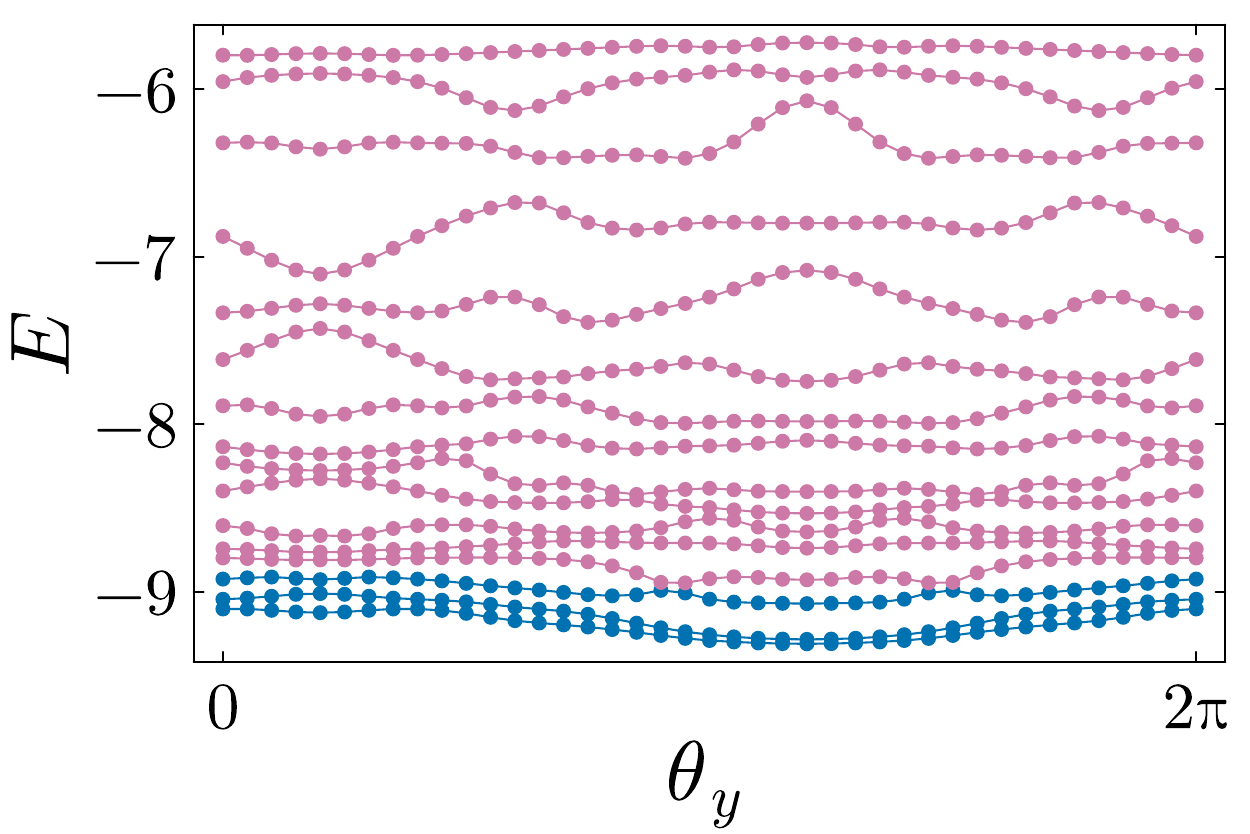}
    \end{subfigure}\\
    \begin{subfigure}{0.49\linewidth}
        \centering
         \textbf{(c) $N=5,U=3.0$}
        \includegraphics[width=\linewidth]{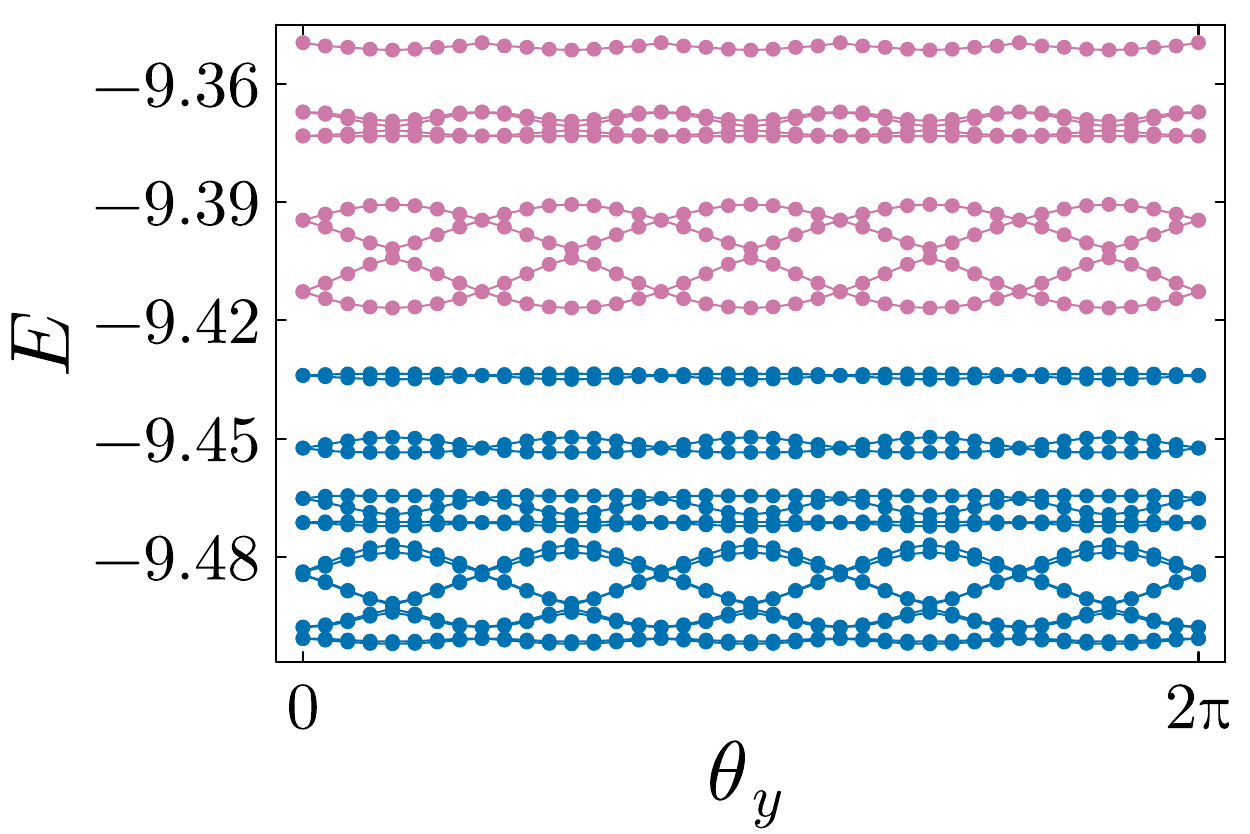}
    \end{subfigure}
    \begin{subfigure}{0.49\linewidth}
        \centering
          \textbf{(d) $V_{p}=10.0$}
        \includegraphics[width=\linewidth]{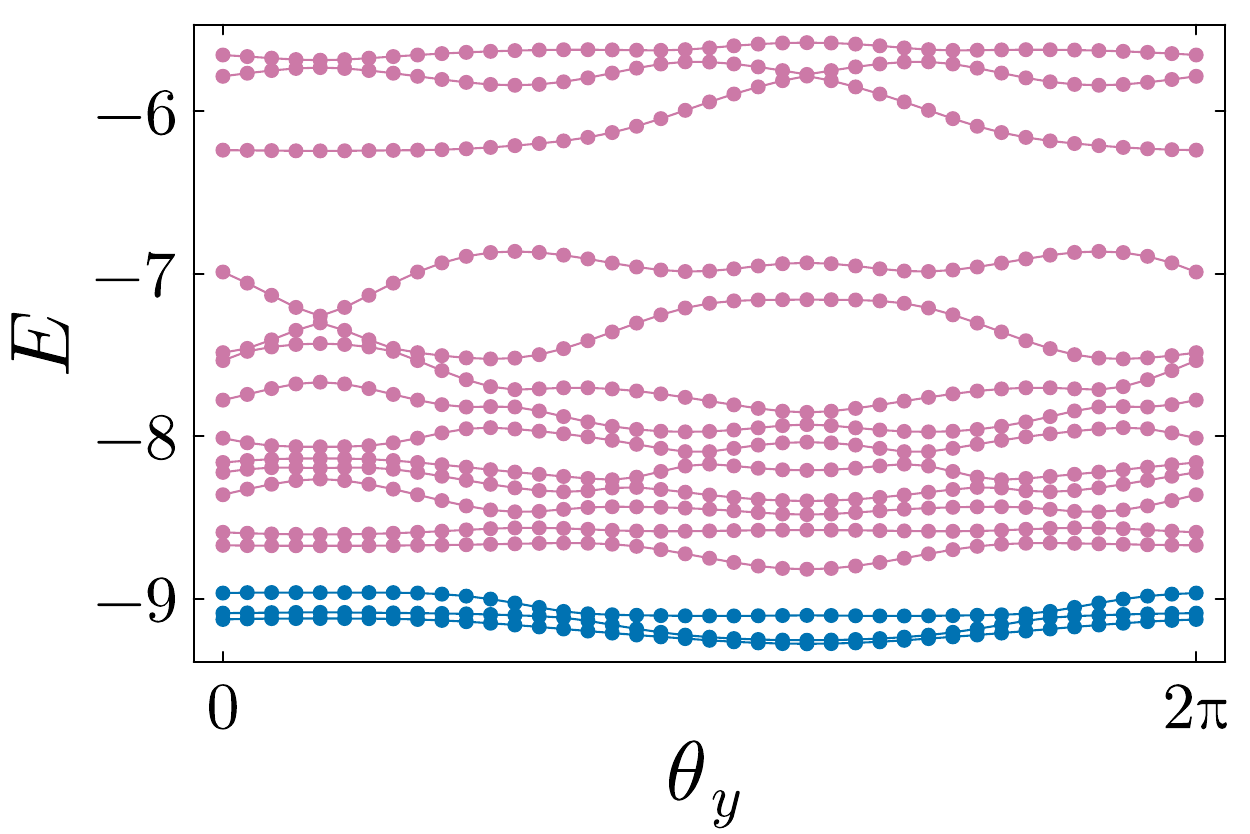}
    \end{subfigure}\\
    \begin{subfigure}{0.49\linewidth}
        \centering
         \textbf{(e) $N=4,U=3.0$}
        \includegraphics[width=\linewidth]{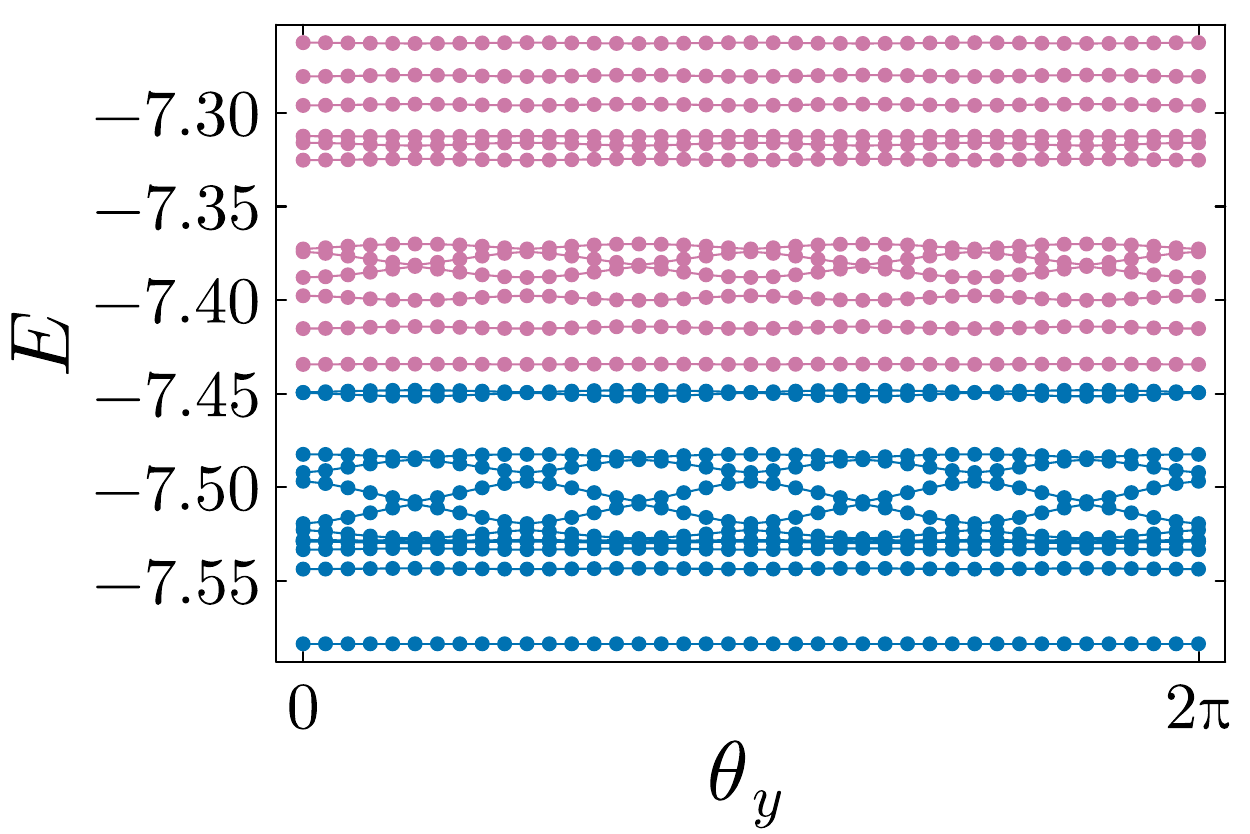}
    \end{subfigure}
    \begin{subfigure}{0.49\linewidth}
        \centering
          \textbf{(f) $V_{p}=10.0$}
        \includegraphics[width=\linewidth]{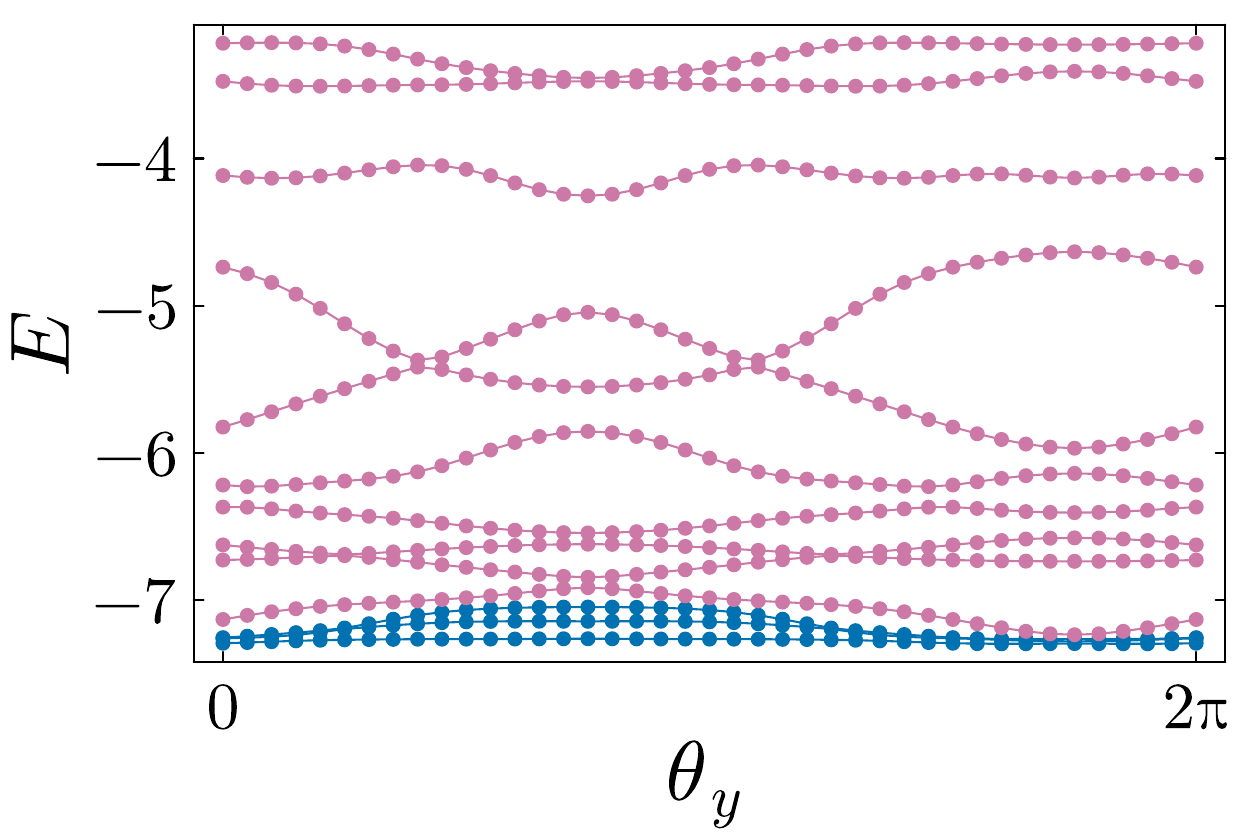}
    \end{subfigure}\\
    \begin{subfigure}{0.49\linewidth}
        \centering
         \textbf{(g) $N=4,U=8.0$}
        \includegraphics[width=\linewidth]{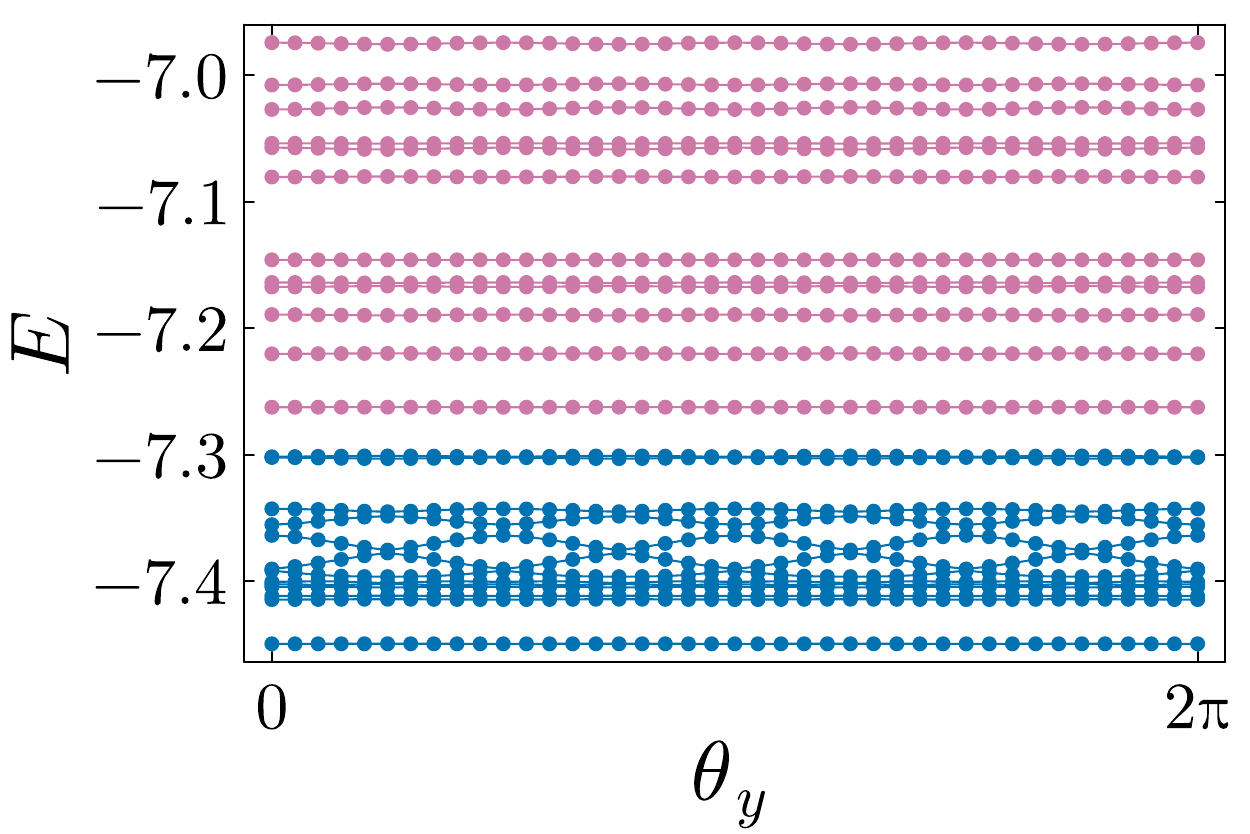}
    \end{subfigure}
    \begin{subfigure}{0.49\linewidth}
        \centering
          \textbf{(h) $V_{p}=10.0$}
        \includegraphics[width=\linewidth]{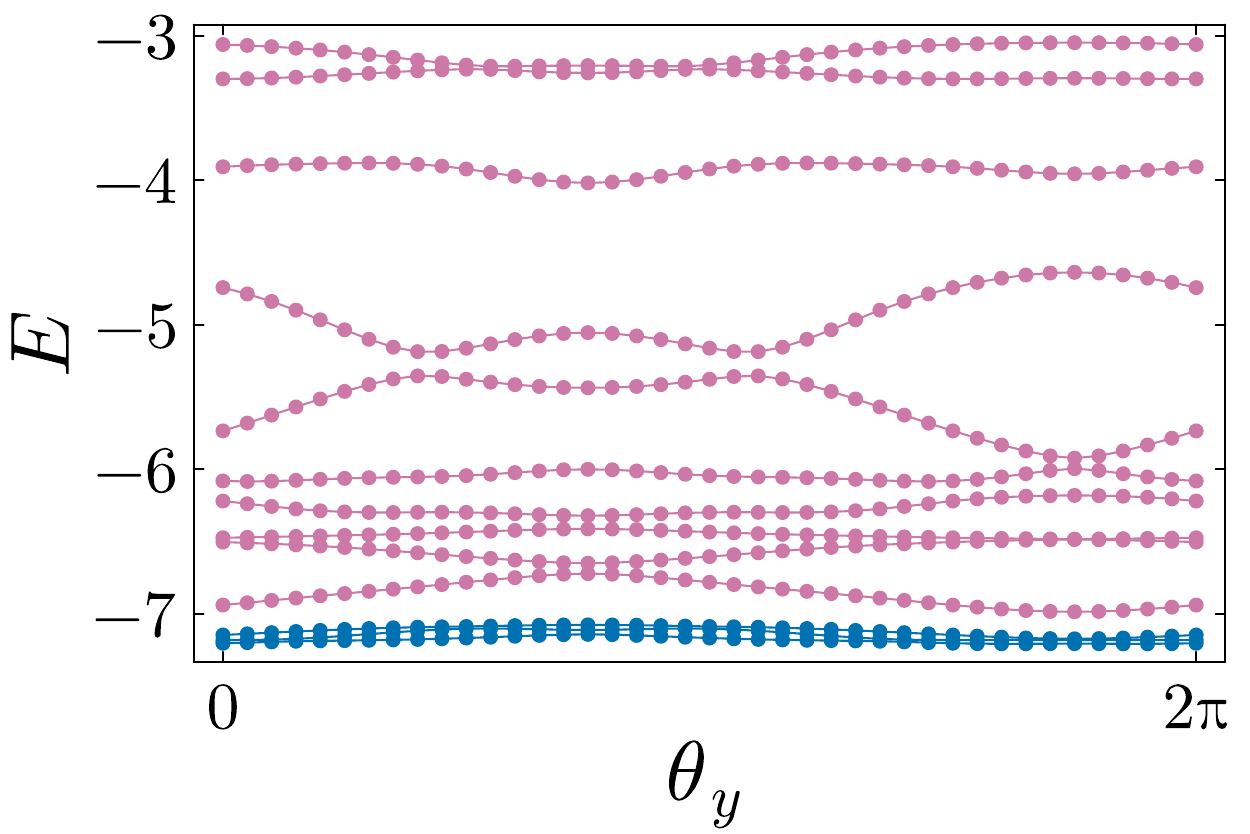}
    \end{subfigure}\\
    \caption{Spectral flow in the checkerboard model for the system under $N_{\phi}=-1$. 
    The system sizes are (a)$\sim$(d)($N_s=30$,$N=5$,$\#_{\text{qh}}=16$) and (e)$\sim$(h)($N_s=24$,$N=4$,$\#_{\text{qh}}=13$). 
    The blue curves for $V_p=0$ are the states expected from the counting rule, and those for $V_p=10$ are the three lowest energy levels
    as in Fig.\ref{fig:KSSF1}. }
    \label{fig:KSSF2}
\end{figure}
We show the spectral flows for $N=4,5$ in $N_{\phi}=-1$ in the checkerboard model in Fig.~\ref{fig:KSSF2}. 
For $N=5$, the counting rule $\#_{\text{qh}}=16$ does not hold at $U=0\sim2.0$ because there is no clear gap between the low energy 16 states and other excited states (Fig.~\ref{fig:KSSF2}(a)). 
The three fold quasi-degeneracy also does not hold when the pinning potential $V_p$ is introduced in Fig.~\ref{fig:KSSF2}(b), which indicates that the quasihole cannot be created and a non-FCI state is formed. 
The phase transition to the FCI with the pinned quasihole occurs around $U=2.0\sim3.0$. 
The counting-rule and the three fold quasi-degeneracy hold under $U\ge3.0$ with the clear energy gap as seen in Fig~\ref{fig:KSSF2}(c),(d), which is consistent with the result of the variance of the one-plaquette Chern number. 
Similarly for $N=4$, energy gaps above low energy states develop between $U=3.0\sim 8.0$ (Fig.~\ref{fig:KSSF2}(e)$\sim$(h)) and the variance $\mathcal{V}[\mathcal{C}]$ is small for $U\ge 5.0$. 

\begin{figure}[b]
    \centering
    \begin{subfigure}{0.49\linewidth}
        \centering
         \textbf{(a) $N=5,U=1.0$}
        \includegraphics[width=\linewidth]{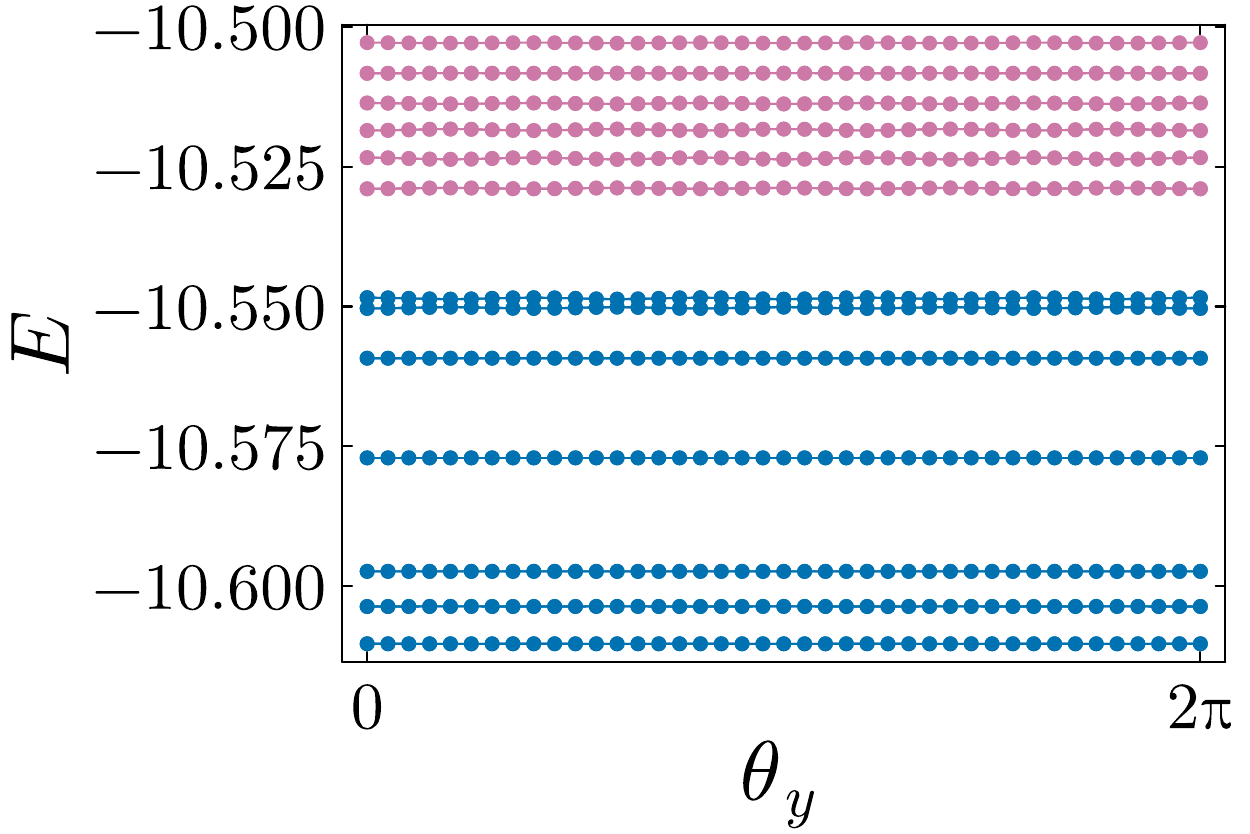}
    \end{subfigure}
    \begin{subfigure}{0.49\linewidth}
        \centering
         \textbf{(b) $V_{p}=-1.0$}
        \includegraphics[width=\linewidth]{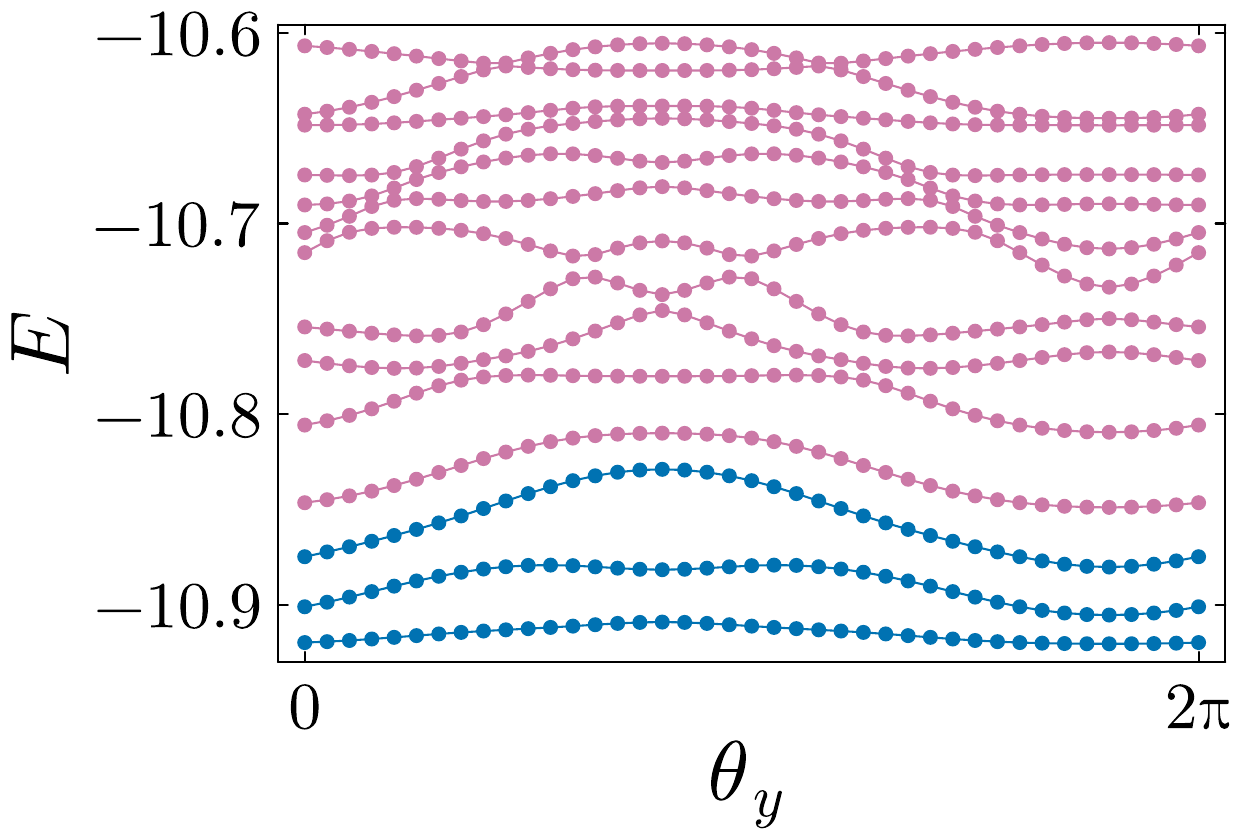}
    \end{subfigure}\\
    \begin{subfigure}{0.49\linewidth}
        \centering
         \textbf{(c) $N=5,U=2.0$}
        \includegraphics[width=\linewidth]{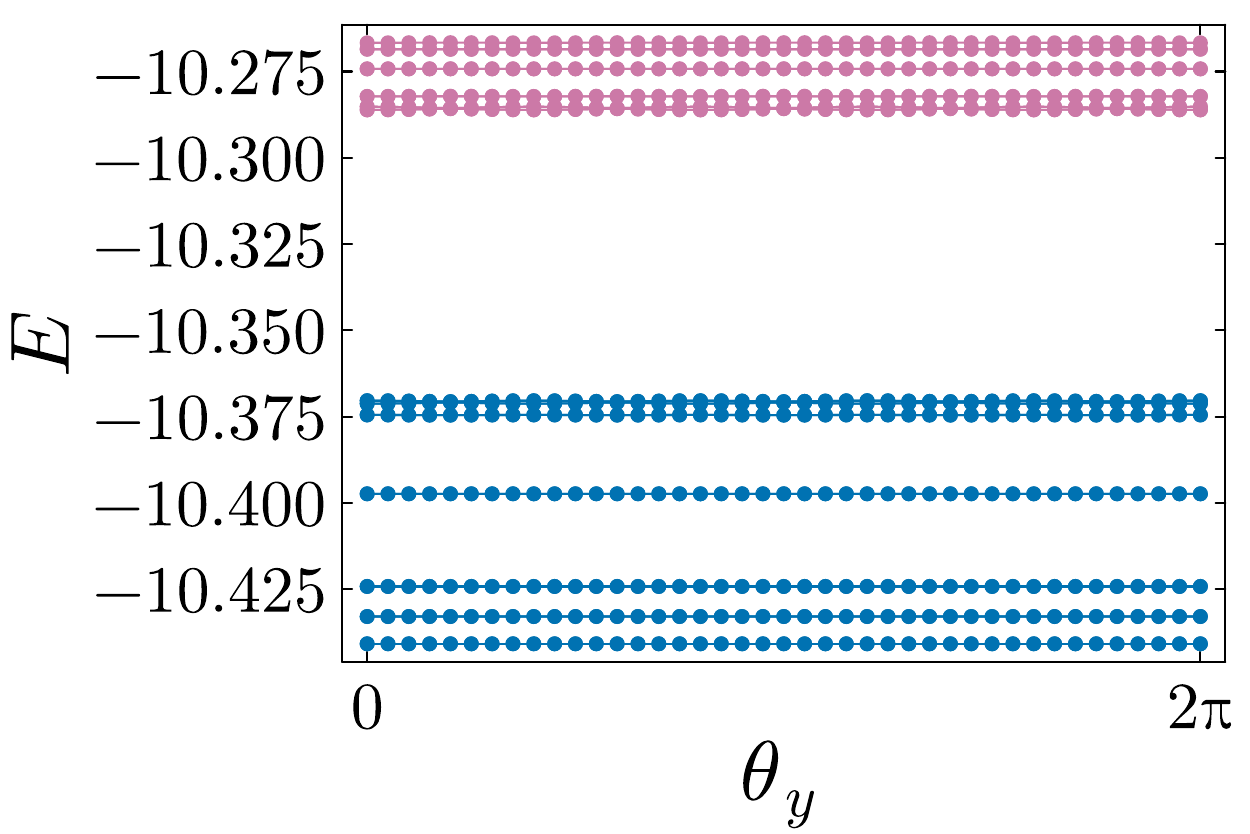}
    \end{subfigure}
    \begin{subfigure}{0.49\linewidth}
        \centering
         \textbf{(d) $V_{p}=-1.0$}
        \includegraphics[width=\linewidth]{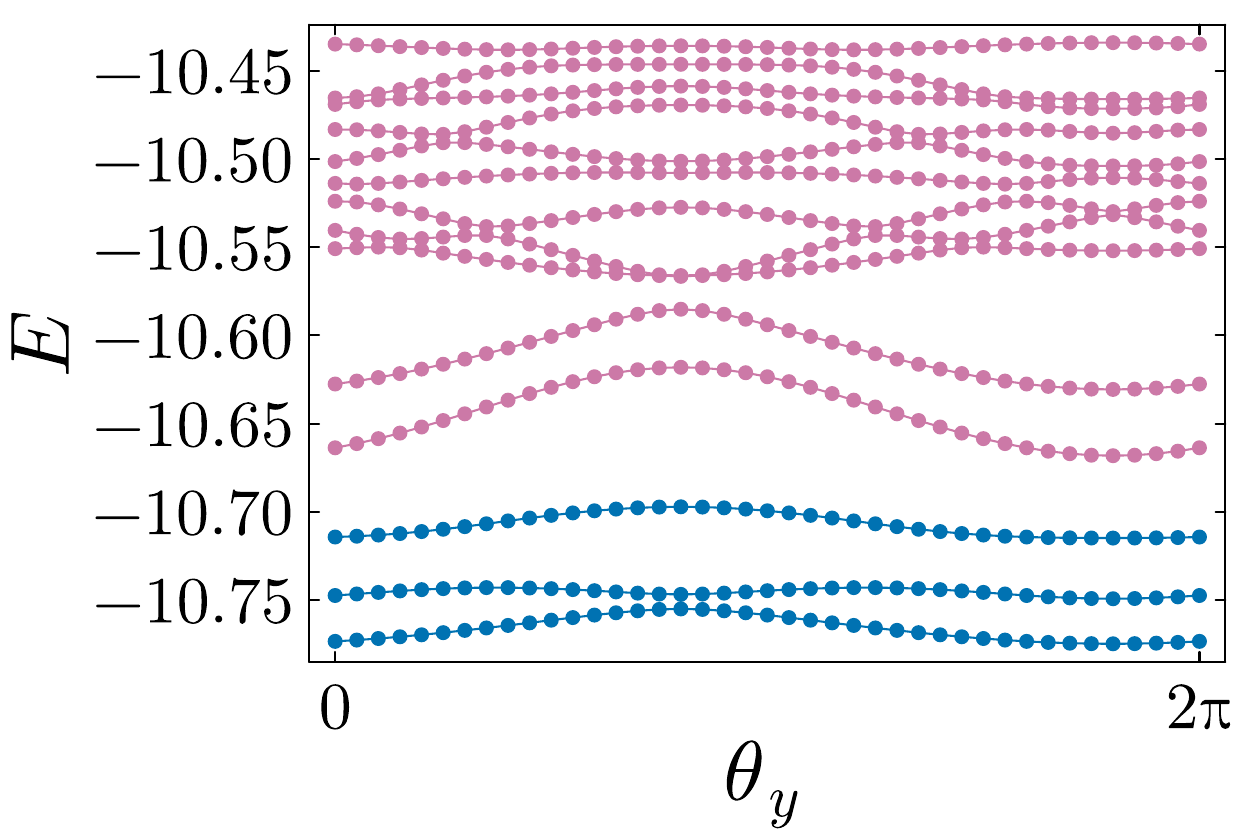}
    \end{subfigure}\\
    \begin{subfigure}{0.49\linewidth}
        \centering
         \textbf{(e) $N=4,U=1.0$}
        \includegraphics[width=\linewidth]{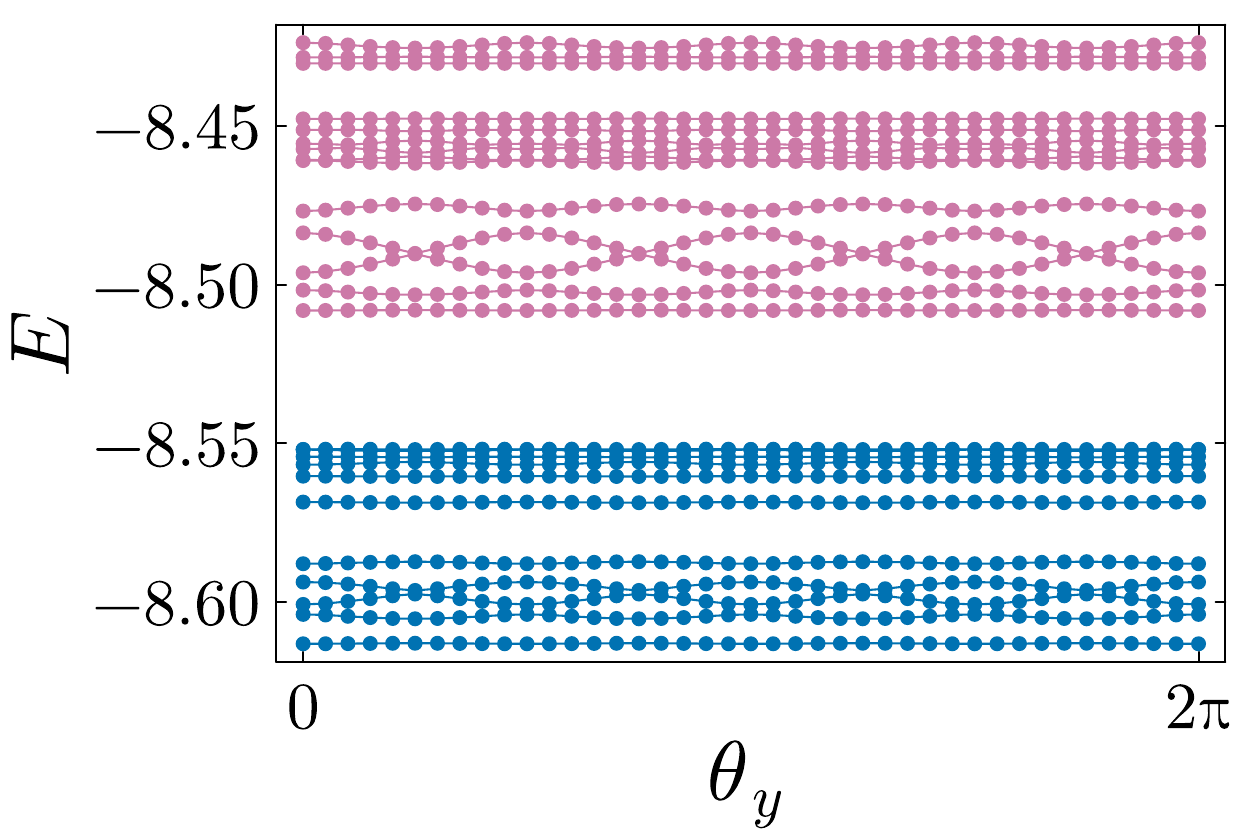}
    \end{subfigure}
    \begin{subfigure}{0.49\linewidth}
        \centering
         \textbf{(f) $V_{p}=-1.0$}
        \includegraphics[width=\linewidth]{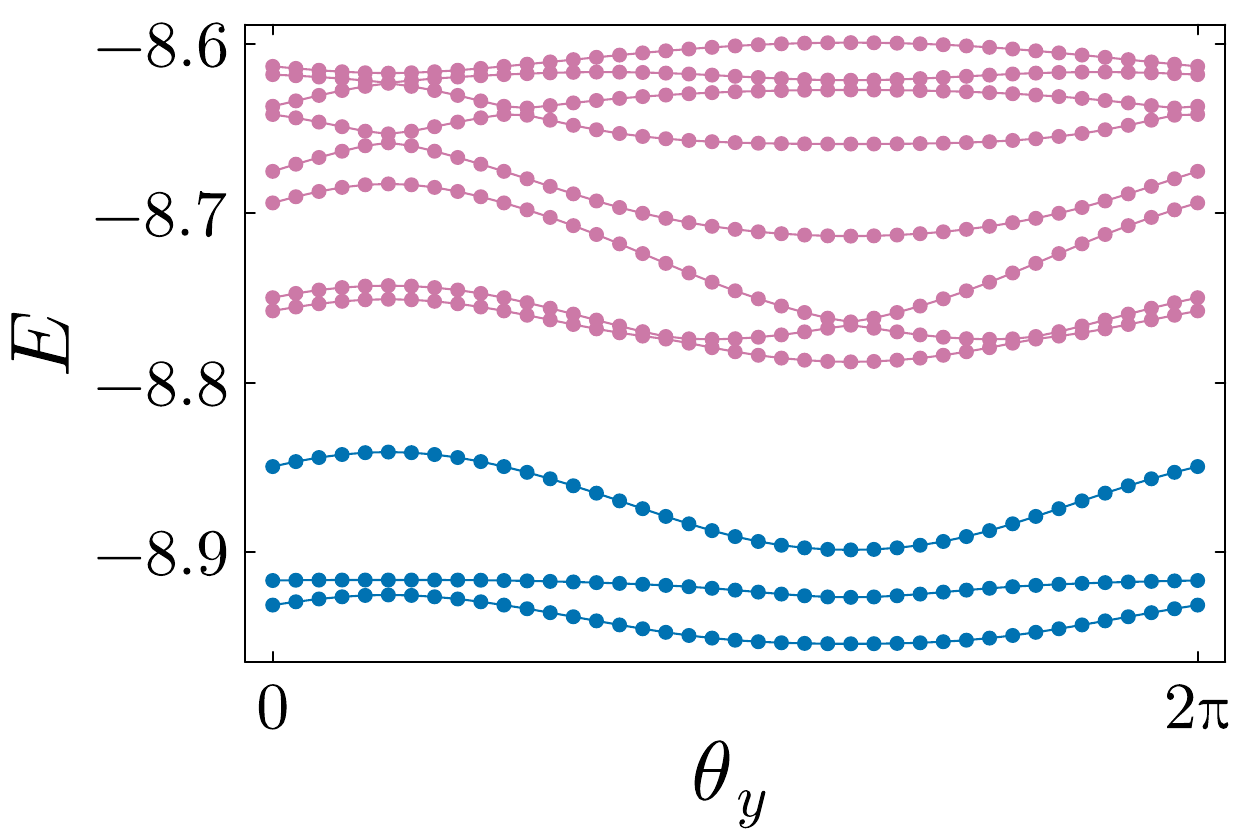}
    \end{subfigure}\\
    \begin{subfigure}{0.49\linewidth}
        \centering
         \textbf{(g) $N=4,U=2.0$}
        \includegraphics[width=\linewidth]{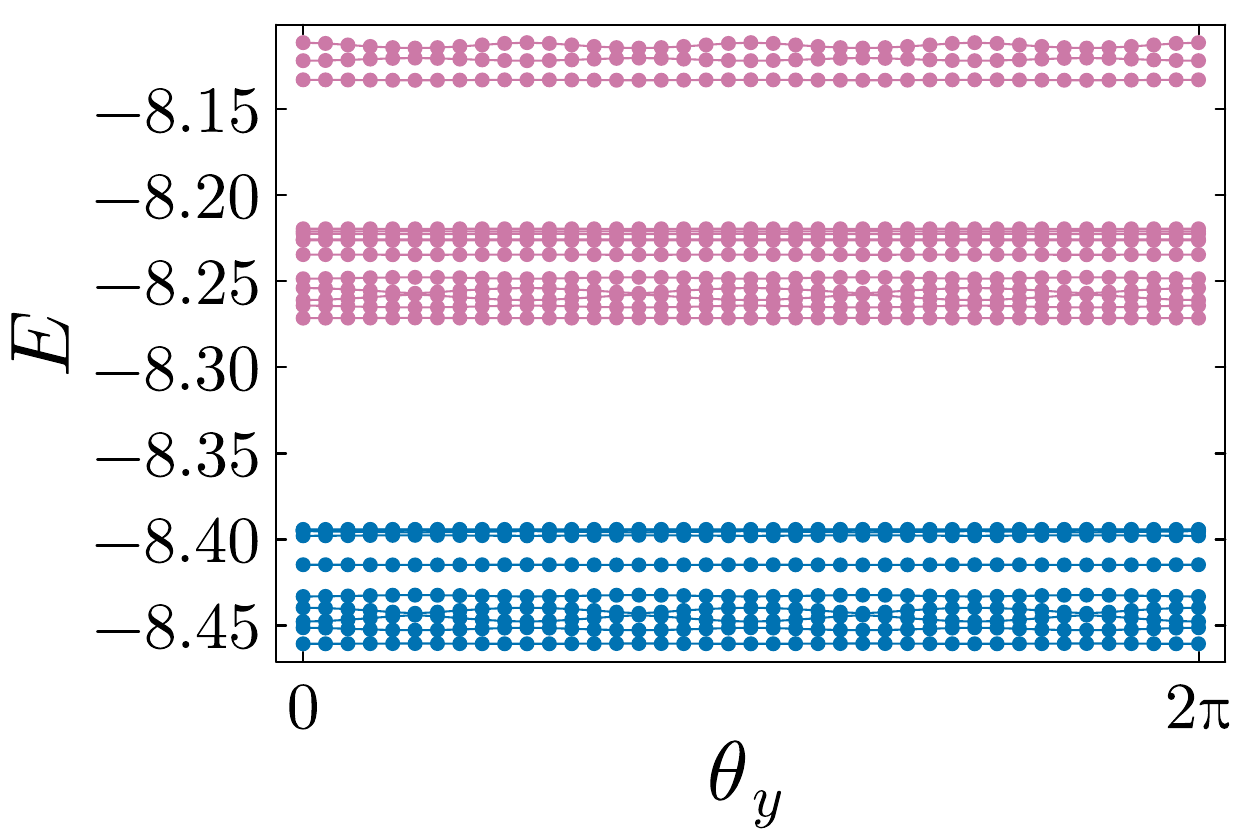}
    \end{subfigure}
    \begin{subfigure}{0.49\linewidth}
        \centering
         \textbf{(h) $V_{p}=-1.0$}
        \includegraphics[width=\linewidth]{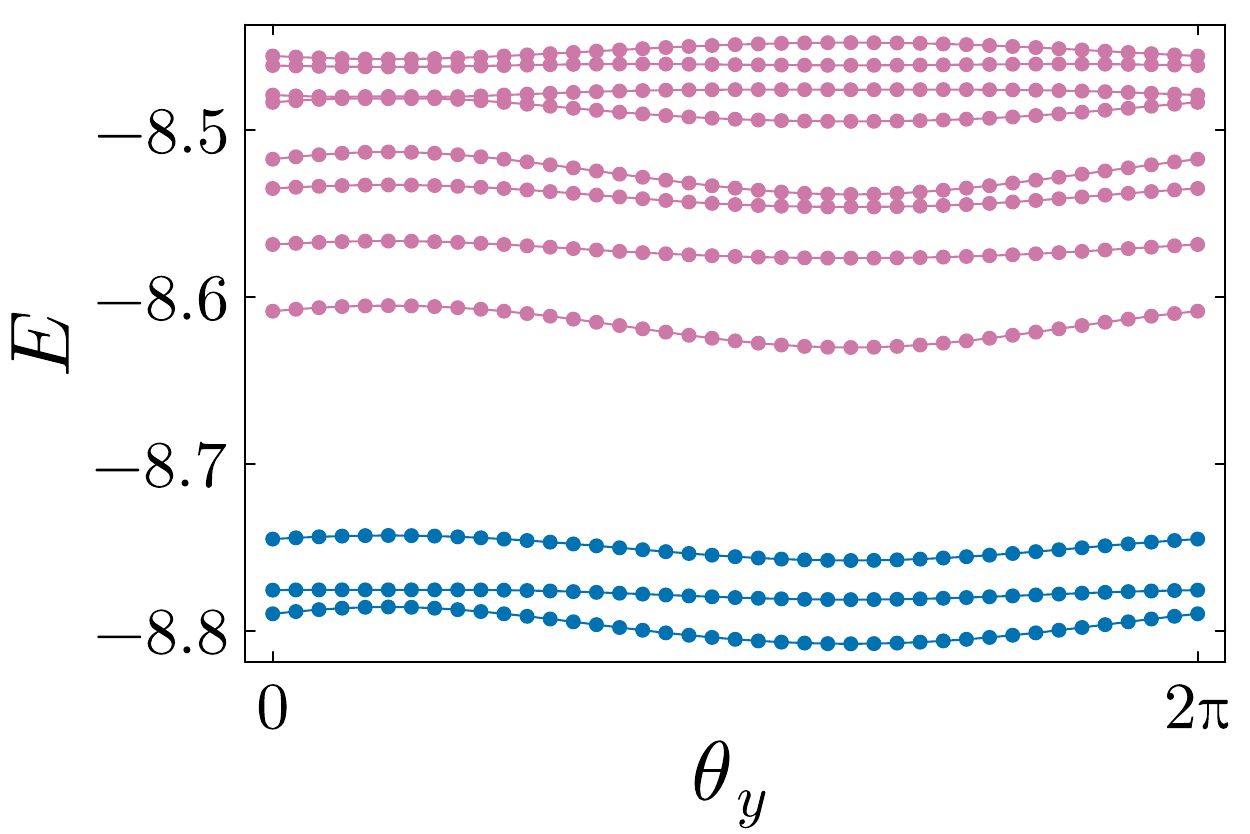}
    \end{subfigure}\\
    \caption{Spectral flow in the checkerboard model for the system under $N_{\phi}=1$. The system sizes are (a)$\sim$(d)$(N_s=30,N=5,\#_{\text{qe}}=14)$, and (e)$\sim$(f)$(N_s=24,N=4,\#_{\text{qe}}=11)$. 
    The blue curves for $V_p=0$ are the states obeying the counting-rule, and those for $V_p=-1$ are the three lowest energy levels.}
    \label{fig:KSSF4}
\end{figure}

For $N_{\phi}=1$, energy gaps above low energy states open even at $U=1.0$ for both $N=4,5$ as shown in Fig.~\ref{fig:KSSF4}. 
The energy gap develops and the FCI with the pinned quasielectron becomes more stable as the interaction increases. 
The stabilization of the FCI is seen also in the variance $\mathcal{V}[\mathcal{C}]$. 
We concluded that the quasielectron creation is stable in a wide range of the interaction for $N=4,5$.
{Note that the stability of the quasielectron creation only weakly depends on the strength of the pinning potential, and we have confirmed that the results are basically unchanged for $V_p=-0.1,-1,-10$. Essentially same results can also be obtained without the projection.}

\section{Flatness of Berry curvature}
\label{app:flatness}

\begin{figure}[ht]
    \centering
        \includegraphics[width=0.99\linewidth]{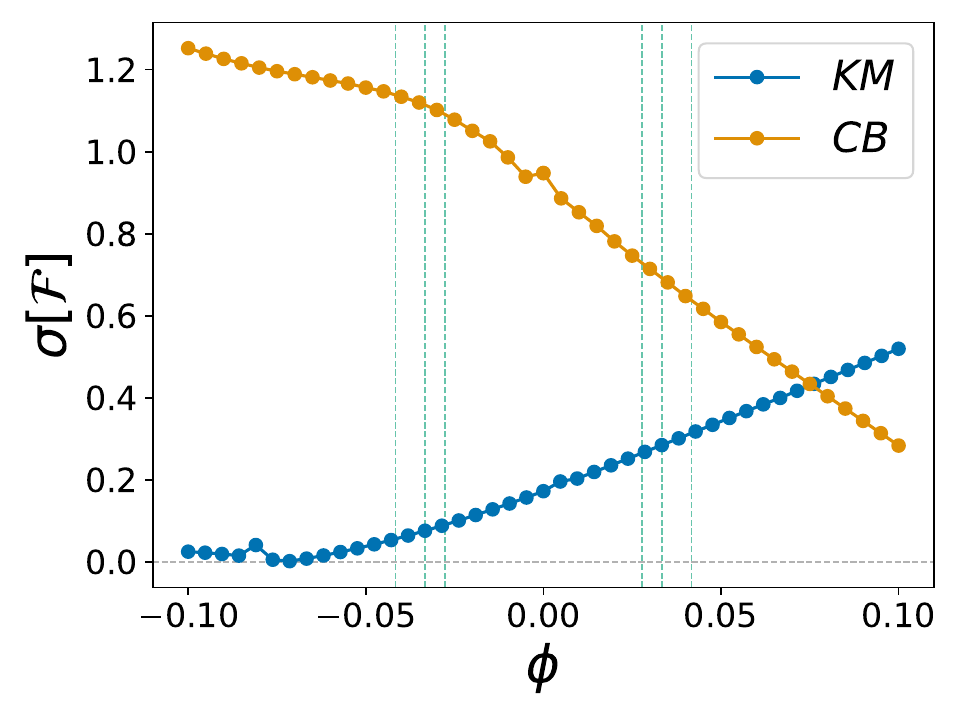}
    \caption{$\sigma[\mathcal{F}]$ for the Kapit-Mueller model (KM) and the checkerboard model (CB). The dashed vertical lines correspond to the values of $\phi$ used in the exact diagonalization.}
    \label{fig:MB4}
\end{figure}
We discuss another multiband geometric property under magnetic fields, namely uniformity of the multiband Berry curvature. 
The uniformity is a condition for the emergent Girvin-MacDonald-Platzman (GMP) algebra~\cite{PhysRevB.33.2481,parker2021fieldtunedzerofieldfractionalchern}. 
The precise condition for the GMP algebra is given by
\begin{equation}
        \mathcal{F} = 2\pi C_{\text{s}} \text{Id}_{N_{\text{b}}},\quad \text{for }\bm{k} \in \text{BZ}, 
\end{equation}
where $\text{Id}_{N_{\text{b}}}$ is the $N_{\text{b}}$ dimensional identity matrix. 
The multiband trace condition~(Eq.~\eqref{trcond}) is also necessary for the GMP algebra.
Relation of the GMP algebra to stability of an FCI is still under debate~\cite{10.21468/SciPostPhys.12.4.118}.

The breakdown of the GMP algebra for a non-uniform Berry curvature can be characterized by the following quantity introduced in the previous studies~\cite{parker2021fieldtunedzerofieldfractionalchern}, 
\begin{equation}
    \sigma[\mathcal{F}] \coloneqq  \text{Tr}\qty[\qty(\frac{1}{2\pi C_{s}}{\mathcal{F}}-\text{Id}_{{N_{\text{b}}}})^{2}]^{1/2},
\end{equation}
which is the Frobenius norm for the Hermitian matrix $\mathcal{F}/(2\pi C_s)-\text{Id}_{{N_{\text{b}}}}$.
We show $\sigma[\mathcal{F}]$ for the two models in Fig.~\ref{fig:MB4}.
$\sigma[\mathcal{F}]$ in the Kapit-Muller model is enhanced for positive magnetic fields. 
However, the breakdown of the GMP algebra does not affect the momentum-space holomorphicity of the Bloch function, because the holomorphicity is guaranteed by the trace condition.
In the checkerboard model, $\sigma[\mathcal{F}]$ decreases with the addition of positive magnetic fields $\phi>0$, whereas they increase with $\phi<0$. 
This is similar to $T[\eta]$ under magnetic fields discussed in the main text.

\newpage
\nocite{*}
\bibliography{letter}

\end{document}